# Limiting Phase Trajectories: a new paradigm for the study of highly non-stationary processes in Nonlinear Physics

Leonid I. Manevitch[a], Agnessa Kovaleva[b1], Yuli Starosvetsky[c]

[a]*Institute of Chemical Physics, Russian Academy of Sciences, Moscow 117997, Russia*
[b]*Space Research Institute, Russian Academy of Sciences, Moscow 119991, Russia*
[c]*Technion -Israel Institute of Technology,Technion City, Haifa 32000, Israel*

This Report discusses a recently developed concept of *Limiting Phase Trajectories* (LPTs) providing a unified description of *resonant energy transport* in a wide range of classical and quantum dynamical systems with constant and time-varying parameters. It is shown that strongly modulated non-stationary processes occurring in a nonlinear oscillator array under certain initial conditions may characterize either maximum possible energy exchange between the clusters of oscillators (effective particles) or maximum energy transfer from an external source of energy to the system. The trajectories corresponding to these processes are referred to as LPTs. The development and the use of the LPT concept are motivated by the fact that highly non-stationary resonance processes occurring in a broad variety of finite-dimensional physical models are beyond the well-known paradigm of the nonlinear normal modes (NNMs), fully justified only for stationary processes, their stability and bifurcations, as well as for non-stationary processes described approximately by some combinations of non-resonating normal modes. Thus, the role of LPTs in understanding and description of non-stationary energy transfer is similar to the role of NNMs for the stationary processes. Several applications of the LPT concept to significant nonlinear problems and a scenario of the transition from regular to chaotic behavior with the LPTs implication are presented. In order to highlight the novelty and perspectives of the developed approach, we place the LPT concept into the context of complex dynamical phenomena related to energy transfer problems.

[*]Corresponding author. Tel.:+7 916 9824740; fax: +7 495 913 3040.
*E-mail address:*agnessa_kovaleva@hotmail.com (A. Kovaleva).

# Content

# 1. Introduction

This Report presents the concept of *Limiting Phase Trajectories* (LPTs), depicts their intrinsic properties, and demonstrates the role of LPTs in intense resonance energy transfer in physical systems of different nature.

By definition, motion along the LPT corresponds either to maximum possible energy exchange between the clusters of oscillators (effective particles) or to maximum energy transfer from an external source of energy to the system. In both cases, the system under consideration exhibits strongly modulated non-stationary processes. The derivation of explicit analytical solutions describing non-stationary dynamics of nonlinear systems is, in general, a daunting task. The well-developed techniques of *Nonlinear Normal Modes* (NNMs), which is effective in the analysis of stationary or non-stationary non-resonant oscillations, cannot be applied to the study of highly non-stationary resonant energy exchange between nonlinear oscillators or clusters of oscillators (see Sections 1.1, 7.2). Recent authors' works have resolved this difficulty by introducing the notion of *Limiting Phase Trajectory*, which is an alternative to *Nonlinear Normal Mode*. Convenient asymptotic procedures for constructing explicit approximations of the LPTs have been suggested.

In this introductory part we briefly describe main approaches to the study of nonlinear interactions between oscillators and present an example, which gives a preliminary idea about the LPTs, their properties and advantages in the study of resonance energy transport. A more detailed study of LPTs in systems with one and two degrees of freedom is postponed until Sections 2 and 3. Sections 4 - 7 demonstrate extensions and applications of the LPT concept to the analysis of resonance energy transport in physical systems of different nature, and the role of LPTs in the transition from regular to chaotic behavior. A comprehensive exposition of the reported results can be found in the works cited in this review.

## 1.1. Nonlinear normal modes in finite oscillatory chains

Non-stationary resonance processes are of the great interest in various fields of modern physics and engineering dealing with energy transfer and exchange. However, their particularities have not been fully recognized for a long time because of the domination of the common idea about the possible representation of any non-stationary process as a combination of stationary oscillations or waves. Actually, in the linear theory the superposition principle implies the representation of every dynamical process as a sum of normal vibrations or normal waves. This representation is approximately valid even in the nonlinear case *in the absence of the intermodal resonance*. Thus, main conventional analytical tools for the study of stationary and non-stationary nonlinear processes, their stability and bifurcations in *finite* dimensional systems have been associated with the NNM concept. Rosenberg [157-159] defined a *NNM* of an undamped discrete multi-particle system as a synchronous periodic oscillation wherein the displacements of all material points of the system reach their extreme values and pass through zero simultaneously. Then the NNM concept was extended to *finite* continuum models [202], to systems subjected to external forcing (e.g., [116, 202] and references therein), as well as to auto-generators, for which NNM can be considered as an attractor corresponding to synchronized motion of oscillators [152]. An important feature that distinguishes NNMs from their linear counterpart, besides an amplitude dependence of the frequency, is that they can exceed in number the degrees of freedom of an oscillatory system. In this case, essentially nonlinear localized NNMs are generated through NNM bifurcations, breaking the symmetry of the dynamics and resulting in the nonlinear energy localization phenomenon (*stationary energy localization*).

The mechanism of the formation of the localized NNMs is directly related to the existence of asymmetry caused by nonlinearity in the conditions of internal resonance. It is well known that spatially localized excitations are considerably important in different fields of nonlinear physics and determine elementary mechanisms of many physical processes. Representative examples are the localized normal modes existing in molecules and polymer crystals.

## 1.2. Solitonic excitations in the infinite models

Over last decades, the significant attention has also been paid to spatially localized excitations studied in the framework of the soliton theory [36, 37,120,190]. This theory was predominantly developed in applications to *infinite continuous* models described by nonlinear partial differential equations (PDE), and then it was extended to *infinite discrete models* (e.g., [1, 190]). The existence of the envelope solitons (breathers) with internal oscillatory degrees of freedom allows one to find a connection with the theory of localized NNMs [202].

One of the widely used models of nonlinear physics is the Fermi-Pasta-Ulam (FPU) oscillator chain [41] that became a starting point for the discovery of solitons. Main results in this field can be found, e.g., in [20, 24, 36, 41, 43, 49, 60, 88, 102, 150, 176, 186, 208]. It is well known that in the long-wave limit the dynamics of the *infinite FPU discrete chain* can be approximated by the Korteweg–de Vries (KdV) equation, which possesses a special family of particular solutions - dynamic solitons. In the short-wave limit, the dynamics of *infinite FPU chain* can be effectively described by the nonlinear Schrödinger equation (NSE) supporting the envelope solitons (breathers) as its particular solutions. As for the *finite FPU chains*, their nonlinear dynamics was intensively studied in terms of the NNMs [150].

## 1.3. Discrete breathers

There exists an additional class of localized excitations in discrete systems referred to as intrinsic localized modes (ILMs), also known as discrete breathers (DBs) [6,7,13,16, 21-23, 31, 42-45, 51,53,66,76,118,119,142,143,147,173,189,195], or discrete solitons (DSs). DB describes time-periodic solution which is mainly localized in very narrow domain (one or several particles). These solutions appear in the lattice dynamics of solids, dynamics of the coupled arrays of Josephson junctions, in the description of photonic crystals with nonlinear response, the homogeneous FPU chains [7, 42, 173], the discrete nonlinear Schrödinger (DNLS) equation [13], the Ablowitz-Ladik model [1], the DB model by Ovchinnikov and Flach [143], etc. (see, for instance, [16, 22, 23, 31, 189, 195]). In relatively recent studies, the DB concept was extended to localized time periodic solutions in the normal mode space, or the so-called *q*-breathers [43, 44, 66, 147]. Also, explicit analytic solutions for discrete breathers emerging in the vibro-impact lattices in conservative and forced damped systems were derived [51, 53].

## 1.4. Topological solitons

The changes in the system state due to the propagation of the wave disturbances are associated with irreversible transitions between different states in a classical nonlinear system or between neighboring energy levels in quantum systems. In the first case, the transition can be realized, e.g., by topological solitons, which not only carry the energy, but also change the state of the system providing the transfer from one potential well to another one. If the potential wells correspond to the same energy levels, the topological soliton represents a spatially localized wave without any disturbances behind its front. In the opposite case of the non-degenerate potential, the inter-well transition from a metastable state to the stable one can be realized; the released energy can lead to some deformation behind the front. In such systems, energy transfer is described by a new type of soliton, which includes not only the wave front, but also the post-front region [120, 174]. Relaxation to the steady state occurs on a sufficiently large distance (compared to the inter-particle separation) from the front.

## 1.5. LPTs of resonant non-stationary processes in a finite chain: a preliminary consideration

As mentioned earlier in this section, a rather comprehensive theory has been developed for the analysis of stationary processes, their stability and bifurcations. Furthermore, combinations of *non-resonant* NNMs in quasi-linear finite models can be also used for an asymptotic description of non-stationary processes. In these approaches, a difference between the linear and nonlinear cases consists only in the amplitude dependence of the natural frequencies of nonlinear systems. As for the energy transfer and the changes in the system state in the *infinite models*, they can be described by solitonic excitations.

However, non-stationary energy transfer and exchange, as well as the change in the system state under the condition of resonance *in a finite oscillatory system* are characterized by complicated non-stationary behaviour, for which any analytical investigation is prohibitively difficult. In this Report we discuss the LPT concept, which provides a unified description of highly non-stationary resonance processes for a wide range of classical and quantum dynamical systems with constant and time-variant parameters. As it was mentioned above, in the absence of the LPT concept, the fundamental meaning of the strongest nonlinear beats in energy transport could not be recognized, because in the linear theory this process can be formally presented as a combination of normal modes. The significance of the LPT as a specific type of regular non-stationary motion becomes apparent in the following manifestations:

- transitions from energy exchange to *non-stationary energy localization* due to the global LPT instability (similarly to the transition from the cooperative NNM to the localized NNM resulting from the local instability of the cooperative NNM, i.e., the transition to *stationary energy localization*);

- transitions from stationary to highly non-stationary processes (autoresonance, nonlinear quantum tunneling, etc.) as a result of transitions between the LPTs of different types;

- an analytical prediction and description of non-conventional synchronization, in which the LPT turns out to be attractor;

- transitions to chaotic behavior through the LPT instability;

-transitions to effective particles in multi-particle systems in the framework of the extended LPT concept;

- an analytical presentation of intensive energy transfer and exchange in a simple and physically evident form.

The inherent properties of the LPT are illustrated below using the simplest example of two identical weakly coupled oscillators. This introductory part makes a reader aware of main ideas without sophisticated mathematics. Nevertheless, we indicate in each chapter that the significant processes under consideration correspond to motion along the LPT, thus providing maximum energy transfer or exchange.

### 1.5.1. LPTs in a symmetric system of two identical oscillators

In this paragraph, we briefly illustrate the notion and the properties of LPTs in the Hamilton system of two weakly coupled identical oscillators (an asymmetric case will be investigated in detail in Section 3).

The equations of motion in this case are are given by

$$\frac{d^2U_1}{d\tau_0^2} + U_1 + 2\varepsilon\beta(U_1 - U_2) + 8\varepsilon\alpha U_1^3 = 0,$$

$$\frac{d^2U_2}{d\tau_0^2} + U_2 + 2\varepsilon\beta(U_2 - U_1) + 8\varepsilon\alpha U_2^3 = 0,$$

$$(1.1)$$

where the small parameter $\varepsilon \ll 1$ is introduced as a relative weak coupling coefficient (see [105-107, 123] for more details).According to the assumption $\varepsilon \ll 1$ not only the coupling is weak but also the nonlinearity has to be small. We recall that in the non-resonance case the NNMs represent appropriate basic solutions for the analysis of this system under arbitrary initial conditions. However, if coupling between the oscillators is weak and the intermodal resonance takes place, the NNMs approach is completely adequate if and only if initial conditions are close to a point on the stable NNM.

In contrast to strong coupling, the presence of the small parameter in (1.1) implies the existence of the slow time scale and the emergence of slowly modulated fast oscillations. Our purpose is to find slow envelopes and motivate a choice of an envelope corresponding to maximum energy exchange between the oscillators. To this end, we employ the multiple time scale methodology [139]. In the first step, we introduce the complex amplitudes by formulas:

$$\varphi_j = (V_j + iU_j)e^{i\tau_0}, \varphi_j^* = (V_j - iU_j)e^{-i\tau_0}, V_j = \frac{dU_j}{d\tau_0}, j = 1,2, \tag{1.2}$$

where the asterisk denotes complex conjugate. There are two reasons for transformations (1.2): first, it allows clearly seeing useful analogies between classical and quantum oscillators; secondly, each of the introduced complex amplitudes provides the leading-order approximation for both potential and kinetic energy of the corresponding oscillator. In the next step, we introduce the *slow time scale* $\tau_1 = \varepsilon\tau_0$ and then use the power expansions of the variables $\varphi_j$ and their time-derivatives

$$\begin{aligned}
\phi_j(\tau_0, \varepsilon) &= \phi_j^{(0)}(\tau_1) + \varepsilon\phi_j^{(1)}(\tau_0, \tau_1) + O(\varepsilon^2), \\
\frac{d\phi_j}{d\tau_0} &= \frac{\partial\phi_j^{(0)}}{\partial\tau_0} + \varepsilon\frac{\partial\phi_j^{(1)}}{\partial\tau_1} + O(\varepsilon^2), j = 1, 2.
\end{aligned} \tag{1.3}$$

It is important to note that expansions (1.3) include the slow main terms $\varphi_j^{(0)}(\tau_1)$. We then define $\varphi_j^{(0)}(\tau_1)$ as

$$\varphi_j^{(0)} = f_j e^{i\beta\tau_1}. \tag{1.4}$$

Substituting (1.2) – (1.4) into (1.1) and reproducing standard transformations of the multiple scale approach [139], we obtain the following equations for the complex amplitudes $f_1, f_2$:

$$\begin{aligned}
\frac{\partial f_1}{\partial\tau_1} + i\beta f_2 - 3i\alpha|f_1|^2 f_1 &= 0, \\
\frac{\partial f_2}{\partial\tau_1} + i\beta f_1 - 3i\alpha|f_2|^2 f_2 &= 0.
\end{aligned} \tag{1.5}$$

Note that Eqs. (1.5) are similar to the equations of two-level quantum system [151]. This system is integrable, and has two integrals of motion

$$H_1 = \beta(f_2 f_1^* + f_1 f_2^*) - \tfrac{3}{2}\alpha(|f_1|^4 + |f_2|^4),$$
$$N = |f_1|^2 + |f_2|^2 = (U_1^2 + U_2^2) + (V_1^2 + V_2^2). \tag{1.6}$$

The change of variables $f_1 = \sqrt{N}\cos\theta e^{i\delta_1}$, $f_2 = \sqrt{N}\sin\theta e^{i\delta_2}$, with the real angle variables

$\theta$ and $\Delta = \delta_1 - \delta_2$ transforms the complex-valued system (1.5) into the equations

$$\frac{d\theta}{d\tau_1} = \beta\sin\Delta,$$
$$\sin 2\theta\frac{d\Delta}{d\tau_1} = 2\beta\cos 2\theta\cos\Delta + \frac{3}{2}\alpha N\sin 4\theta. \tag{1.7}$$

It now follows that system (1.7) has the following integral of motion:

$$H_2 = N(\cos\Delta + k\sin 2\theta)\sin 2\theta, \; k = 3\alpha N/(4\beta). \tag{1.8}$$

Figure 1 presents the sets of phase trajectories described by integral (1.8) for different values of $k$.

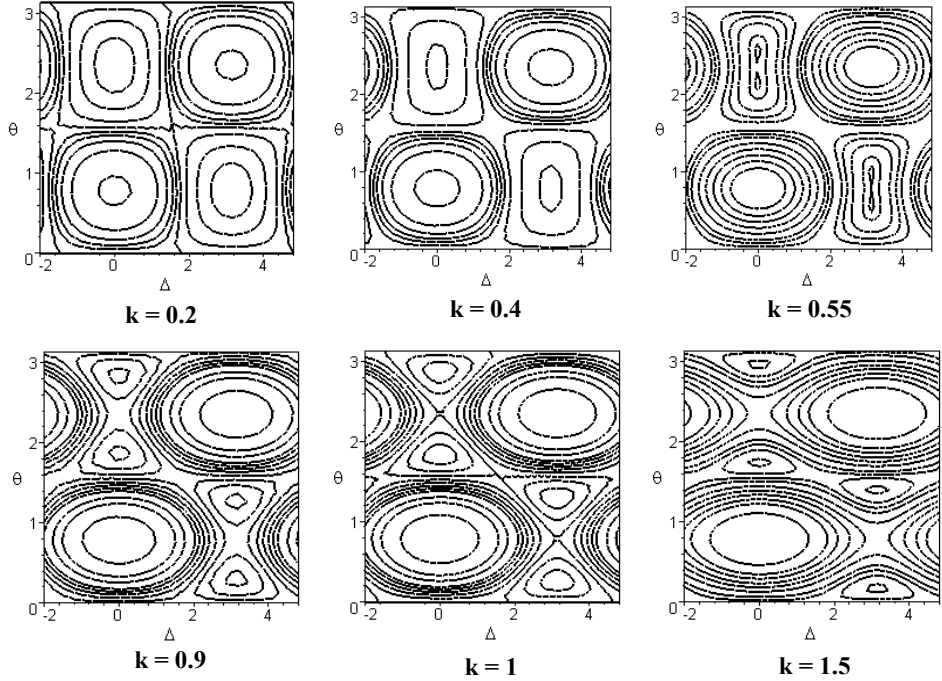

**Fig. 1**. Phase plane: evolution of NNMs and LPT with growth of $k$.

Due to the periodic structure of the phase plane, only two neighboring (in the horizontal direction) quadrants have to be considered. Stationary points in the phase plots correspond to NNMs and can be found from the conditions $d\theta/d\tau_1 = 0$, $d\Delta/d\tau_1 = 0$. These points are associated with the quasi-equilibrium states (in the slow time). Therefore, in the resonance case, an explicit definition of the stationary processes can be obtained directly from Eqs. (1.7).

The sets of trajectories encircling the stationary points correspond to non-stationary regimes. Since system (1.7) is integrable, one can formally find its analytical solution for arbitrary initial conditions. However, this approach cannot be extended to more complicated models with many degrees of freedom, for which the existence of two integrals is insufficient to find an analytical solution, as well as to non-conservative systems having no integrals. In such a case conventional approach is usually connected with the NNMs concept.

This approach is justified even for the considered simplest 2DOF model if the boundary conditions are close to those for the NNMs themselves. Then, a chosen trajectory in the phase plane is at a relatively small distance from the stable stationary point. The NNM concept becomes non-adequate only when dealing with highly non-stationary resonance processes because of the *strong intermodal interaction*. However, namely in the latter case there exist the basic processes of other type –the *LPTs* - which are alternative to the NNMs. The LPT contains the segments $\theta = 0$ and $\theta = \pi/2$. Since this trajectory represents an outer boundary for a set of trajectories encircling the basic stationary points, we refer to it as the limiting phase trajectory. The motion along this trajectory leads to full reversible energy transfer, and therefore, corresponds *to maximum possible energy exchange between weakly interacting nonlinear oscillators* (Fig.1). This implies that LPTs represent *highly non-stationary resonance processes.* Both NNMs and LPTs together with close phase trajectories form a class of *regular motions.*

Note that the LPT is defined by a certain set of initial conditions. The equality $\theta(0) = 0$ implies that $f_2(0) = 0$, and, in virtue of (1.2) – (1.4), $U_2 = 0$, $V_2 = 0$ at $\tau_0 = 0$. Therefore, the condition $\theta(0) = 0$ defines the LPT of the system, where the first oscillator is initially excited but the second oscillator stays initially at rest. On the contrary, the condition $\theta(0) = \pi/2$ defines the LPT in the system with the excited second oscillator, while the first oscillator is initially at rest.

In the linear limit ($k = 0$) both branches of the LPT represent the rectangles. In the resonance conditions this is a single case when every solution, including LPTs, can be formally calculated as a combination of NNMs. However, a direct determination of LPTs from Eqs.(1.7) leads to more descriptive analytical expressions and, that is the most essential, admits a natural extension to nonlinear beats. At $0 < k < 0.5$, the shape of the LPT is slightly changed but the LPT preserves the same topology (Figs. 1(a, b)). Two branches of the LPT encircling the stationary points corresponding to in-phase and out-of-phase NNMs, separate the phase plane into two domains. However, in general, the LPT does not contain stationary points; therefore, it does not coincide with the separatrix. This means that motion along the LPT requires a finite time.

When the parameter $k$ grows, the evolutions of NNMs, corresponding to stationary processes, and LPTs, corresponding to highly non-stationary resonant processes, become different. The evolution of the out-of phase NNM demonstrates the well-known instability at $k = 0.5$ and the birth of new two stable NNMs with the encircling separatrix (Fig. 1(c)). This local instability is crucial for the stationary dynamics, but full energy exchange between weakly interacting oscillators described by the LPT still exists. However, there is a qualitative change of the phase plane caused by the confluence of LPT and the separatrix at $k = 1$, which underlines the impossibility of full energy exchange between the oscillators.

It follows from (1.8) that an increase of $k$ corresponds to increasing nonlinearity $\alpha$ or to weakening of coupling $\beta$ under fixed intensity of excitation $N$. The problem of energy transfer and exchange in weakly coupled nonlinear oscillators is discussed in details in Section 3.

An alternative demonstration of the LPT evolution can be obtained by excluding the variable $\Delta$ from (1.7). Taking into account that the LPT contains the segment $\theta = 0$ and, therefore, $H_2=0$ on the LPT, one can express $\cos\Delta$ as a function of $\theta$ and, finally, derive the following pendulum equationvalid on the LPT only, but with significant restriction

$$\frac{d^2\theta}{d\tau_1^2} + k^2 \sin 4\theta = 0, \ 0 \le \theta \le \pi/2. \tag{1.9}$$

It follows from (1.9) that the LPTs situated far from the pendulum separatrix(and outside it) are close to the straight segments corresponding to almost linear beats at $k = 0$ (Fig.2). Because of presented restrictions such segments form a broken line close to so-called saw-tooth function. When approaching the separatrix, these segments acquire a noticeable curvature (essentially nonlinear beats). The separatrix itself is at the same time one of the LPTs (the confluence of the LPT and the separatrix was mentioned above). The LPT inside the separatrix corresponds to non-stationary energy localization.

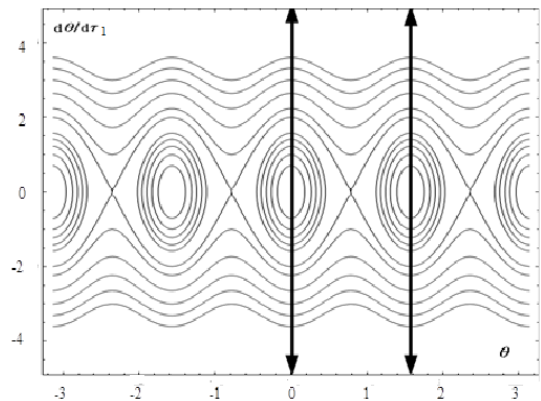

**Fig. 2.** Phase plane $(\theta,\ d\theta/d\tau_1)$

### 1.5.2. Non-smooth transformations

Exact (numerical) solutions for the LPTs in terms $(\theta, \Delta)$ for $k = 0.5$ and $k = 0.9$ are presented in Fig. 3.

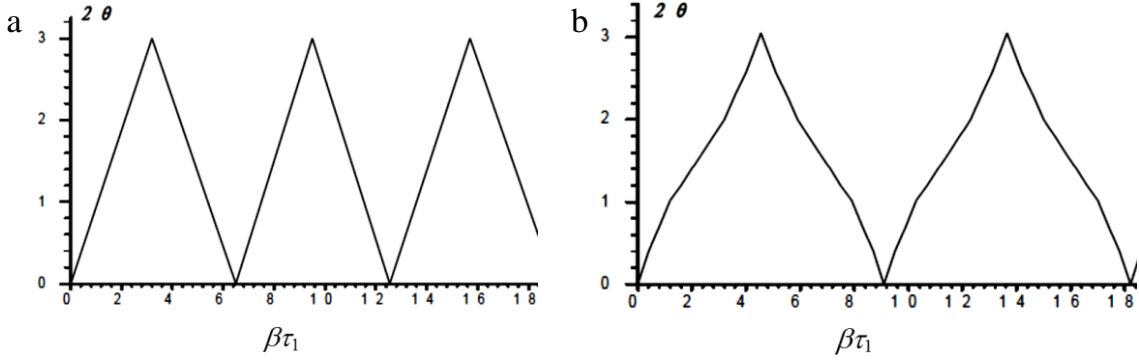

**Fig. 3**. Exact solutions for $k = 0.5$ (a) and $k = 0.9$ (b).

Both LPTs are obviously close to the so-called saw-tooth functions introduced for the study of strongly nonlinear processes similar to the vibro-impact responses of a free particle between the rigid walls [149]. Applications of non-smooth transformations to the analysis of the LPTs of highly–non-stationary resonance processes are discussed in Section 2. A detailed study of non-smooth approximations can be found in [103, 105-107, 110-113, 123, 149] and references therein.

It is important to underline that the existing techniques [149] allows one to overcome mathematical difficulties caused by non-smoothness of the LPTs. A special change of variables [105-107] transforms Eqs (1.7) into *smooth equations for non-smooth variables*. It is easy to obtain that the envelope of linear beats is exactly described by the saw-tooth function. The shape of the envelope of nonlinear beats is slightly different from a $\Delta$-shape saw-tooth function (Fig. 4); the curvature can be taken into account with the help of the power series (e.g., [105]). The period of beats $T$ can be easily calculated in any order of approximations from the condition of complete energy exchange, or from the condition of achievement of the maximum admissible value of $\theta$ at $\tau_1 = T/2$. Examples are discussed in Section 2 and investigated in detail in [79-87,105-117,121-124,149,175,185,186].

### 1.5.3. Weakly dissipative systems

We now briefly examine the dynamics of a weakly dissipated system. The equations of motion with weak dissipative terms are given by

$$\frac{d^2 U_1}{d\tau_0^2} + U_1 + 2\varepsilon n \dot{U}_1 + 2\beta(U_1 - U_2) + 8\alpha\varepsilon U_1^3 = 0,$$

$$\frac{d^2 U_2}{d\tau_0^2} + U_2 + 2\varepsilon n \dot{U}_2 + 2\beta(U_2 - U_1) + 8\alpha\varepsilon U_2^3 = 0. \tag{1.10}$$

Using the changes of variables (1.2) – (1.4) and introducing the additional transformations

$$\Phi_j = e^{n\tau_1} f_j, \tag{1.11}$$

we obtain the following equations for the slow complex amplitudes $\Phi_j$, $j = 1{,}2$:

$$\frac{\partial \Phi_1}{\partial \tau_1} + i\beta\, \Phi_2 - 3i\alpha e^{-2n\tau_1} |\Phi_1|^2 \Phi_1 = 0,$$

$$\frac{\partial \Phi_2}{\partial \tau_1} + i\beta\, \Phi_1 - 3i\alpha e^{-2n\tau_1} |\Phi_2|^2 \Phi_2 = 0. \tag{1.12}$$

It now follows from (1.6), (1.11), that a constant $N$ can be interpreted as a measure of the initial energy

$$N = |\Phi_1(0)|^2 + |\Phi_2(0)|^2 = (U_1^2(0) + U_2^2(0)) + (V_1^2(0) + V_2^2(0)). \tag{1.13}$$

The change of variables $\Phi_1 = \sqrt{N}\cos\theta e^{i\delta_1}$, $\Phi_2 = \sqrt{N}\sin\theta e^{i\delta_2}$, with the real angle variables $\theta$ and $\Delta = \delta_1 - \delta_2$ transform the complex-valued system (1.13) into the real-valued equations similar to (1.7)

$$\frac{d\theta}{d\tau_1} = \beta\sin\Delta,$$

$$\sin 2\theta\, \frac{d\Delta}{d\tau_1} = 2\beta\cos 2\theta\cos\Delta + \frac{3}{2}\alpha N e^{-2n\tau_1}\sin 4\theta. \tag{1.15}$$

Figure 4 depicts damped beating in the system with the initially excited first oscillator. One can note a significant difference between two presented plots. Figure 3(a) demonstrates well-defined damped beats, while in Fig. 3(b) predominant energy localization on the first oscillator caused by strong initial excitation is changed by damped beats. However, in both cases the approximate analytical solution of Eqs. (1.15) can be obtained in terms of non-smooth basic functions [105].

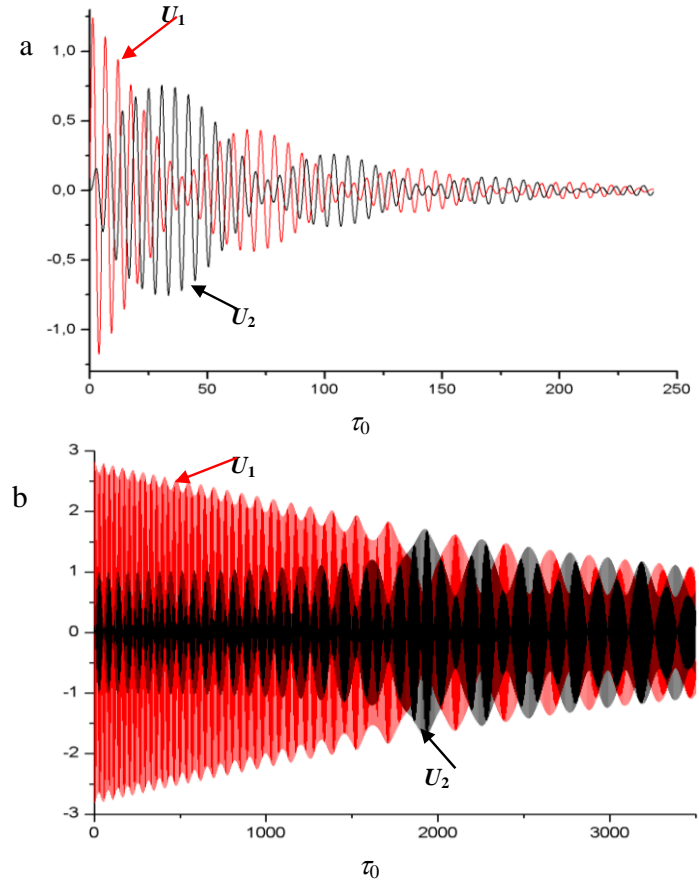

**Fig. 4.** Displacements $U_1(\tau_0)$, $U_2(\tau_0)$ in the systems with different initial excitations: a - $N$ =2.1; b - $N$ =8.

## 1.6. Further extensions of the LPT theory

### 1.6.1. Linear classical and quantum systems

The principal question is: whether it is possible and necessary to extend the above-mentioned results to more complicated cases, or not. In this connection, we briefly discuss existing approaches to the description of strongly non-stationary resonance processes.

In the framework of linear physics, energy transfer is usually associated with the propagation of the wave packets, that is, of the combinations of normal modes with close frequencies. Typically, a wave packet eventually spreads due to dispersion, which leads to the dependence of the velocity on the frequencies of normal modes. In quantum mechanics, the wave packet formed by a combination of stationary states belonging to the continuous spectrum is the image of a free particle (e.g., an electron). The quantum harmonic oscillator, for which the Schrödinger equation has the solution in the form of a wave packet, provides a unique example of a non-spreading non-stationary process (in the presence of the force field). Such a non-spreading wave packet describes the non-stationary coherent state as similar to the classical harmonic oscillator. The combination of the wave functions with close energies may result in their mutual compensation everywhere except for a relatively narrow area in which the "amplification" of the excitation occurs forming the localized profiles. Thus, when dealing with the linear dynamics, the superposition principle allows the formal description of the localized non-stationary resonance processes through the combinations of normal modes. It will be shown that intermodal coherence provides an alternative possibility to describe such processes on the basis of the LPT concept, which can be, however, extended to nonlinear classical and quantum systems.

### 1.6.2. Nonlinear infinite-dimensional models

In the nonlinear dynamics of multi-particle infinite systems only the soliton theory gave birth to an appropriate approach to the non-stationary resonance problems among which one can mention an energy transfer and a change of the system's state.

It was shown that the nonlinear effects play the central role in the soliton theory because they compensate the inevitable spreading of the wave packet (or, vice versa, the dispersion compensates its inevitable breaking (distortion) of the wave packet due to nonlinearity). Note that, although energy transfer in ordered systems can also be realized by the mobile "vibration solitons", or breathers, the physical reason for the mutual compensation of nonlinear and dispersive effects are hidden behind a powerful mathematical tool - the inverse scattering method. This method reveals a high degree of symmetry in certain classes of nonlinear systems, which enables the formation of the stable solitons [36, 37, 167, 190]. We will show that a physical nature of the energy localization in this case may be clarified while considering the localized LPTs in the finite systems.

The most important point is that, as we mentioned above, the soliton theory is directly applied only to infinite continuum systems as well as to discrete systems with an *infinite number* of particles. It seems that the spatial localization of vibrational solitons predetermines their survival in fairly extensive but finite discrete systems. However, the clarification of the physical nature of the soliton formation and survival requires an analysis perfectly different from that one in the infinite-dimensional case. The adequate analysis aims to explain the occurrence of the transition from intense energy exchange to energy localization in nonlinear classical and quantum systems (the two-level and multilevel quantum models) with a *finite number* of particles. This issue is closely connected with an extension of the beating problem to multi-particle models which can be solved in the frameworks of the LPT concept.

As for the change in the state of the system, described in infinite models by topological solitons, it can be realized in a finite model due to variation of its parameter only. The most significant examples are passage of a classical oscillator through resonance and quantum tunneling. An application of the LPT concept to these problems is shown in the Section 6 of the Report

### 1.6.3. Autoresonance

Note, that a purely nonlinear phenomenon – autoresonance - can be also related to such class of the problems. Autoresonance (AR) in nonlinear oscillators, which manifests as the resonance amplitude growth caused by periodic forcing with slowly varying frequency, has numerous applications in different fields of physics, see, e.g., [3,10,11,17,25,27,40,46,68, 127,129,206]. After first studies for the purposes of particle acceleration, AR has become a very active field of research. It was noticed in the early work [17, 95, 129, 206] that the physical mechanism underlying autoresonance can be interpreted as adiabatic nonlinear phase-locking between the system and the driving signal; it is directly connected with the inherent property of nonlinear systems to remain in resonance when the driving frequency varies in time. However, an analytical investigation of the transition from bounded oscillations to AR has not been reported in the literature. The only known critical threshold was obtained only numerically. Even omitting a discussion of the applicability of a numerically found threshold to a large class of physical problems, it is necessary to note that this threshold is unusable in the problem of adiabatic AR [83, 84].

*1.6.4. Targeted energy transfer*

A special attention should be given to the problem of targeted energy transfer (TET), where energy of some form is directed from a source (donor) to a receiver (recipient) in a one-way irreversible fashion. Any process in nature involves to a certain extent some type of energy transfer. It is not surprising then that TET phenomena have received much attention in applied mathematics, applied physics, and engineering Applications of energy localization and TET in diverse areas are given, e.g., in [35, 205], including analytical, computational and experimental results. Representative examples are early works on TET between nonlinear oscillators and/or discrete breathers [8, 75], wherein the notion of TET was first introduced, and following works on the TET manifestation in physical systems [125, 126, 131, 135]. The dynamical mechanisms considered in these works are based on imposing conditions of nonlinear resonance between interacting dynamical systems in order to achieve TET from one to the other, and then 'breaking" this condition at the end of the energy transfer to make it irreversible. This principle was further employed in a wide variety of mechanical applications. TET from an impulsively loaded linear substructure to a strongly nonlinear lightweight mass attachment referred to as a nonlinear energy sink (NES) was intensively studied over past two decades; a large number of examples can be found in the monographs [198, 200], in which the extensive bibliography can be found. The results of [200] demonstrated that *almost irreversible energy transfer in the systems coupled with NES could not be analytically described in the framework of the traditional approaches.* Recent research [82,86,105-113,117,122-124,175,178,185,196,203,204] proved that explicit analytical results for impulsively loaded structures as well as for nonlinear oscillators with an external source of energy can be obtained with the help of the LPT-methodology

An analogy between classical and quantum energy transport is an interesting aspect for discussing. Examples of this analogy can be found in [59,78,81,85,115,125,151]. In quantum mechanics, the transition between neighboring levels is realized by quantum tunneling, first analyzed contemporaneously but independently in [90, 100, 187, 210]. Of particular interest is the generalization of the tunneling theory to nonlinear quantum systems, such as superconducting Josephson junctions and Bose-Einstein condensates [94, 191, 209, 213]. However, existing analytical techniques that enable the calculation of the asymptotic values of transitional probabilities at large times are insufficient to describe tunneling on the finite times with arbitrary initial conditions.

### 1.6.5. Synchronization of autogenerators

Synchronization is a fundamental problem discussed over centuries, starting from Huygens' famous observations; an extensive bibliography and discussion of old and recent results can be found in, e.g., [148]. Synchronization in the chains of coupled van der Pol or van der Pol-Duffing oscillators has numerous applications in different fields of physics and biophysics [9, 39, 87, 114, 132, 145, 152, 146, 148, 155]. Despite the long history, there still exist a number of open questions even for these benchmark models. In particular, we note that previous studies considered synchronization of these systems in the regimes similar to nonlinear normal modes (NNMs) and did not discuss a possibility of alternative types of synchronization.

In the Report we present new concepts and methods of analysis, which provide the answers to the following open questions:

1) What is an adequate description of nonlinear beating in conservative, damped and forced nonlinear systems?

2) Is it possible to extend the concept of beats on multi-particle systems?

3) What are the conditions of the transition from intense energy transfer to energy localization in finite discrete systems, including the formation of mobile breathers?

4) What conditions provide irreversible energy transfer in time-dependent classical and quantum systems?

5) What are the conditions of transition from irreversible energy transfer (tunneling) to autoresonance in the oscillator with the slowly varying frequency of the external excitation?

6) Can the non-conventional synchronization of weakly coupled auto-generators exist?

7) What is the role of LPTs in the transition from regular to chaotic behavior?

Recent studies proved that the answers to the above-listed questions can be obtained in the framework of the LPT concept.

The Report is organized as follows.

In the first part of the Report we discuss the results related to the oscillators and oscillator arrays with constant parameters. The considered models are related to

- transient vibrations of the forced nonlinear oscillator;

- intense energy exchange and transition to energy localization in the conservative and dissipative systems (e.g. coupled nonlinear oscillators);

- a new type of self-sustained oscillations characterized by the LPT synchronization in the system of coupled oscillators.

The second part of this report is concerned with applications of the LPT concept to systems with slowly-varying parameters. First, it is shown that the necessary conditions for the emergence of auto-resonance, which can be attributed to a purely nonlinear phenomenon, can be completely understood and depicted in the framework of the LPT-concept. Then, a direct mathematical analogy between the targeted energy transfer in the classical systems of weakly coupled oscillators and nonlinear quantum tunneling is demonstrated. It is shown that this analogy allows a unified study of both classical and quantum problems. The main attention is given to

-the targeted energy transfer in nonlinear systems with time-varying parameters and nonlinear Landau-Zener tunneling;

- the emergence of auto-resonance in the nonlinear systems with a time-dependent frequency detuning.

In the final part, we briefly discuss the applications of the LPT concept to the description of the transition from regular to chaotic behavior and to the analysis of energy exchange, localization and transfer in multi-particle oscillatory chains.

## 2. Limiting Phase Trajectories of a single oscillator

In this section, we illustrate the role of LPTs in the analysis of nonlinear non-stationary oscillations by a simple example of a periodically forced single-degree-of-freedom (SDOF) Duffing oscillator. The main difference from the conservative system described in the preliminary section is that we deal here with resonance energy flow from the source of energy instead of internal (intermodal) resonance.

In the first part of this section, we define the stationary states and the LPTs for this model and then employ the LPT concept to describe salient features of the non-stationary dynamics of the forced Duffing oscillator near 1:1 resonance. It will be shown that LPTs can be considered as borderlines between different types of trajectories, associated with maximum targeted energy transfer from the source of energy to the oscillator and related, depending on the parameters, to quasi-linear, moderately nonlinear and strongly nonlinear regimes of oscillations. At the same time, steady (stationary) oscillations of a non-autonomic system play the role of NNMs. We also extend the notion of the LPT to oscillators subjected to bi-harmonic external excitations.

Since nonlinear nonstationary problems seldom yield closed-form analytical solutions, a well-known method of approximate solving for nonlinear equations is the method of multiple scales [139]. Throughout this work, we will make an extensive use of thisasymptotic technique based on the complexification of the dynamics and the separation of the slow dynamics.

### 2.1. Duffing oscillator with harmonic forcing near 1:1 resonance

### 2.1.1. Formulation of the problem and main equations

We investigate a dimensionless weakly nonlinear oscillator subject to a periodic excitation in the neighborhood of 1:1 resonance. The equation of motion is given by

$$\frac{d^2u}{d\tau_0^2} + 2\varepsilon\gamma\frac{du}{d\tau_0} + u + 8\alpha\varepsilon u^3 = 2\varepsilon F\sin(1+\varepsilon s)\tau_0, \qquad (2.1)$$

where $\gamma$, $\alpha$, $F$, $s$ are positive parameters, $\varepsilon > 0$ is a small parameter of the system. In non-stationary case the oscillator is assumed to be initially at rest; this assumption is equivalent to initial conditions $u = 0$, $v = du/d\tau_0 = 0$ at $\tau_0 = 0^+$. Below, we write $\tau_0 = 0$ instead of $\tau_0 = 0^+$, except as otherwise noted. In earlier work [105] the trajectory with these initial conditions was introduced as a *Limiting Phase Trajectory* (LPT) corresponding to maximum possible energy flow from the source of energy to the oscillator. This extreme property will be illustrated below.

We recall again that closed-form analytical solutions of nonlinear non-stationary problems are, in general, unavailable. To construct an explicit asymptotic solution, we make use of the multiple scales analysis [139]. This approach, especially convenient in neighborhoods of resonances will be used throughout this work.

The first step for applying this method is to introduce the complex-valued variables

$$Y = (v + iu)e^{-i\tau_0}, Y^* = (v - iu)e^{i\tau_0}, i = \sqrt{-1}, \qquad (2.2)$$

where asterisk denotes complex conjugate. Substituting the representation (2.2) into (2.1) yields the following alternative (still exact) equation of motion::

$$\frac{dY}{d\tau_0} = 3i\varepsilon\alpha\,|\,Y\,|^2\,Y - \gamma Y - i\varepsilon Fe^{i\tau_1} + i\varepsilon G_0(\tau_0, \tau_1, Y, Y^*), Y(0) = 0, \qquad (2.3)$$

where $\tau_1 = \varepsilon s \tau_0$ is the leading order slow time scale. The coefficient $G_0$ is given by

$$G_0(\tau_0, \tau_1, Y, Y^*) = [Fe^{-i\tau_1} + i\gamma Y^* - 3\alpha Y^*\,|\,Y\,|^2]e^{-2i\tau_0} - \alpha Y^3 e^{2i\tau_0} + (Y^*)^3 e^{-4i\tau_0}, \qquad (2.4)$$

$\tau_1 = \varepsilon s \tau_0$ is the leading-order slow time scale. Equation (2.3) is exact, as it is derived from the original equation of motion without omitting any terms in the process. As in Introduction, we apply the multiple scales method [139] to Eq. (2.3). Since Eq. (2.3) depends on two time scales, the solution is sought in the form of the expansion [135, 165]:

$$Y(\tau_0, \varepsilon) = \varphi^{(0)}(\tau_1) + \varepsilon\varphi^{(1)}(\tau_0, \tau_1) + O(\varepsilon^2),$$

$$\frac{dY}{d\tau_0} = \frac{d\varphi_j^{(0)}}{d\tau_1} + \varepsilon\frac{\partial\varphi_j^{(1)}}{\partial\tau_0} + O(\varepsilon^2) \qquad (2.5)$$

with the slow leading-order term $\varphi^{(0)}(\tau_1)$. After substituting (2.5) into (2.3) and eliminating the sum of non-oscillating terms (responsible for appearance of growing constituents in solution) from the resulting equation, we obtain that the following equation for the slow variable $\varphi^{(0)}(\tau_1)$:

$$s\frac{d\varphi^{(0)}}{d\tau_1} + \gamma\varphi^{(0)} - 3i\alpha|\varphi^{(0)}|^2\varphi^{(0)} = -iFe^{i\tau_1}, \quad \varphi^{(0)}(0) = 0. \qquad (2.6)$$

This equation approximately governs the slow evolution of the complex amplitude with time. Rescaling of the variables and the parameters

$$\varphi^{(0)}(\tau_1) = \Lambda\varphi(\tau_1)e^{i\tau_1}, \ \Lambda = (s/3\alpha)^{1/2}, f = F/s\Lambda = F\sqrt{3\alpha/s^3}, \ \gamma_1 = \gamma/s \qquad (2.7)$$

reduces Eq. (2.6) to the form

$$\frac{d\varphi}{d\tau_1} + \gamma_1\varphi + i(1-|\varphi|^2)\varphi = -if, \ \varphi(0) = 0. \qquad (2.8)$$

After introducing a polar decomposition of $\varphi = ae^{i\Delta}$, Eq. (2.8) is rewritten as:

$$\frac{da}{d\tau_1} + \gamma_1 a = -f\sin\Delta, \qquad (2.9)$$

$$a\frac{d\Delta}{d\tau_1} = -a + a^3 - f\cos\Delta,$$

with initial condition $a(0) = 0$. The second initial condition including $\Delta(0)$ will be derived below. It now follows from (2.2), (2.5), (2.7) that

$$u(\tau_0, \varepsilon) = \Lambda a(\tau_1)\sin(\tau_0 + \Delta(\tau_1) + \tau_1) + O(\varepsilon), \qquad (2.10)$$

$$v(\tau_0, \varepsilon) = \Lambda a(\tau_1)\cos(\tau_0 + \Delta(\tau_1) + \tau_1) + O(\varepsilon).$$

It is well known that the difference between a precise solution of Eq. (2.1) and its approximation (2.10) is of $O(\varepsilon)$ in large time intervals $\tau_0 \sim O(1/\varepsilon)$ [139]. However, a refined analysis [134] shows that the interval of convergence depends on the properties of higher approximations and may tend to infinity. Moreover, relatively large values of $\varepsilon$ in particular problems do not necessarily imply that the derived analytical approximations will be poor at larger times; numerical examples are given in [202].

### 2.1.2. Stationary states, LPTs, and critical parameters of non-dissipative systems

In this subsection, we consider the non-dissipative system

$$\frac{da}{d\tau_1} = -f \sin \Delta , \qquad (2.11)$$

$$a \frac{d\Delta}{d\tau_1} = -a + a^3 - f \cos \Delta.$$

with initial condition $a(0) = 0$. It is easy to prove that system (2.11) is integrable, yielding the following integral of motion

$$K = a(\frac{1}{4} a^3 - \frac{1}{2} a - f \cos \Delta). \qquad (2.12)$$

Since $a(0) = 0$, then $K = 0$ on the LPT. This implies that there exist two branches of the LPT, the branch $a \equiv 0$ for any $\Delta$ corresponds to an instant change of the phase shift (Fig.5); the non-trivial branch solves the cubic equation

$$\frac{1}{4} a^3 - \frac{1}{2} a - f \cos \Delta = 0. \qquad (2.13)$$

Equality (2.13) determines the second initial condition $a(0^+) = 0$, $\cos \Delta(0^+) = 0$. Suppose that $da/d\tau_1 > 0$ at $\tau_1 = 0^+$; under this assumption, $\Delta(0^+) = -\pi/2$. Hence, initial conditions $a = 0$, $\Delta = -\pi/2$ at $\tau_1 = 0$ corresponds to the LPT of system (2.11).

Our purpose is to find critical values of the parameter $f$ dictating different types of the dynamical behavior.

#### a). Stationary states in slow time scale and their evolution in the parametric space

Since any stable orbit encircles a corresponding stationary point (the latter is an analogue of NNM in conservative case), the first step is to find all such points of Eqs (2.11) from the equations $da/d\tau_1 = 0$, $d\Delta/d\tau_1 = 0$, or,

$$-a + a^3 - f\,\text{sgn}\,(\cos\Delta) = 0, \quad \cos\Delta = \pm 1. \tag{2.14}$$

This stationary problem was repeatedly considered in different versions, and the conditions of transition (in the parametric space) from one to three stationary states in the slow time scale. Let us present shortly corresponding results.

Due to periodicity, only two stationary points $\Delta = 0$ and $\Delta = -\pi$ can be considered. The corresponding stationary states are denoted by $C_+$ and $C_-$, respectively. The number of the roots of the algebraic equation (2.14) depends on the properties of its discriminant (see [74] for more details)

$$D_2 = 4 - 27f^2.$$

If $D_2 < 0$, then Eq. (2.14) has 3 different real roots; if $D_2 > 0$, then there exist the single real and two complex conjugate roots; if $D_2 = 0$, two real roots merge [74]. The latter condition determines the critical value which is crucial for stationary dynamics

$$f_2 = 2/\sqrt{27} \approx 0.3849. \tag{2.15}$$

A straightforward investigation of Eq. (2.14) proves that the system has a single stable centre $C_+$: $(0, a_+)$ if $f > f_2$ (Fig. 5c).Note that the parameter $f = F\sqrt{3\alpha/s^3}$ reflects the effect of all parameters on the system dynamics, and the condition $f > f_2$ implies not only strong nonlinearity or large excitation amplitude but also intense excitation of the oscillator with small frequency detuning.

If $f < f_2$, then there exist two stable centres $C_-$: $(-\pi, a_-)$, $C_+$: $(0, a_+)$, and an intermediate unstable hyperbolic point $O$: $(-\pi, a_0)$ (Figs. 5a and 5b).

*b). LPTs and their evolution in parametric space*

The LPTs in the considered were first introduced in [117] where their significance for highly non-stationary resonance dynamics has been shown. Earlier analytical studies of non-stationary dynamics were restricted by vicinities of stationary states in the connection with the problem of their stability.

In both above mentioned cases, the LPT begins at $a = 0$, $\Delta = -\pi/2$ but its direction depends on the value of $f$. Note that the stable center $C_-$ and the small LPT near $C_-$ exist provided Eq. (2.13) is solvable at $\Delta = -\pi$. In order to find a critical value $f_1 < f_2$ ensuring the transition from small to large non-stationary oscillations, we analyze the discriminant $D_1$ of Eq. (2.13) at $\Delta = -\pi$

$D_1 = 16(2 - 27f^2)$.

If $D_1 = 0$, the hyperbolic point in the axis $\Delta = -\pi$ coincides with the maximum point of the small LPT defined by the conditions $da/d\tau_1 = 0$, $\sin\Delta = 0$. Thus the critical value $f_1$ is given by the condition $D_1 = 0$, or

$$f_1 = \sqrt{2/27} \approx 0.2722. \tag{2.16}$$

The threshold $f_1$ corresponds to a boundary between small and large non-stationary oscillations: at $f = f_1$ the LPT of small oscillations coalesces with the separatrix going through the homoclinic point on the axis $\Delta = -\pi$. This implies that the transition from small to large non-stationary oscillations occurs due to loss of global stability of the LPT of small oscillations. At $f = f_2$, the stable center on the axis $\Delta = -\pi$ vanishes due to the coalescence with the homoclinic point, and only a single stable center remains on the axis $\Delta = 0$.

Conditions $f < f_1$, $f_1 < f < f_2$, and $f > f_2$ characterize *quasi-linear*, *moderately nonlinear*, and *strongly nonlinear* dynamical behaviour, respectively. These definitions are consistent with the plots presented in Fig. 5.

Figure 5 clearly demonstrates the "limiting" property of the LPTs in the time-invariant system. It is seen that the LPT represents an outer boundary for a set of closed trajectories encircling the stable center in the phase plane $(\Delta, a)$. In particular, this property proves that motion along the LPT possesses the maximum amplitude, and therefore, the maximum energy among all other closed trajectories characterising oscillatory motion.

It is important to note that the value $f_2$ is independent of the initial conditions; it was earlier (see, e.g., [140]) as a boundary between the regimes with one or three stationary points. However, the boundary $f_1$ between small and large non-stationary oscillations strictly depends on the choice of the initial point which has to belong to LPT. This boundary was first revealed in [105, 113]

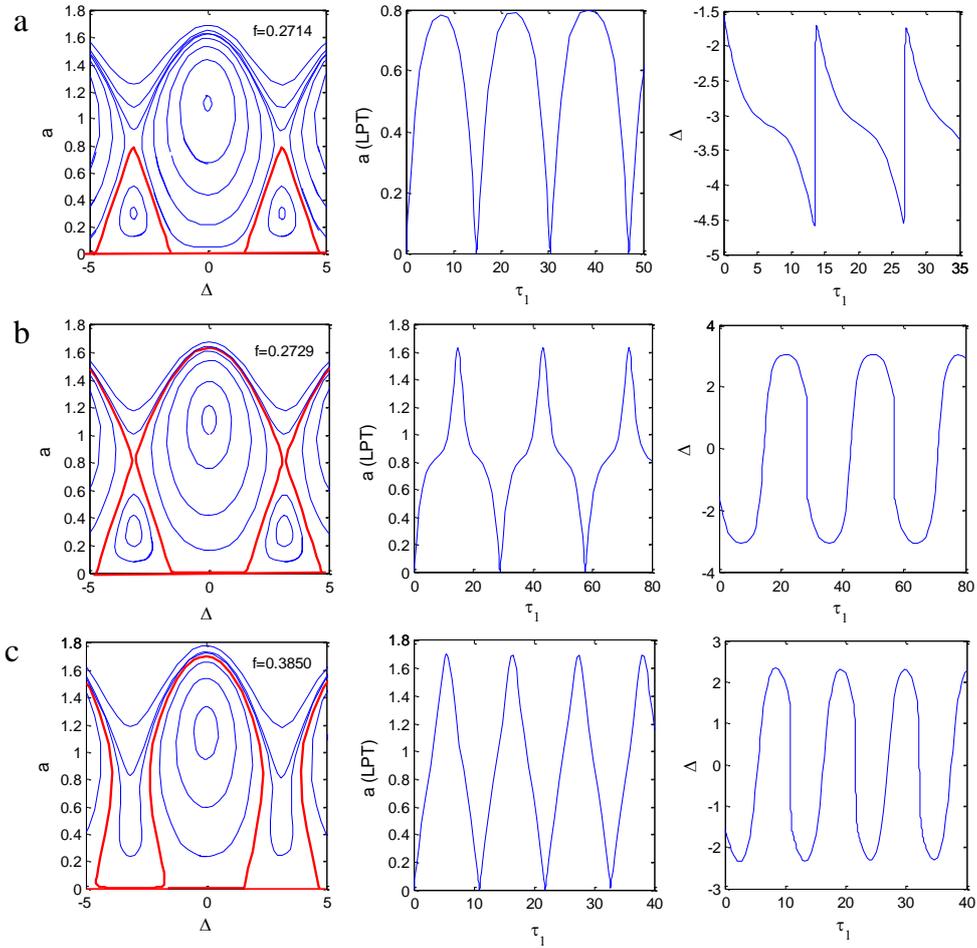

**Fig. 5.** Phase portraits, slow amplitudes and phases for weakly nonlinear (a), moderately nonlinear (b) and strongly nonlinear (c) modes of oscillation

Similarly to the case of two identical conservative oscillators considered briefly in Introduction, one can derive here the autonomic nonlinear equation valid only for the LPTs. To do this, we exclude the phase $\Delta(\tau_1)$ from (2.11) to obtaim

$$\frac{d^2a}{d\tau_1^2} = -f\cos\Delta\frac{d\Delta}{d\tau_1},$$

on the LPT. Taking into account that, by virtue of (2.13), $f\cos\Delta = a(a^2-2)/4$, we obtain the following second-order equation for the LPT:

$$\frac{d^2a}{d\tau_1^2} + \frac{dU}{da} = 0, \qquad (2.17)$$

with initial conditions $a(0) = 0$, $v(0) = da/d\tau_1 = f$. The potential $U(a)$ is defined by the condition of the energy conservation, namely, $E = \frac{1}{2}v^2 + U(a) = \frac{1}{2}v^2(0)$. It follows from (2.11), (2.13) that

$$v = -f\sin\Delta = \pm[f^2 - \tfrac{1}{16}a^2(a^2-2)^2]^{1/2},$$

and, therefore,

$$U(a) = \frac{a^2(a^2-2)^2}{32}$$
$$u(a) = \frac{dU}{da} = \frac{a}{4}(\frac{a^2}{2}-1)(\frac{3a^2}{2}-1). \qquad (2.18)$$

The function $32U(a)$ and the phase portraits of oscillator (2.17) in the plane $(a, v)$ are shown in Fig. 6. It is seen that, dependent on the initial energy, motion varies from small to large non-stationary oscillations with observable deceleration near the saddle points and then to motion with an almost constant velocity up to the reflection from the wall of the potential well (Fig. 6). In the latter case, $a(\tau_1)$ tends to a saw-tooth function (Fig. 7), and the phase portrait of system (2.17) is consistent with motion of a free particle between two rigid walls (Section 2.1.3).

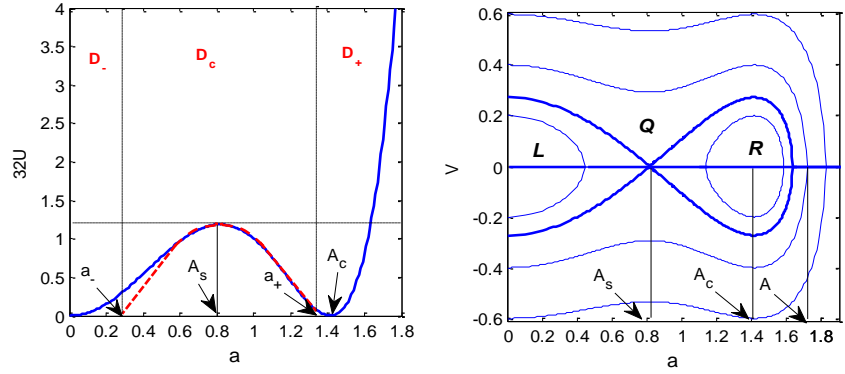

**Fig. 6.** Normalized potential and phase portraits of system (2.17) corresponding to different initial energy; $A_s$ – the saddle point; $A_c$ – the stable center; $A$– the amplitude of oscillations; red dashed lines in the left plot represent the quadratic approximation of the potential in the domain $D_C$

The coordinates of the saddle point $A_s$ and the stable center $A_c$ are defined by the equality $u(a) = 0$, that is, $A_s = \sqrt{2/3} \approx 0.816$, $A_c = \sqrt{2} \approx 1.414$. The amplitude of oscillations $A$ may be calculated by Eq. (2.13) at $\Delta = 0$ and $\Delta = \pi$, that is $A|A^2 - 2| = 4f$; the half-period $T(A)$ is calculated by formula [161]

$$T(A) = \int_0^A \frac{da}{\sqrt{f^2 - 2U(a)}} \,. \tag{2.19}$$

### 2.1.3. Non-smooth approximations

We pay the maximum attention to highly-energetic non-stationary regimes with large amplitudes of oscillations occurring at $f > f_2$. In this case, motion along the LPT is characterized by a saw-tooth envelope $a(\tau_1)$ (Fig. 5(c)). This underlines the fact that important essentially nonlinear phenomena (such as this one) may be missed when resorting to perturbation techniques based on linear (harmonic) generating functions, whose range of validity is restricted to stationary or non-stationary, but non-resonance motions. The similarity of the LPT to a saw-tooth function admits an approximation of strongly-nonlinear highly-energetic smooth oscillations by a non-smooth response of a free particle moving with constant velocity between two rigid walls $a = 0$ and $a = A$, where $A$ is a maximum amplitude of oscillations. A detailed exposition of a connection between smooth and vibro-impact modes of motion in a smooth nonlinear oscillator can be found in [149, 200], and references therein.

Referring to the method of non-smooth transformations [149], we introduce the pair of non-smooth basic functions $\tau(\phi)$, and $e(\phi) = d\tau/d\phi$, $\phi = \Omega\tau_1$ by formulas

$$\tau(\phi) = \frac{2}{\pi} |\arcsin(\sin\frac{\pi\phi}{2})|,$$

$$e(\phi) = d\tau/d\phi = \operatorname{sgn}[\sin(\pi\phi)], \qquad\qquad (2.20)$$

$$\frac{d}{d\tau_1} = \Omega[e(\phi)\frac{\partial}{\partial\tau} + \frac{\partial}{\partial\phi}],$$

where $\Omega = 1/T$; the half-period $T$ will be defined below. Plots of non-smooth functions (2.20) are presented in Fig. 7.

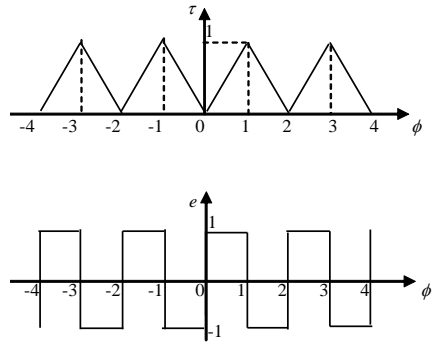

**Fig. 7.** Functions $\tau(\phi)$ and $e(\phi)$

It was shown in earlier works [111-113] that the solutions of Eqs. (2.11) can be expressed through the non-smooth functions by formulas

$$a(\tau_1) = X(\tau), \Delta(\tau_1) = e(\phi)Y(\tau), \qquad\qquad (2.21)$$

$$\frac{d}{d\tau_1} = \Omega(e\frac{\partial}{\partial\tau} + \frac{\partial}{\partial\phi}).$$

where $X(\tau)$, $Y(\tau)$ are "smooth functions of non-smooth variable". Although the functions $\Delta(\tau_1)$, $da/d\tau_1$, and $d\Delta/d\tau_1$ are formally continuous at $\phi = 2n - 1$, $\tau = 1$ and discontinuous at $\phi = 2n$, $\tau = 0$, Eqs. (2.21) yield discontinuity at both $\phi = 2n - 1$, $\tau = 1$ and $\phi = 2n$, $\tau = 0$. Singularity at $\phi = 2n - 1$ vanishes if $Y(\tau) = 0$ at $\tau = 1$. Formally, one can impose this smoothening condition in order to eliminate singular terms from the resulting equations but, as shown below, this condition holds by virtue of the dynamical equations.

To derive the equations for $X$, $Y$, we insert (2.21) into (2.11) and then separate the terms with and without the coefficient $e$. This yields the set of equations

$$\Omega \frac{dX}{d\tau} = -f \sin Y,$$

$$X \frac{dY}{d\tau} + X - X^3 = -f \cos Y,$$

(2.22)

with initial conditions $X = 0$, $Y = -\pi/2$ at $\tau = 0^+$. It is obvious that

$$\frac{dX}{d\tau} = 0 \text{ at } Y = 0.$$

It now follows that equality $Y(1) = 0$ has a simple physical meaning: it means that $X_1$ is maximal at $Y = 0$, $\tau = 1$.

In order to assess the amplitude and the period of oscillations, we reduce (2.22) to the second-order form similar to (2.17). Transformations of the same sort as above reduce (2.22) to the second-order equation

$$\Omega^2 \frac{d^2X}{d\tau^2} + u(X) = 0,$$

(2.23)

with initial conditions $X = 0$, $dX/d\tau = f/\Omega$ at $\tau = 0^+$; the function $u(X) = dU/dX$ is given by (2.18). Using the vibro-impact approximations, we calculate successive iterations by formulas

$$X = x_0 + x_1 + \dots, \ Y_2 = y_0 + y_1 + \dots,$$

(2.24)

where it is assumed that $|x_1(\tau)| << |x_0(\tau)|$, $|y_1(\tau| << |y_0(\tau)|$ in the interval $0 \leq \tau \leq 1$. The leading-order approximation $x_0$ is chosen as the solution of the equation of free particles moving between the walls, namely,$d^2x_0/d\tau^2 = 0$,with initial conditions $x_0 = 0$, $dx_0/d\tau = f/\Omega$ at $\tau = 0^+$. This yields the following non-smooth approximations:

$$x_0(\phi) = a_0(\phi) = A_0\tau(\phi), \ v_0(\phi) = A_0\text{sgn}(\sin 2\phi),$$

(2.25)

$$y_0(\phi) = \Delta_0(\phi) = -(\pi/2)e(\phi), \quad \phi = \Omega_0\tau_1, \quad \Omega_0 = 1/T_0.$$

By construction, the inverse transformation $a_0 = A_0\tau(\Omega_0\tau_1)$ produces the saw-tooth periodic solution associated formally with the vibro-impact process. The generating half-period $T_0 = 1/\Omega_0$ is defined as $T_0 = A_0/f$, where $A_0$ is calculated after the substitution the latter equality into expression (2.19).

The first-order term $x_1$ is governed by the equations

$$\frac{d^2 x_1}{d\tau^2} = -\Omega_0^{-2}u(a_0), \quad x_1(\tau) = -\Omega_0^{-2}\int_0^\tau (\tau-\xi)u(A_0\xi)\mathrm{d}\xi.$$

Integration by parts together with formulas (2.24) gives

$$a_1(\tau) = a_0(\tau) + x_1(\tau) = A_0\tau - \frac{A_0\tau^3}{8\Omega_1^2}\left[\frac{(A_0\tau)^4}{28} - \frac{(A_0\tau)^2}{5} + \frac{1}{3}\right], \tag{2.26}$$

$$\Delta_1(\tau) = \Delta_0(\tau) + y_1(\tau) = e(\phi)\left[-\frac{\pi}{2} + \frac{\tau}{\Omega_1}\left(-\frac{1}{2} + \frac{(A_0\tau)^2}{4}\right)\right],$$

(the detailed derivation of (2.26) can be found in [111, 113]). Note that the solution (2.26) is constructed as a function of $\tau$ but the inverse transformation $\tau \to \tau_1$ by formulas (2.20) automatically yields the solution periodic in $\tau_1$ (Fig. 8). A more detailed consideration of this approach as well as an analysis of numerical solutions can be found in [111-113].

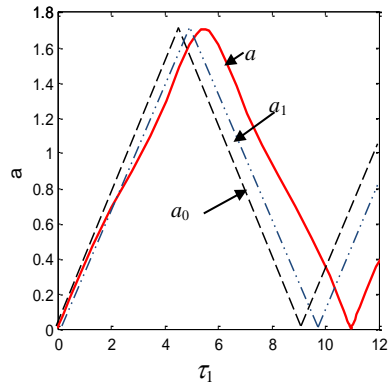

**Fig. 8**. Exact solution $a(\tau_1)$ (solid), approximations $a_0(\tau_1)$ (dash) and $a_1(\tau_1)$ (dash-dot) for Eq. (2.17) with $f = 0.385$.

*2.1.4. Transient dynamics of the dissipative system*

In this section we apply the LPT concept to examine high-energy non-stationary oscillations of a weakly damped oscillator. The system with parameters $f = 0.385$, $\gamma_1 = 0.05$ is considered as an example. In Fig. 9, one can observe two stages of motion: in the initial interval $0 \leq \tau_1 \leq \tau^*$ the trajectory is close to the LPT of the conservative system ($\gamma_1 = 0$), whereas in the second interval $\tau_1 > \tau^*$, motion is similar to quasi-linear oscillations and then converges to stationary oscillations of constant amplitude; an instant $\tau^*$ corresponds to the first maximum of the amplitude $a(\tau_1)$. Thus the first part of the trajectory may be approximated by the previously obtained segment of the LPT for the non-dissipative system; the matching point is $a(\tau^*) = A$ at $\tau^* = T_0 = A/f$, with amplitude $A$ calculated form (2.19).

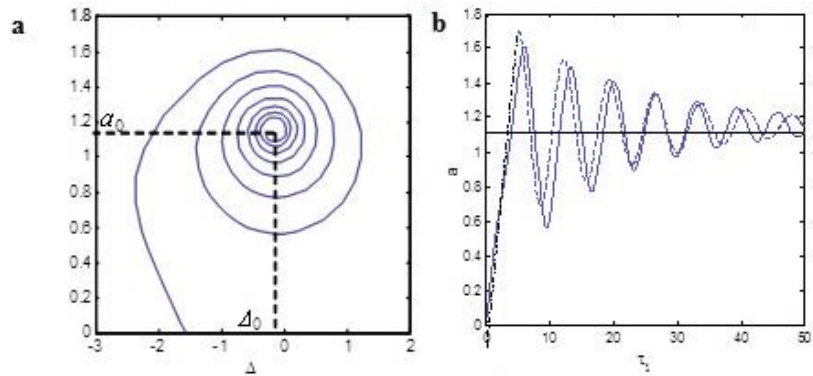

**Fig. 9**. (a) - phase portrait of weakly dissipative nonlinear oscillator with parameters $f = 0.385$, $\gamma_1 = 0.05$; (b) - plot of $a(\tau_1)$:exact (numerical) solution (solid); segment (2.25) (dot-and-dash); segment (2.27) (dash)

In the second interval of motion the trajectory of the dissipative system tends to the steady state $O$: $(a_0, \Delta_0)$ as $\tau_1 \to \infty$. The point $O$: $(a_0, \Delta_0)$ is defined by the equality $a^2[(1 - a^2)^2 + \gamma_1^2] = f^2$, or, for sufficiently small $\gamma_1$,

$\gamma_1 a_0 = -f \sin\Delta_0$, $a_0 - a_0^3 = -f \cos\Delta_0$,

$\Delta_0 \approx -\gamma_1 a_0/f + O(\gamma_1^3)$, $a_0(1 - a_0^2) = -f + O(\gamma_1^2)$.

Let $\xi = a - a_0$, $\delta = \Delta - \Delta_0$ be deviations from the steady state satisfying the matching conditions $a_0 + \xi = A$, $d\xi/d\tau_1 = 0$ at $\tau_1 = T_0$. Assuming negligible contributions of nonlinearity in small oscillations near $O$, the system linearized near $O$ can be considered. The linearized equations are given by

$$\frac{d\xi}{d\tau_1} + \gamma_1\xi + f\delta = 0, \frac{d\delta}{d\tau_1} - k\xi + \gamma_1\delta = 0.$$

where $k = 2a_0 + f/a_0^2$. This yields

$$\xi(\tau_1) = c_0 e^{-\gamma_1(\tau_1 - T_0)}\cos\kappa(\tau_1 - T_0), \ \delta(\tau_1) = rc_0 e^{-\gamma_1(\tau_1 - T_0)}\sin\kappa(\tau_1 - T_0), \ \tau_1 > T_0, \qquad (2.27)$$

where $c_0 = A - a_0$, $\kappa = (fk)^{1/2}$, $r = \kappa/f$. Figure 5 demonstrates a good agreement between the numerical solution of Eq. (2.9) (solid line) and its approximations. Despite a certain discrepancy in the initial interval, the numerical and analytic solutions closely approach as $\tau_1$ increases. These simplest matched approximations are sufficient to describe a complicated near-resonance dynamics. Note that the simplicity of the obtained solutions, which results from the attention to the physical properties of the system and effective treatment of the LPT theory is contrasted with the daunting complexity of the traditional analysis of transient nonlinear processes.

## 2.2. *Duffing oscillator subject to biharmonic forcing near the primary resonance*

In this section we perform an analytical investigation of non-stationary processes in the Duffing oscillator subjected to bi-harmonic forcing under conditions of a primary resonance. First, we employ the earlier presented LPT methodology to investigate a non-dissipated system with a bi-harmonic excitation. We demonstrate that the presence of an additional harmonic with a slowly changing frequency entails recurrent transitions from one type of LPT to another one. Next, we investigate the occurrence of relaxation oscillations in a lightly damped system. It is also demonstrated that the mechanism of relaxations may be approximated and explained through the existence LPTs characterized by a strong energy exchange between a single oscillator and an external source of energy. It is shown that the results of analytical approximations and numerical simulations are in a quite satisfactory agreement.

### 2.2.1. Equations of fast and slow motion

We investigate dimensionless weakly nonlinear oscillator subjected to a bi-harmonic excitation in the neighborhood of 1:1 resonance. The equation of motion is given by

$$\frac{d^2u}{d\tau_0^2} + 2\varepsilon\gamma\frac{du}{d\tau} + u + 8\alpha\varepsilon u^3 = 2\varepsilon[\tilde{F}_1\sin(1+\varepsilon s_1)\tau_0 + \tilde{F}_2\sin(1+\varepsilon s_2)\tau_0]. \tag{2.28}$$

where $s_2 = (1 + \varepsilon\sigma)s_1$. External forcing in this system consists of two distinct harmonic components with close frequencies. We will refer to this type of excitation as a biharmonic (quasi-periodic) one.

We recall that the maximum energy transfer from the source of energy to the oscillator happens if the oscillator is initially at rest, that is $u = 0$, $v = du/d\tau_0 = 0$ at $\tau_0 = 0$. These initial conditions define the above-introduced *limiting phase trajectory* (LPT) of the oscillator with biharmonic forcing.

To derive an analytical solution of Eq. (2.28), we invoke the complex-valued transformation $(u, v) \rightarrow (Y, Y^*)$ and then apply the multiple scales decomposition $Y = \varphi^{(0)} + \varepsilon\varphi^{(1)} + \dots$. Taking into account that the excitation directly depends on the fast time scale $\tau_0$, the slow time scale $\tau_1 = \varepsilon s\tau_0$, and the super-slow time scale $\tau_2 = \varepsilon\tau_1$, the leading-order slow term $\varphi^{(0)}$ as well as and the expansion of the full time-derivative are presented as

$$\varphi^{(0)} = \varphi^{(0)}(\tau_1, \tau_2), \ d\varphi^{(0)}/d\tau_1 = \partial\varphi^{(0)}/\partial\tau_1 + \varepsilon\partial\varphi^{(0)}/\partial\tau_2. \tag{2.29}$$

Therefore, the leading-order equation for $\varphi_0$ includes the slow and super-slow time scales:

$$s\frac{\partial\varphi^{(0)}}{\partial\tau_1} + \gamma\varphi^{(0)} - 3i\alpha|\varphi^{(0)}|^2\varphi^{(0)} = -i\tilde{F}_1 e^{i\tau_1} - i\tilde{F}_2 e^{i(\tau_1 + \sigma\tau_2)} \ , \ \varphi^{(0)}(0) = 0. \tag{2.30}$$

Introducing rescaling of the variables and the parameters

$$\delta(\tau_2) = \sigma\tau_2, \ \varphi^{(0)}(\tau_1) = \Lambda\varphi(\tau_1) e^{i\tau_1}, \ \Lambda = (s/3\alpha)^{1/2}, \ F_j = \tilde{F}_j/s\Lambda = \tilde{F}_j\sqrt{3\alpha/s^3} \ , \ \gamma_1 = \gamma/s, j = 1, 2,$$

we transform (2.30) into a more convenient form similar to (2.6)

$$\frac{\partial \varphi}{\partial \tau_1} + \gamma_1 \varphi + i\varphi(1 - |\varphi|^2) = -i(F_1 + F_2 \exp(i\delta(\tau_2))), \varphi(0) = 0. \tag{2.31}$$

### 2.2.2. LPTs of slow motion in a non-dissipative system

In this section, we consider a non-dissipative model with the "frozen" phase $\delta$. The equation of motion is given by

$$\frac{\partial \varphi}{\partial \tau_1} + i\varphi(1 - |\varphi|^2) = -i(F_1 + F_2 \exp(i\delta), \quad \varphi(0) = 0. \tag{2.32}$$

After introducing a polar decomposition $\varphi = a\, e^{i\Delta}$, we transform(2.32) into the equations for the real valued amplitude $a \geq 0$ and phase $\Delta$

$$\frac{\partial a}{\partial \tau_1} = -F_1 \sin\Delta + F_2 \sin(\delta - \Delta) = 0, a(0) = 0. \tag{2.33}$$

$$a\frac{\partial \Delta}{\partial \tau_1} = -a + a^3 - F_1 \cos\Delta - F_2 \cos(\delta - \Delta).$$

As in Section 2.1, we investigate the system dynamics through the analysis of LPTs. It is easy to prove that system (2.33) possesses the integral of motion (with respect to the time scale $\tau_1$)

$$K = a(\frac{a^3}{4} - \frac{a}{2} - F_1 \cos\Delta - F_2 \cos(\delta - \Delta)), \tag{2.34}$$

and $K = 0$ on the LPT. As in the time-invariant case, the LPT has two branches; the first branch is trivial ($a = 0$), while the second branch satisfies the cubic equation similar to (2.13)

$$g(a, \Delta) = \frac{1}{4}a^3 - \frac{1}{2}a - F_1 \cos\Delta - F_2 \cos(\delta - \Delta) = 0. \tag{2.35}$$

Note that the above equation determines the phase $\Delta(a)$ on the LPT for any $\delta$. Since the point $a(0) = 0$ belongs to the LPT, we obtain from (2.35) the following expression for the initial phase $\Delta_0$ corresponding to a non-trivial branch of the LPT:

$$F_1 \cos \varDelta_0 + F_2 \cos(\delta - \varDelta_0) = 0,$$

$$\varDelta_0 = -\arctan \frac{F_1 + F_2 \cos \delta}{g_2 \sin \delta} + n\pi. \tag{2.36}$$

In the next step, we study bifurcations of the slow motion. We begin with the analysis of the stationary (in $\tau_1$) points of system (2.33). By letting $da/d\tau_1 = 0$, $d\varDelta/d\tau_1 = 0$, one obtains the following algebraic equations for the stationary points $(a_s, \varDelta_s)$:

$$F_1 \sin \varDelta - F_2 \sin(\delta - \varDelta) = 0, \tag{2.37}$$

$$a - a^3 + f = 0,$$

where $f = \pm(F_1 \cos \varDelta_s + F_2 \sin(\delta - \varDelta_s))$. It follows from the first equation in (2.37) that the coordinates $\varDelta_s$ of the stationary points in the phase plane are given by

$$\varDelta_s = \arctan \frac{F_2 \sin \delta}{F_1 + F_2 \cos \delta} + n\pi. \tag{2.38}$$

It is important to note that the second equation in (2.37) formally coincides with Eq. (2.14) and thus determines quasi-stationary points depending on the "frozen" phase $\delta$. This implies that one can consider the critical values (2.15), (2.16) of the parameter $f$ as the boundaries separating different types of motion.

As in Section 2.1, we obtain two critical relationships determining locations of the centers and the shape of the phase orbits of system. We recall that the critical value $f_1 = \sqrt{2/27}$ characterizes the boundary between quasi-linear oscillations with relatively small amplitude ($|f| < f_1$) and moderately nonlinear oscillations with larger amplitude at $f_1 < |f| < f_2$. The threshold $f_2 = 2/\sqrt{27}$ corresponds to the transition from moderately nonlinear ($f_1 < |f| < f_2$) to strongly nonlinear ($|f| > f_2$) regimes with large amplitudes and energy.

It is worth mentioning that, in virtue of (2.37), (2.38), the reduced forcing amplitude $f$ directly depends on the parameters $F_1$, $F_2$, $\delta$, and a proper choice of these parameters may provide zero forcing (in $\tau_1$) such that $\partial a/\partial \tau_1 = 0$ in (2.33). In this case, the bifurcation analysis presented in this section becomes unacceptable, as it formally corresponds to an unforced response of the nonlinear oscillator.

### 2.2.3. Super-slow model

Next we consider the super-slow evolution of the LPT due to monotonous variations of the parameter $\delta(\tau_2)$. It was mentioned that in the system under investigation the parameter $f$ is $\delta$-dependent. Furthermore, since the phase $\delta(\tau_2)$ monotonously varies with respect to the super-slow time scale $\tau_2$, all bifurcation parameters are also time-varying. This also means that global changes in the system dynamics may arise as the time increases. We show that at $\delta = 0$ there exists the LPT of the second type (strongly nonlinear high-energy oscillations with a single stable center in the phase plane) which, after a certain time interval, will bifurcate to the LPT of the first type (moderately nonlinear oscillations) due to the parametric switching from $|f(\delta(\tau_2))| > f_2$ to $|f(\delta(\tau_2))| < f_2$.

As shown in Fig. 10, the LPT starting at $\delta = 0$ undergoes global bifurcations. It is evident that, as the parameter $\delta$ increases, the LPT slowly changes, and the left and right corner points of the LPT meet at the saddle point at $\delta_{cr} = 1.99$. The coincidence of the corner points brings about the global bifurcation that finally results in the disappearance of strongly nonlinear regime and the transition to the LPT of moderately nonlinear regime (cf. Fig. 5).

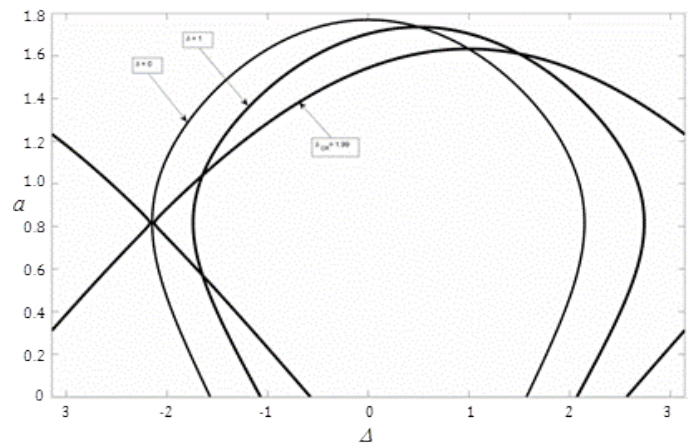

**Fig. 10.** Evolution of the LPT of the second type (strongly nonlinear oscillations) up to the transition to the LPT of the first type (moderately nonlinear oscillations)

The sequence of LPTs arising right after the bifurcation is illustrated in Fig. 11 for several values of $\delta$. It is clearly seen that an increase in phase detuning $\delta(\tau_2)$ is equivalent to deviations of the excitation frequency from resonance, which, in turn, entails a decreasing amplitude and passage from one type of oscillations to another one.

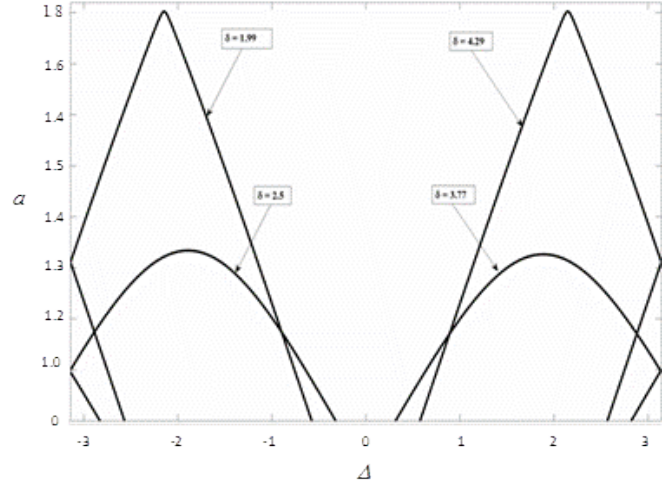

**Fig. 11.** Formation of the LPTs of the first type and their super-slow evolution till the time point of additional transition to the LPT of the second type

One may observe in Fig. 11 that the aforementioned bifurcation leads to the transition from a strongly nonlinear regime to a moderately nonlinear one. The LPT gradually evolves with the variation of $\delta$ in time until it reaches a certain critical point beyond which another transition from the moderately to strongly nonlinear regime is observed. These transitions occur recurrently in large time intervals.

We now estimate analytically the critical values $\delta_{cr}$, at which the transitions may occur. First, we use Eq. (2.35) to find the values of the phase $\Delta$ and the amplitude $a_{cr}$, at which $\partial\Delta/\partial a = 0$. The critical amplitude $a_{cr}$ is defined as follows:

$$\frac{\partial g(a, \Delta(a))}{\partial a}\bigg|_{\frac{\partial\Delta}{\partial a}=0} = \frac{3}{2}a^2 - 1 = 0, \ a_{cr} = \sqrt{2/3} \ . \tag{2.39}$$

Substituting $a = a_{cr}$ into (2.35) yields:

$$2F_1\cos\Delta + 2F_2\cos(\delta - \Delta) = \frac{1}{2}a_{cr}^3 - a_{cr} \ . \tag{2.40}$$

Now we find the critical value $\delta_{cr}$ at which two stationary points collide. To this end, we derive the condition for Eq. (2.40) to have a single solution. This can be achieved by equating the amplitude of the left-hand side "oscillating part" (with respect to $\Delta$) to that of the right-hand side. This implies that the critical value $\delta = \delta_{cr}$ satisfies the following equations:

$$\delta_{cr} = \pm \arccos \frac{(\frac{a_{cr}^3}{2} - a_{cr})^2 - 4(F_1^2 + F_2^2)}{8F_1F_2}, a_{cr} = \sqrt{\frac{2}{3}}. \tag{2.41}$$

### 2.2.4. Relaxation oscillations in a lightly-damped system

In this section we demonstrate the possibility of relaxation oscillations in the lightly damped system (2.31) and a connection between the trajectories of relaxation oscillations and the LPTs.

We recall that relaxation oscillations are characterized by the occurrence of recurrent segments of fast and slow motion. In order to highlight the super-slow motion, we rewrite Eq. (2.31) as a singular equation

$$\varepsilon \frac{\partial \varphi}{\partial \tau_2} + \gamma_1 \varphi + i\varphi(1 - |\varphi|^2) = -i(F_1 + F_2 \exp(i\delta(\tau_2))), \varphi(0) = 0. \tag{2.42}$$

The limit as $\varepsilon \to 0$ gives the following algebraic equation

$$i\Phi(1 - |\Phi|^2) + \gamma_1 \Phi = -i(F_1 + F_1 e^{i\sigma\tau_2}). \tag{2.43}$$

with trajectories depending on the super-slow time $\tau_2$. It follows from (2.42), (2.43) that $\Phi(\tau_2) = \lim \varphi(\tau_1, \tau_2)$ as $\varepsilon \to \infty$. Thus, $\Phi(\tau_2)$ can be interpreted as a quasi-stationary value of the complex amplitude $\varphi_0$ with "frozen" parameter $\tau_2$.

The transformation $|\Phi| = \sqrt{Z}$ reduces Eq. (2.43) to the real-valued form:

$$\gamma_1^2 Z + (1 - Z)^2 Z = Q(\tau_2), \tag{2.44}$$

where $Q(\tau_2) = F_1^2 + F_2^2 + 2F_1F_2\cos\sigma\tau_2$ represents the quadratic amplitude of excitation. The function $|\Phi| = \sqrt{Z}$ obviously evaluates the amplitude of oscillations $a = |\varphi_0|$ as $\tau_1 \to \infty$.

The plot of $\Phi$ versus $Q$ is presented in Fig. 12. It is seen that the plot may be folded at certain parameters. This folded structure contains two stable branches and one unstable branch (Fig. 12).

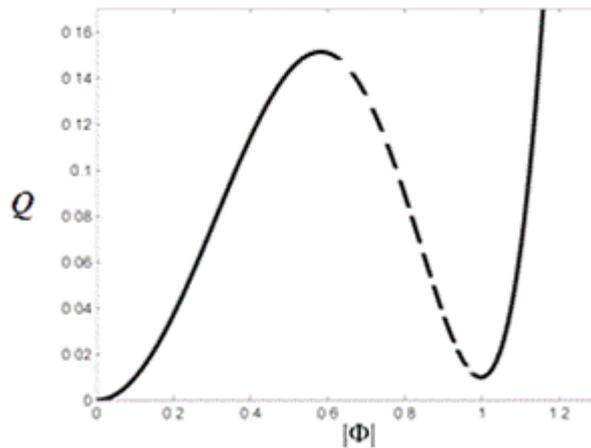

**Fig. 12.** LPT of the second type (before bifurcations occur).

It was shown in earlier works [3,52,54,57,58,179-183,188] that relaxation oscillations are characterized by consequent jumps from one stable branch to another one, accompanied by a super-slow evolution on each of the stable branches. To find the conditions of jumps, expression (2.44) is differentiated with respect to $Z$. Equating the derivative to zero, we obtain the equation

$$3Z^2 - 4Z + \gamma_1^2 + 1 = 0 \tag{2.45}$$

which determines extreme points. It is easy to deduce that a real-valued solution $Z > 0$ exists if $\gamma_1^2 < 1/3$. Therefore, above critical damping $\gamma_{cr} = \sqrt{1/3}$, relaxation is impossible.

It follows from Eq. (2.44) that motion on the stable branches is fully governed by the forcing amplitudes $F_1, F_2$. It is easy to deduce that, if $F_1 \neq 0$, $F_2 = 0$, then there exist only stationary points on the stable branches of the manifold. If $F_1 \neq 0$, $F_2 \neq 0$, then there are three possibilities: the first two types of motion correspond to continuous oscillations on each of the stable branches; the third type corresponds to the aforementioned relaxation oscillations.

The conditions of the occurrence of relaxation oscillations can be found from (2.44), (2.45). Solutions of Eq. (2.45) define the values of $Z$ corresponding to the fold points:

$$Z_{1,2} = \tfrac{2}{3}(1 \pm \sqrt{1 - \tfrac{3}{4}(\gamma_1^2 + 1)}). \tag{2.46}$$

Substituting (2.46) into (2.44), we find the corresponding excitation amplitude $Q$

$$Q_j = \gamma_1^2 Z_j + (1 - Z_j)^2 Z_j, \, j = 1,2. \tag{2.47}$$

Now one may formulate the following necessary conditions ensuring the regime of relaxation:

$$F_1^2 + F_2^2 + 2F_1 F_2 > Q_1, \, F_1^2 + F_2^2 - 2F_1 F_2 < Q_2 \tag{2.48}$$

The effect of relaxation is illustrated in Figs. 13, 14. In Fig. 13 one can compare numerical results and analytical approximations for the amplitude of oscillations on the stable branches of the manifold provided conditions (2.48) are not fulfilled ($F_1 = F_2 = 0.065$).

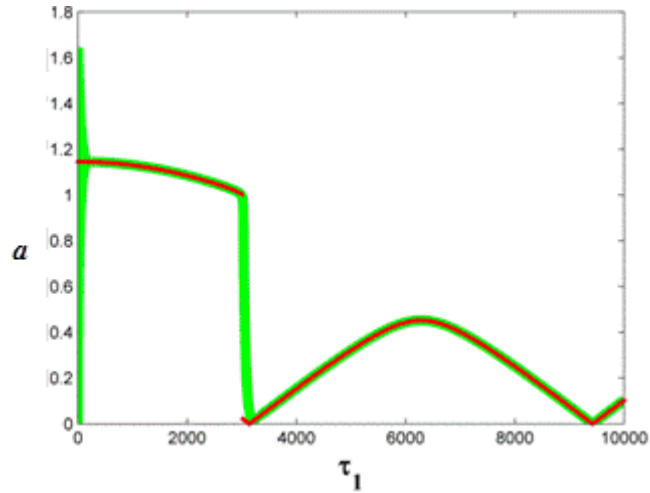

**Fig. 13.** Comparison of numerical and analytical results in the absence of the stable relaxation regime: green line – analytical approximations, brown lines – numerical simulations.

As shown in Fig. 13, there exists an initial phase of relaxation oscillations with a super-slow change of the amplitude corresponding to the upper stable branch, and then there is a single jump to the lower stable branch, wherein the system continues evolving. No additional relaxation oscillations are possible for this case, since conditions (2.48) are not fulfilled. If the forcing parameters satisfy (2.48), then there exist relaxation oscillations on the upper and lower stable branches (Fig. 14).

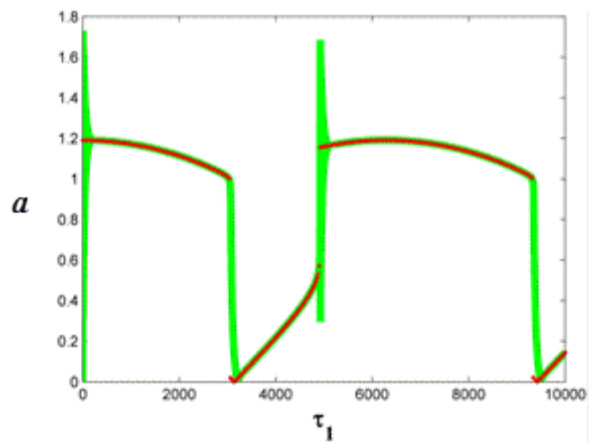

**Fig. 14.** Comparison of numerical and analytical results for the case of relaxation oscillations: green lines – analytical approximations, blue lines – numerical simulations.

The super-slow flow analysis does not demonstrate a global dynamical picture of the lightly damped system. To make the analysis complete, we need to study the slow flow dynamics, which corresponds to the phase of relaxations from one stable branch of the super-slow manifold to another one followed by quasi-linear damped oscillations. We study the dynamics of a weakly damped oscillator, in which $0 < \gamma < \gamma_{cr}$. As in the previous sections, it is rather natural to assume that during the relaxation period $T^*$ (the time required for the trajectory emanating from the fold of a super-slow surface to reach its first peak) the weakly damped trajectory runs sufficiently close to the non-dissipative one, and motion during this period is very close to motion along the corresponding LPT of the non-dissipative system.

In order to illustrate the correspondence of the LPT of strongly nonlinear oscillations to the initial phase of relaxation ( $0 < t < T^*$ ), we plot the LPT corresponding to the value of $\delta$ at the point of relaxation on the phase plane of the damped system (Fig. 15).

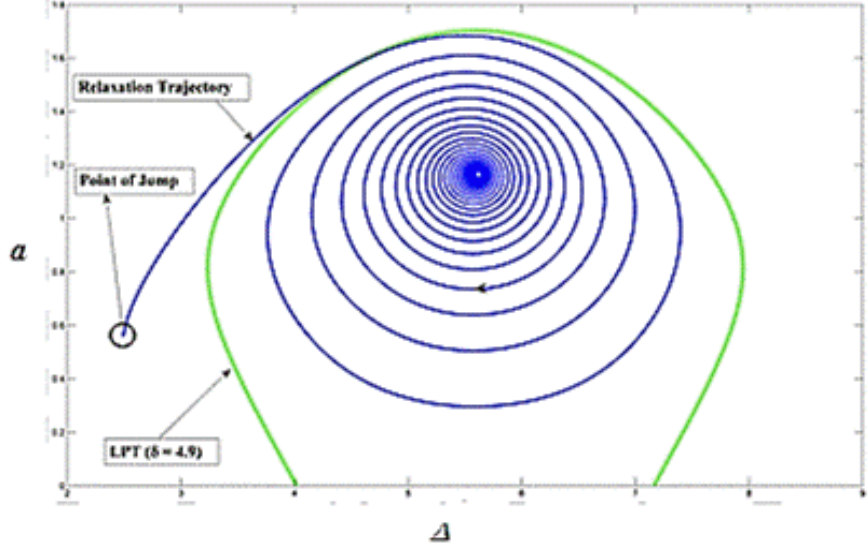

**Fig. 15.** Comparison of the damped trajectory of relaxation with the LPT of the corresponding Hamiltonian system

It is evident from Fig. 15 that the damped trajectory corresponding to a jump from the lower branch of the super slow manifold wraps up the LPT of the corresponding non-dissipative system, thus reaching the maximum value in the close vicinity of the LPT. This also means that the LPT of the non-dissipative system predicts fairly well the maximum amplitude of oscillations. It is also clear from Fig. 15 that after reaching the peak of the response, the amplitude of oscillations diminishes and the response becomes quasi-linear. As in Section 2.1, this response may be described with the help of the model linearized around the upper stable branch.

We now calculate the critical values of $\delta$ in the lower folds, corresponding to the minimal relaxation amplitude. It follows from (2.46) that the amplitude of the lower fold equals to:

$$Z_1 = \tfrac{2}{3}(1 - \sqrt{1 - \tfrac{3}{4}(\gamma_1^2 + 1)}).$$ (2.49)

Thus, according to (2.44), one obtains

$$Q_1 = F_1^2 + F_2^2 + 2F_1 F_2 \cos\delta_{cr},$$
$$\delta_{cr} = \pm\arccos(\frac{Q_1 - F_1^2 - F_2^2}{2F_1 F_2}).$$ (2.50)

The amplitudes of oscillations and the sequences of the LPTs in the corresponding undamped system at the points of jump are demonstrated in Fig.16. As one may observe, the LPTs are reconstructed accordingly and provide fairly good estimations for jumps.

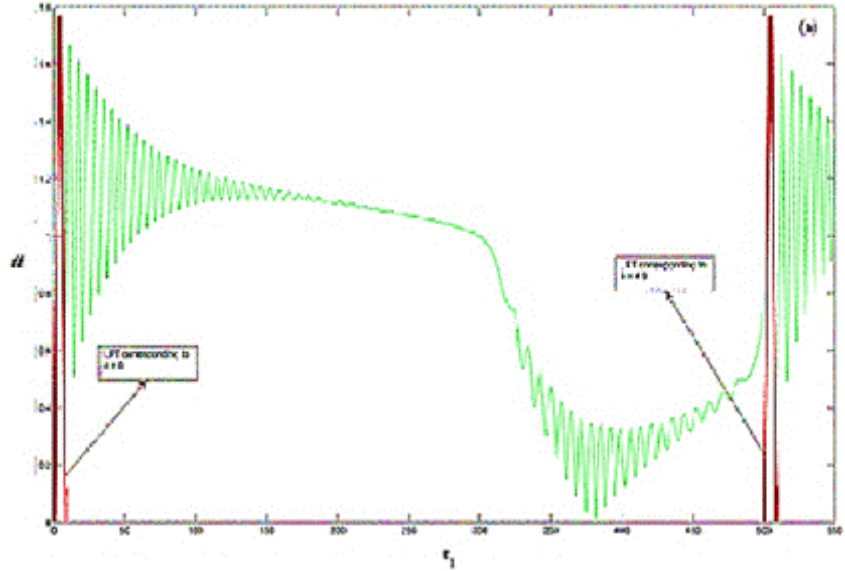

**Fig. 16.** Sequence of jumps of the LPT for relaxation oscillations; the actual response is marked with the green solid line.

As mentioned above, at the initial stage of motion, wherein the effect of damping is negligibly small, the trajectory of a lightly damped system is close to the LPT of a corresponding non-dissipated system (Fig. 15) but at the second stage, after reaching the first peak of the relaxation phase with the coordinates $a(T^*) = a_m$, $\Delta(T^*) = \Delta_m$, motion of a lightly damped system may be described fairly well by linearizing (2.45) near the upper stable branch of the super-slow flow manifold. Thus, assuming small deviations near the upper stable branch, one may suggest the following approximation:

$$\varphi(\tau_1, \tau_2) = \Phi(\tau_2) + \phi(\tau_1, \tau_2).$$ (2.51)

Substituting (2.51) into (2.31), we obtain the linearized equation

$$\frac{\partial \phi}{\partial \tau_1} + i(1 - 2|\Phi|^2)\phi + \gamma\phi - i\Phi^2\phi^* = 0.$$ (2.52)

Equation (2.52) describes the last stage of damped oscillations near the stable attractor. Initial conditions are taken at the point $a(T^*) = a_m$, $\Delta(T^*) = \Delta_m$.

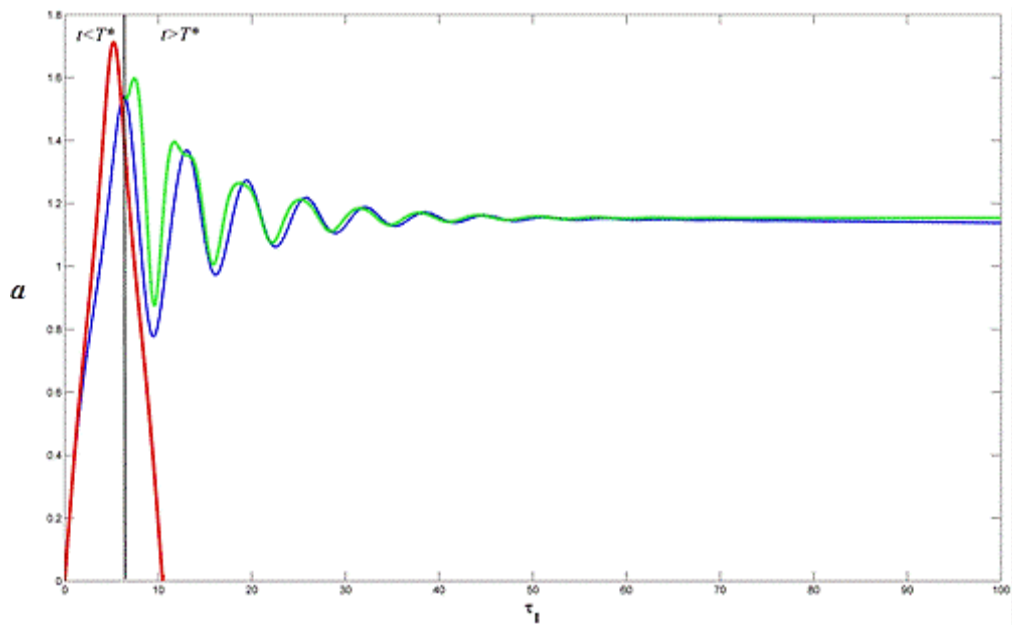

**Fig. 17.** Comparison of analytical approximations and numerical simulations

Comparison of analytical approximations with the results of numerical simulation is given in Fig. 17. It is seen that the interval of the slow-time scale oscillations can be analytically approximated to a quite satisfactory extent.

We now return to the consideration of the lightly damped system. From Fig. 12, it is seen that the upper fold point is slightly distant from the horizontal axis. Arguing as above, one can state that the value of damping parameter dictates the distance of the upper fold from the horizontal axis. At the same time, for sufficiently light damping the critical parameter $\delta_{cr}$ (corresponding to a point of jump) may be related to the non-dissipative case of the LPT of the first type.

Thus, if the jump from the upper stable branch to the lower one occurs far away from the LPT of the first kind (which is most likely to happen when forcing is small enough and thus hardly affects the damped response of large amplitude), its trajectory may be roughly approximated by the equation of free oscillations

$$\frac{\partial \varphi}{\partial \tau} + i\varphi - i|\varphi|^2 \varphi + \gamma\varphi \cong 0 \qquad (2.53)$$

This immediately yields the exponential decay of the response amplitude:

$$|\varphi| = \sqrt{Z_2} \exp(-\gamma\tau) \qquad (2.54)$$

We underline that Eq. (2.54) is valid only if the response amplitude far exceeds the amplitude of forcing. Therefore, if the trajectory of relaxation starts far away from the LPT of the first type, then the response is of simple exponential decay type. A comparison of the analytical approximation (2.54) with the actual (numerical) response is presented in Fig. 18. As shown in Fig. 18, relaxation oscillations may be approximated by the response of the damped unforced weakly nonlinear oscillator.

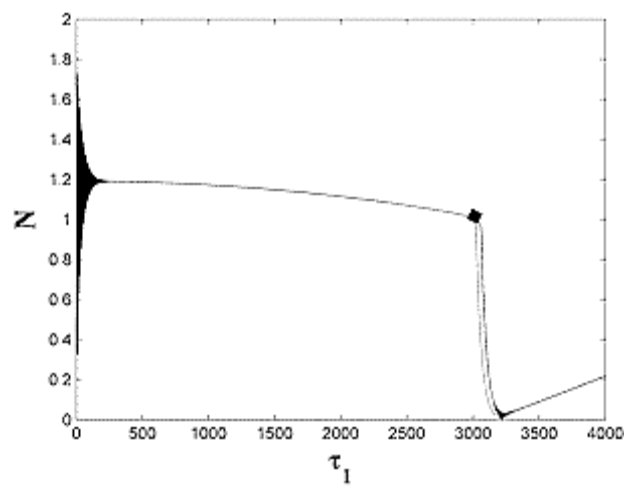

**Fig. 18.** Comparison of an analytical approximation of relaxation with the result of numerical simulation. Actual system response is denoted with the bold solid line; analytical approximation is denoted with thin line.

# 3.  Limiting Phase Trajectories of weakly coupled nonlinear oscillators. Classical-quantum analogy

In this section, we consider in details energy transfer between two weakly coupled oscillators. A special example of energy exchange between two identical oscillators was briefly discussed in Introduction. In this section we define the LPTs in a two-degree of freedom (2DOF) asymmetric system and examine the role of LPTs in the resonance energy exchange and localization. Targeted energy transfer in coupled oscillators was considered in a number of earlier published works [75, 99, 105, 106, 109, 117, 123, 125, 126, 151, 174, 193, 194]. We suggest a thorough dynamical analysis, which helps us highlight a similarity between classical and quantum systems and find the sets of parameters corresponding to quasi-linear, moderately nonlinear and strongly nonlinear dynamical behavior of the coupled systems. Besides, the suggested analysis underlies further study of coupled oscillators with slowly-varying parameters.

It is important to note that earlier work on nonlinear tunneling [94, 213] outlined a single parametric boundary associated with the change of the number of the stationary points and corresponding, in our terminology, to the boundary between quasi-linear and moderately nonlinear regimes. We derive the second parametric boundary corresponding to the boundary between moderately nonlinear and strongly nonlinear non-stationary dynamical behavior and then show that this dynamical transition is closely related to the behavior of Limiting Phase Trajectories (LPTs). This implies that the dynamical analysis cannot be reduced to the study of the stationary points and their stability.

## 3.1. Equations of motion and explicit approximate solutions

We consider two coupled oscillators with equal mass $m$; denoting by $c_1$ and $c_2$, the coefficients of linear stiffness of the corresponding oscillators; by $c$, the coefficient of cubic nonlinearity; by $c_{12}$, the stiffness of linear coupling; $u_1$ and $u_2$ are the absolute displacements of the first and second oscillators, respectively. As in the case of single oscillator, the small parameter $\varepsilon$ is defined through relative stiffness of weak coupling: $c_{12}/c_1 = 2\varepsilon << 1$. Then, assuming weak nonlinearity and taking into account resonance properties of the system, we redefine the system parameters as follows:

$$c_1/m = \omega_0^2, \ c_2/m = \omega_0^2(1 + 2\varepsilon g), \ \tau_0 = \omega_0 t, \ c/c_1 = 8\varepsilon\alpha, \ c_{12}/c_j = 2\varepsilon\lambda_j, \ j = 1, 2. \tag{3.1}$$

It follows from (3.1) that $\lambda_1 = 1$, $\lambda_2 = (1 + 2\varepsilon g)^{-1}$. The symmetric system case considered in Introduction corresponds to $g = 0$. Using parameters (3.1), we reduce the equations of motion to the form

$$\frac{d^2u_1}{d\tau_0^2} + u_1 + 2\varepsilon(u_1 - u_2) + 8\varepsilon\alpha u_1^3 = 0, \tag{3.2}$$

$$\frac{d^2u_2}{d\tau_0^2} + (1 + 2\varepsilon g)u_2 + 2\varepsilon(u_2 - u_1) + 8\varepsilon\alpha u_2^3 = 0.$$

Selected initial conditions in non-stationary case correspond to a unit impulse imposed to one of the oscillators with the system being initially at rest, i.e., $u_1 = u_2 = 0$; $v_1 = du_1/d\tau_0 = 1$, $v_2 = du_2/d\tau_0 = 0$ at $\tau_0 = 0$ if the impulse applied to the first oscillators, or $v_1 = du_1/d\tau_0 = 0$, $v_2 = du_2/d\tau_0 = 1$ if the initial impulse $v_2 = 1$ is applied to the second oscillator. These sets of initial conditions determine LPTs of system (3.2). The LPT concept for symmetric systems ($g = 0$), briefly discussed in Introduction, was earlier derived and employed in various applications [105,106,117]. An explanation and more detailed discussion of LPT properties and their connection with the resonance energy transfer in an asymmetric system ($g \neq 0$) is given below.

As in the previous section, an asymptotic solution of (3.2) for small $\varepsilon$ is based on complexification of the dynamics and then separation of the slow and fast constituents. To this end, we introduce the change of variables $v_j + iu_j = Y_j e^{i\tau_0}$, $j = 0,1$, and then present the complex amplitude as $Y_j(\tau_0, \tau_1, \varepsilon) = \varphi_j^{(0)}(\tau_1) + \varepsilon\varphi_j^{(0)}(\tau_0, \tau_1) + O(\varepsilon^2)$, $\tau_1 = \varepsilon t$. Then, using the multiple-scale techniques and separating the fast and slow time-scales (see Section 2.1), we derive the equations for the leading-order slow terms analogous to (2.6), from which we deduce that the main slow term are presented in the form

$$\varphi_1^{(0)}(\tau_1) = a(\tau_1)e^{i\tau_1}, \; \varphi_2^{(0)}(\tau_1) = b(\tau_1)e^{i\tau_1}, \tag{3.3}$$

with the complex envelopes $a(\tau_1)$ and $b(\tau_1)$ given by the equations similar to (1.5)

$$\frac{da}{d\tau_1} + ib - 3i\alpha|a|^2 a = 0, \tag{3.4}$$

$$\frac{db}{d\tau_1} + ia - 3i\alpha|b|^2 b - 2ig(\tau_2)b = 0.$$

Details of the derivation of Eqs (3.4) can be found in [105]. Initial conditions $a(0) = 1$, $b(0) = 0$ correspond to the excitation of the first oscillator with the second one being initially at rest; in the opposite case, when the second oscillator is excited but the first one is initially at rest, initial conditions are given by $a(0) = 0$, $b(0) = 1$. We recall that the solutions of Eqs. (3.4) with the above-mentioned initial conditions define not only the LPT of the averaged system but also the LPT of the initial system (3.2).

System (3.4) highlights similarity of the averaged equations of a classical oscillator to the equations of two-level quantum system [151, 174]. This similarity confirms a direct mathematical analogy between quantum and classical transitions, and the applicability of the results derived in this section to a wide class of physical problems.

The polar representation $a = \cos\theta e^{i\delta_1}$, $b = \sin\theta e^{i\delta_2}$, $\Delta = \delta_1 - \delta_2$, yields following real-valued equations:

$$\frac{d\theta}{d\tau_1} = \sin\Delta, \tag{3.5}$$

$$\sin 2\theta \frac{d\Delta}{d\tau_1} = 2(\cos\Delta + 2k\sin 2\theta)\cos 2\theta - 2g\sin 2\theta,$$

where $k = 3\alpha/4$. Initial conditions for system (3.5) are chosen as $\theta = 0$, $\Delta = \pi/2$ (the first oscillator is excited but the second one is initially at rest) or $\theta = \pi/2$, $\Delta = \pi/2$ (the second oscillator is excited but the first one is at rest). Both conditions correspond to the LPTs of system (3.5) in the plane ($\Delta$, $\theta$). Note that system (3.5) conserves the integral of motion

$$K = (\cos\Delta + k\sin2\theta) \sin2\theta + g\cos2\theta = const \qquad (3.6)$$

Properties of integral (3.6) are used in further analysis.

### 3.2. Stationary points and LPTs

The first step in the dynamical analysis is to define the stationary points of system (3.5). The first condition of stationarity $d\theta/d\tau_1 = 0$ yields $\sin\Delta = 0$; this implies that all steady states lie on the vertical axes $\Delta_1 = 0$ or $\Delta_2 = \pi$. The second condition $d\Delta/d\tau_1 = 0$ implies that the stationary values of $\theta$ are given by the equation

$$F_{\pm}(\theta) = (\pm 1 + 2k\sin2\theta)\cot2\theta = g, \qquad (3.7)$$

where the signs "+" and "−" correspond to $\Delta = 0$ and $\Delta = \pi$, respectively.

Figures 19–22 depict phase portraits of system (3.5) in the plane ($\Delta$, $\theta$) for different values of the parameters $k$ and $g$. In the quasi-linear case, when $k \leq 0.5$, there exists a unique solution of Eq. (3.7) on each of the axes $\Delta = 0$ and $\Delta = \pi$. It was recently shown [81] that in this case the solution of the nonlinear system is close to that of the linear system.

Phase portraits of system (3.5) for $k = 0.35$ and different $g$ are shown in Fig.19. It is easy to deduce from (3.5) that the change of the sign of the parameter $g \rightarrow -g$ entails the change of the solution: $\theta \rightarrow \pi - \theta$, $\Delta \rightarrow 2\pi - \Delta$. This allows the construction of the phase portraits only for $g \geq 0$ (Fig. 19). It is seen in Fig. 19 that the LPT represents an outer boundary for a set of closed trajectories encircling the stable center in the phase plane ($\Delta$, $\theta$).

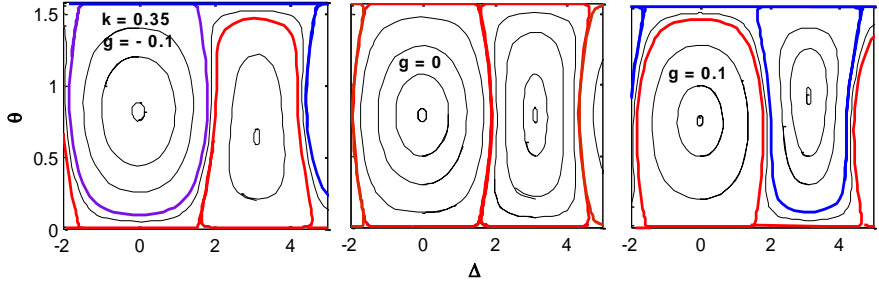

**Fig. 19.**Phase portraits for $k = 0.35$ and different values of $g$

In what follows, we study in detail the dynamics of the system with nonlinearity $k > 0.5$. It is seen in Figs. 20–22 that if $k > 0.5$, then there exists a certain value $g^*$ such that the system possesses 3 stationary points on the axis $\Delta = \pi$ for $|g| < |g^*|$ but it has a single point for $|g| \geq |g^*|$, and the transition occurs through the coalescence of two stationary states at a certain point $\theta_T$ such that $F_-(\theta_T) = g^*$. The condition $dF_-/d\theta = 0$ at a coalescence point of two roots of Eq. (3.7) yields the following expressions for $\theta_T$ and $g^*$:

$$\sin 2\theta_T = (2k)^{-1/3}, \quad g^* = \pm [(2k)^{2/3} - 1]^{3/2}. \tag{3.8}$$

It follows from (3.8) that 4 stationary states exist in the domain $k > 0.5$, $|g| < |g^*|$. Note that the parameter $g^*$ coincides with the critical parameter reported without proof in [94] (in different notations).

As shown below, the system with 4 stationary states exhibits two different types of dynamical behavior corresponding to moderately nonlinear and strongly nonlinear non-stationary regimes. First, we consider the system with $k = 0.65$. It is easy to deduce from (3.5) that the change of the sign of the parameter $g \to -g$ entails the change of the solution: $\theta \to \pi - \theta$, $\Delta \to 2\pi - \Delta$. This allows the construction of the phase portraits only for $g \geq 0$ (Fig. 20). Bold lines in Fig. 20 depict the LPT of system (3.5) in the plane $(\theta, \Delta)$; dotted lines correspond to the homoclinic separatrix. It is seen that with an increase of $g$ the lower homoclinic loop vanishes through the coalescence of the stable and unstable states, and the number of the stationary states changes from 4 to 2. The corresponding critical value $g^* = 0.083$ coincides with the theoretical value given by formula (3.8). It is easy to find that the number of the stationary points changes from 2 to 4 at $g^* = -0.083$.

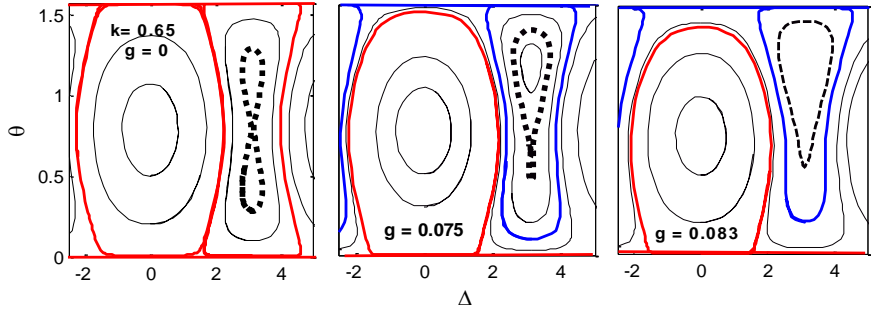

**Fig. 20.** Phase portraits for system (3.5) in the plane $(\Delta, \theta)$; $k = 0.65$, $g = 0, 0.075, 0.083$

Figure 20 illustrates complete energy exchange between the symmetric oscillators ($g = 0$), i.e., the upper level $\theta = \pi/2$ is reached during the cycle of motion along the LPT starting at $\theta = 0$. Motion along the closed orbits within the domain encircled by the LPT obviously provides less extensive energy exchange than motion along the LPT. Vanishing of the homoclinic separatrix and the emergence of a new closed orbit of finite period characterizes *moderately nonlinear non-stationary regime*. The phase portraits for $g < 0$ can be constructed by symmetry. In these systems the localization of energy near the lower center changes to the localization near the upper center. This phenomenon, responsible for the occurrence of nonlinear tunneling in the slowly time-dependent systems [94,191,213] underlies the study of transient processes in Section IV.

From Fig. 21 it is seen that the system with $k = 0.9$ exhibits a more complicated dynamical behavior. For clarity, the phase portraits for both $g < 0$ and $g > 0$ are shown; bold lines correspond to the LPTs; dash-dotted lines depict the homoclinic separatrix coinciding with the LPT; edges of the dashed "beaks" in lie at the points of annihilation of the stable and unstable states.

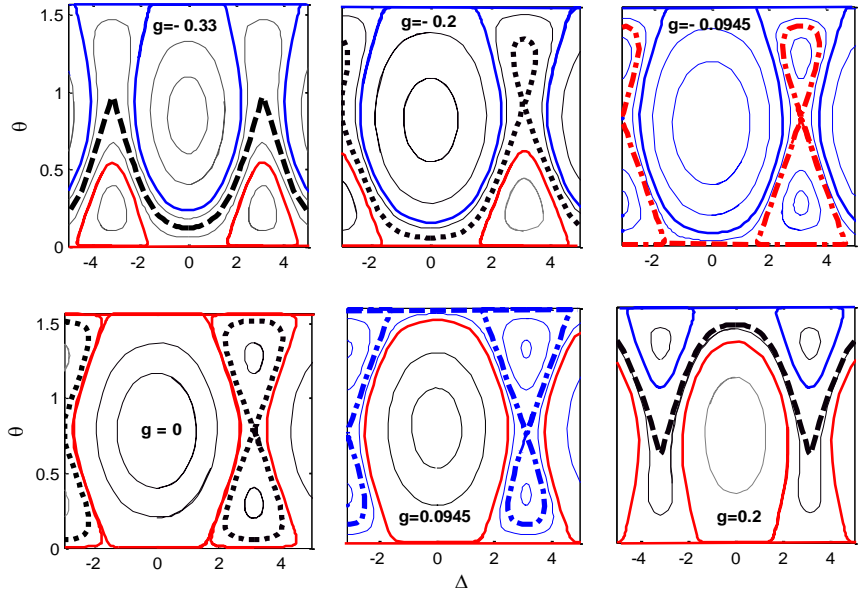

**Fig**. **21**. Phase portraits for system (3.5) in the plane (Δ, θ) for k = 0.9 and different values of g

The change from 2 to 4 fixed points at $g = -0.33$ entails the emergence of a separatrix passing through the hyperbolic point and consisting of the homoclinic and heteroclinic branches (by the heteroclinic separatrix we understand a trajectory from the hyperbolic point on the axis $\Delta = \pi$ to the hyperbolic point on the axis $\Delta = -\pi$ separating locked and unlocked orbits). At $g = -0.0945$, the heteroclinic loop coincides with the LPT at $\theta = 0$. Further increase of $g$ leads to the occurrence of the homoclinic separatrix and then to the confluence of the separatrix with the LPT at $\theta = \pi/2$ for $g = 0.0945$; finally, the coalescence of the lower stable center with the hyperbolic point results in the degeneration of the separatrix and the change from 4 to 2 fixed points. The numerically found value $|g^*| = 0.33$ coincides with the results of calculation by formula (3.8). Complete energy exchange is observed in the symmetric system ($g = 0$) moving along the LPT.

The phase portraits of system (3.5) with the parameters $k = 1$, $g > 0$, $k = 1.3$, $g > 0$ are depicted in Fig. 22; the portraits for $g < 0$ may be constructed by symmetry. Figure 22*b* demonstrates in more detail the transformations of the separatrix and the LPT associated with the transition from energy localization to energy exchange.

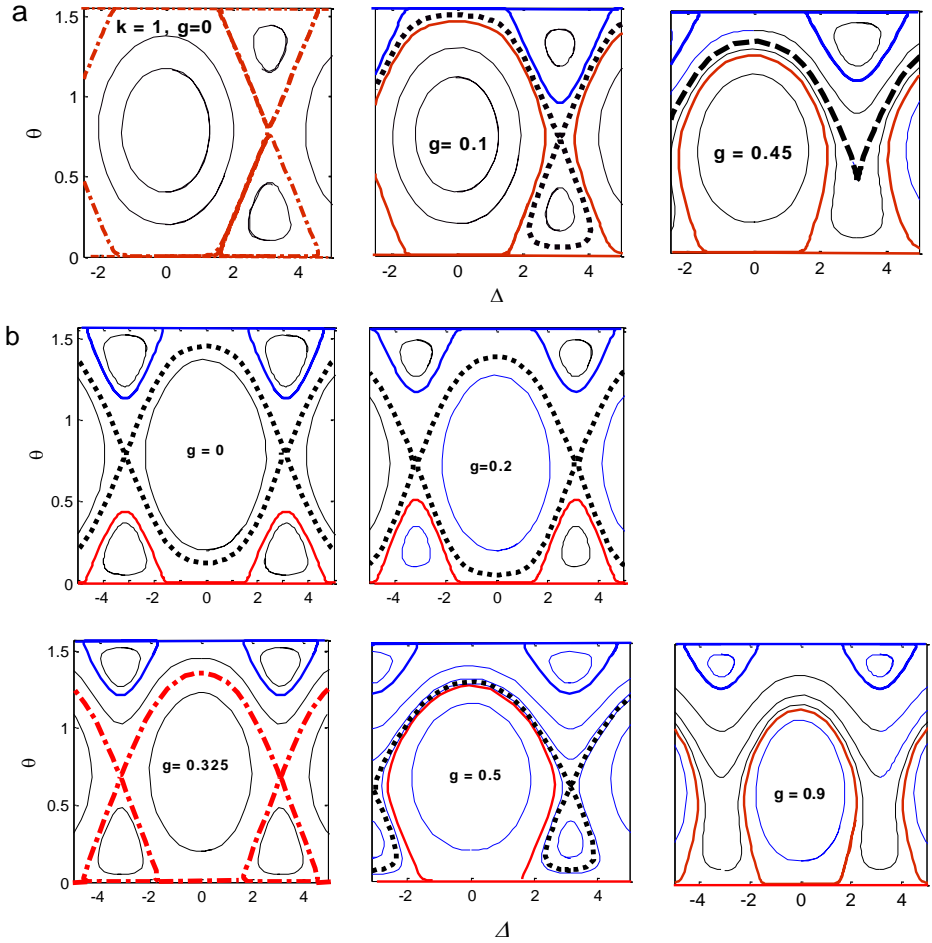

**Fig. 22.**Phase portraits for system (3.5) in the plane ($\Delta$, $\theta$) for $k = 1$ (*a*), $k = 1.3$ (*b*) and different values of $g > 0$.

The confluence of the separatrix with the LPT and the transition from energy localization to energy exchange correspond to *strongly nonlinear regimes.* The coalescence of the stable and unstable states entails the emergence of a new unlocked orbit and an associated transition from weak to strong energy exchange (Fig. 20– 22).

### 3.3. Critical parameters

Transitions from energy localization in the vicinity of the stable state to energy exchange in the asymmetric system were investigated numerically [193, 194], but an analytical boundary between the corresponding domains of parameters was not derived. In this section, we find analytical conditions that ensure the transition from the energy localization on the excited oscillator to strong energy exchange.

As remarked previously, the non-stationary dynamical behavior may be considered as strongly nonlinear if its separatrix coincides with the LPT. Using this condition, we find a set of parameters determining strongly nonlinear non-stationary regimes. To this end, we consider integral of motion (3.6). It follows from the initial condition $\theta = 0$ that $K = g$ on the LPT, and therefore,

$$(\cos\varDelta + k\sin2\theta)\sin2\theta - 2g\sin^2\theta = 0 \qquad (3.9)$$

on the LPT. Now we obtain from Eqs. (3.5) and (3.9) that

$$d\theta/d\tau_1 = V = \sin\varDelta; \quad V = \pm\,[1 - (k\sin2\theta - g\tan\theta)^2]^{1/2}. \qquad (3.10)$$

$(k\sin2\theta - g\tan\theta)^2 = 1$ at $V = 0$

Using equality (3.9) to exclude $\varDelta$, we replace system (3.5) by the following second-order equation which is valid for LPT only

$$\frac{d^2\theta}{d\tau_1^2} + \frac{dU}{d\theta} = 0 \qquad (3.11)$$

with initial conditions $\theta\,(0) = 0$, $V(0) = 1$. The potential $U(\theta)$ in (3.11) can be found from the energy conservation law $E = \frac{1}{2}V_2 + U(\theta) = 1$, which gives

$$U(\theta) = 1 - \tfrac{1}{2}V^2 = \tfrac{1}{2}[1 + (k\sin2\theta - g\tan\theta)^2]. \qquad (3.12)$$

The maximum value $U(\theta) = 1$ is attained at $V = 0$. Potential $U(\theta)$ and phase portraits for system (3.9) with different coefficients $k$ and $g$ are shown in Fig.23.

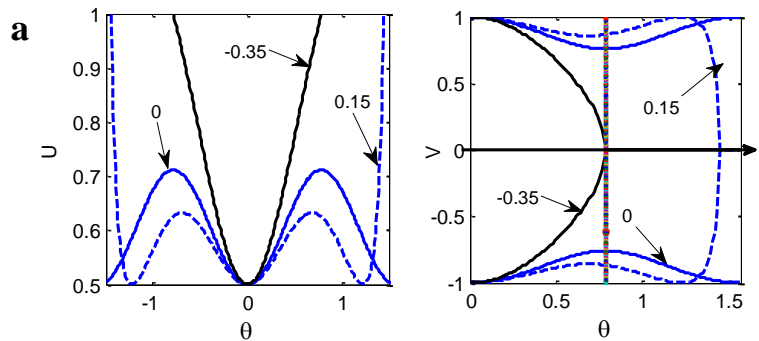

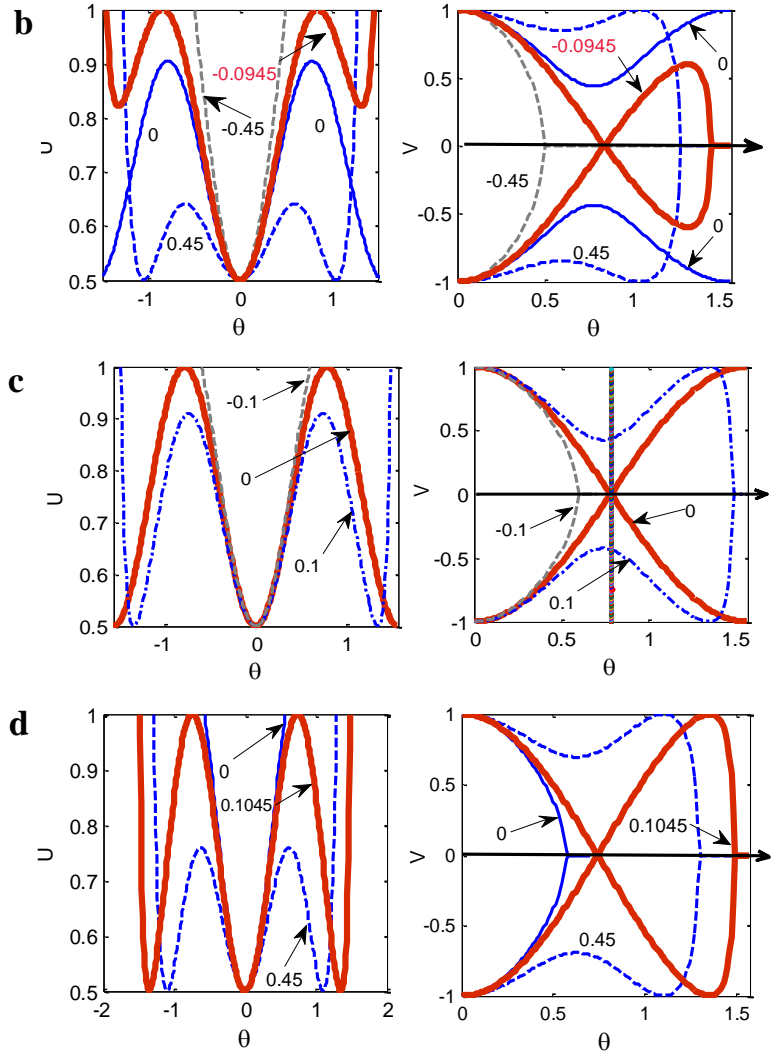

**Fig. 23**.Potential $U(\theta)$ and phase portraits in the plane $(\theta, v)$ for system (3.9) at $k$= 0.65 (a), $k$ = 0.9 (b), $k$ = 1 (c), $k$ = 1.1 (d); detuning is indicated on each curve; bold lines depict critical potentials and corresponding separatrices confluent with the LPT at $\theta = 0$

The sought separatrix may exist if and only if $dU/d\theta = 0$ at $\theta_h \in (0, \pi/2)$, as in this case there exists a potential barrier corresponding to the local maximum $U(\theta_h)$ and attained at $V = 0$; the latter condition is equivalent to $\Delta = 0$. This implies that $(\theta_h, 0)$ is a hyperbolic point. It now follows from (3.12) that the equality $dU/d\theta = 0$ is equivalent to

$$(k\sin 2\theta_h - g\tan\theta_h)(2k\cos 2\theta_h - g/\cos^2\theta_h) = 0. \qquad (3.13)$$

Combining (3.10), (3.13), we find that Eq. (3.13) is reduced to a simple biquadratic equation

$$4k\cos^4\theta_h - 2k\cos^2\theta_h - g = 0, \quad \cos^2\theta_k = \frac{1}{4} \pm \sqrt{\frac{1}{16} + \frac{g}{4k}} \ . \qquad (3.14)$$

It follows from (3.14) that $\cos^2\theta_h \approx 1/2 + g/2k$, $\theta_h \approx \pi/4 - g/2k$ if $|g/k| \ll \frac{1}{4}$. Substituting these approximations into the last equation (3.10), we derive a simple condition providing the existence of the required separatrix

$$|k - g_h| = 1. \qquad (3.15)$$

It is easy to check that the theoretical threshold $g_h$ closely agrees with the results of numerical calculations. In particular, $g_h = -0.1$ for $k = 0.9$, whereas the numerical threshold $g = -0.0945$; $g = g_h = 0.1$ for $k = 1.1$ (Fig. 23), $g_h = 0.3$ for $k = 1.3$ whereas the numerical threshold $g = 0.325$ (Fig. 22).

In a similar way, one can prove that, if the initial condition is taken at $\theta = \pi/2$, then the condition (3.15) is turned into the equality

$$|k + g_h| = 1. \qquad (3.16)$$

If $g < 0$, solution (3.14) exists provided that $|g| \leq k/4$; in the limiting case $g = -k/4$ one may find $\cos\theta_h = \frac{1}{2}$, $\theta_h = \pi/3$, and condition (3.14) becomes

$$\frac{3\sqrt{3}}{4}k = 1, \, k^* \approx 0.77. \qquad (3.17)$$

Inequalities $k \geq k^*$, $|g| \leq k/4$ express the necessary conditions for the existence of the separatrix coinciding with the LPT at $\theta = 0$. Additionally, one needs to calculate the argument $\theta_h$ from (3.14).

Finally, we note that the solutions $\theta(\tau_1)$, $\Delta(\tau_1)$ as explicit functions of the slow time $\tau_1$ can be found with the help of non-smooth transformations similar to presented in Sec. 2 [111]. A special case of symmetric system of two coupled oscillators ($g = 0$) is discussed in [105].

## 4.  Weakly coupled oscillators under the condition of sonic vacuum

In this section we investigate resonance energy transport in a purely nonlinear system, wherein harmonic oscillations are prohibited by the system potential of degree higher than two. In contrast to the models considered in the previous sections, the potential of this system does not contain quadratic terms. Hence, in the considered system there are no constant natural frequencies predetermining resonant properties of the system in the quasi-linear approximation. This implies that the study of the resonance processes and resonance energy transport in a purely nonlinear system ("in the state of sonic vacuum") requires a special approach.

The basic model considered in this section comprises two weakly coupled purely nonlinear oscillators, wherein initial energy is imported to one of them. Numerical simulations reveal the existence of strong classical beats corresponding to full recurrent resonant energy exchanges between the oscillators in the state of sonic vacuum, where no resonance frequencies can be defined. In this study we show that both intense energy exchange and transition to energy localization are adequately described in the framework of the LPT concept.

It is demonstrated that the occurrence of the recurrent energy exchanges in this highly degenerate model strictly depends on the system parameters. For instance, choosing the parameter of coupling below a certain threshold leads to the significant energy localization on one of the oscillators; on the contrary, increasing the strength of coupling above the threshold brings to the formation of a strong beating response.

Analytical studies pursued in this section predict the occurrence of the strong beating phenomenon and provide necessary conditions for its emergence. Moreover, a careful analysis of the beating phenomenon reveals a qualitatively new global bifurcation of highly non-stationary regime. Results of the analytical study are in a very good agreement with numerical simulations.

## 4.1. *Evidence of energy localization and exchange in coupled oscillators in the state of sonic vacuum*

We study energy localization and complete recurrent energy transport in a homogeneous system of two coupled anharmonic oscillators in the state of sonic vacuum. The main goal of this section is to describe the transition from initial energy localization on a single oscillator to complete recurrent energy exchanges (strong beating phenomenon) between the oscillators due to variations of the system parameters

The model under consideration comprises two identical anharmonic oscillators coupled with an anharmonic spring. The non-dimensional equations of motion are given by

$$\frac{d^2x_1}{dt^2} + x_1^n = \chi(x_2 - x_1)^n,$$

$$\frac{d^2x_2}{dt^2} + x_2^n = \chi(x_1 - x_2)^n. \qquad (4.1)$$

where $n = 2k + 1$, $k = 1, 2, 3, \ldots$; the parameter $\chi$ denotes coupling stiffness. Initial conditions

$$x_1 = X_1, v_1 = dx_1/dt = V_1, \ x_2 = 0, \ v_2 = dx_2/dt = 0 \text{ at } t = 0$$

correspond to complete initial localization of the system energy on the first oscillator with the second oscillator being initially at rest.

It is important to emphasize that system (4.1) is homogeneous, and therefore its total energy $E$ can be normalized to unity ($E = 1$) by choosing appropriate rescaling of the dependent and independent variables. This means that the global system dynamics is energy-independent and can be studied for an arbitrary value of the total energy level.

Below we illustrate numerically the existence of two different regimes such as energy localization on the initially excited oscillator and complete energy exchanges between the oscillators. We also show that the first regime corresponds to the non-resonant behavior of the coupled oscillators while the second one is triggered by the formation of permanent 1:1 resonance capture resulting in the complete recurrent energy transport between the oscillators.

Figure 24($a$) depicts instantaneous energy $E_k = v_k^2/2 + (n + 1)^{-1}x_k^{n+1}$ of each of the oscillators in the system with parameters $\chi = 0.12$, $n = 3$ and initial conditions $x_1 = 1$, $v_1 = 0$, $x_2 = 0$, $v_2 = 0$ at $t = 0$. Figures 24$b$, 24$c$ illustrate the fast Fourier transforms (FFT) of $x_1(t)$ and $x_2(t)$, respectively.

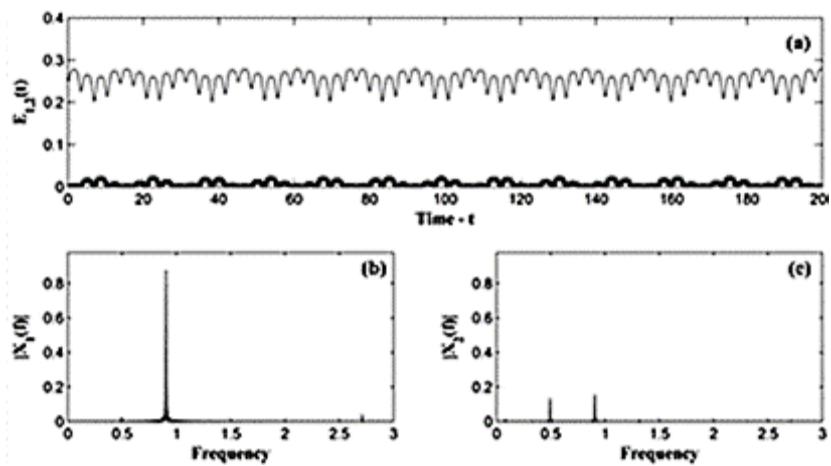

**Fig. 24.** Regime of energy localization (a) Instantaneous energies recorded on the first and the second oscillators ($E_1(t)$- thin solid line, $E_2(t)$ - bold solid line); (b) Fast Fourier Transform of the response of the first oscillator $x_1(t)$; (c) Fast Fourier Transform of the response of the second oscillator $x_2(t)$.

In Fig. 24 (a) one can observe localization of the initial energy on the first oscillator. This result is not surprising, as the system under consideration is purely nonlinear (sonic vacuum), and strong spatial energy localization on certain fragments of the system is expected. Moreover, the FFT diagrams presented in Figs. 24(b), 24(c) show that this type of response exhibits a trivial non-resonant behavior, where the amplitude of a main frequency component of the first oscillator far exceeds the amplitude of the main components of the second oscillator. Also, the second oscillator possesses two comparable components with remote frequencies exhibiting a clear sub-harmonic motion. The detailed analysis of this type of response is performed in Section 4.2.

Slightly increasing the coupling parameter $\chi$, we observe a global change of the response. In particular, instead of energy localization on the first oscillator we observe the formation of a beating response characterized by complete energy exchange between the oscillators. Instantaneous energy is plotted in Fig. 25(a) for each of the oscillators in the system with parameters $\chi = 0.18$, $n = 3$ and initial conditions $x_1 = 1$, $v_1 = 0$, $x_2 = 0$, $v_2 = 0$ at $t = 0$. Figures 25(b), 25(c) illustrate the FFT of $x_1(t)$ and $x_2(t)$, respectively.

Figure 25(a) depicts the beating response characterized by complete recurrent energy exchange between the oscillators. FFT diagrams in Fig. 25 (b, c) reveal the resonant nature of the response ensuring the formation of 1:1 resonance between the oscillators, which, in turn, leads to the strong beating response.

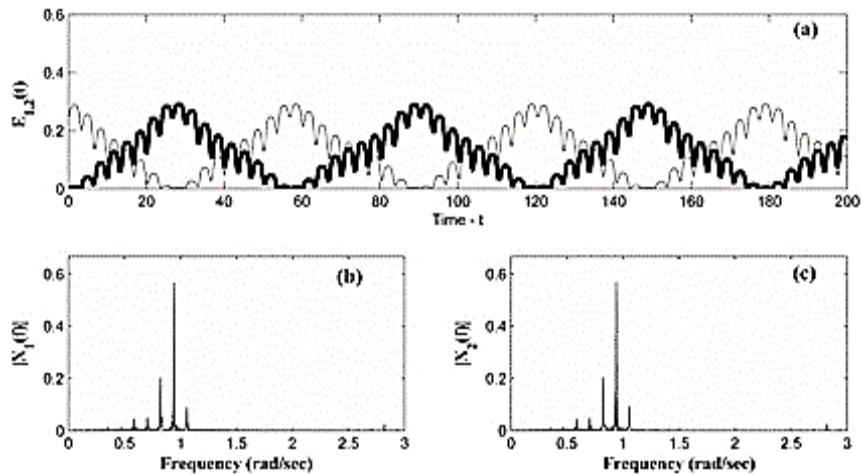

**Fig. 25**. Strong beating response (a) Instantaneous energies recorded on the first and the second oscillators ($E_1(t)$ - thin dotted line, $E_2(t)$ - bold solid line); (b) FFT of $x_1(t)$; (c) FFT of $x_2(t)$.

It is important to note that beating oscillations are usually observed in linear or weakly non-linear systems possessing at least one pair of close natural frequencies [108, 123]. In these cases, the resonance frequency of the response is determined either by the natural frequency of the linear sub-system or by the frequency of a periodic excitation. The system under consideration is, in some sense, degenerate, as it does not include linear components, that is, the resonance frequency is not predetermined. Furthermore, resonance frequencies observed in Fig. 25 (b, c) are obviously amplitude- dependent. This means that an increase of the initial amplitude of the first oscillator results in an increase of the resonant frequency of oscillations.

## 4.2. Asymptotic analysis of resonance motion

This section suggests theoretical analysis of the near-resonant behavior of system (4.1). Special emphasis is given to an analytical description of the formation and the annihilation of a regular beating response, along with the local and global bifurcation analysis of the system dynamics.

Taking into account the pure resonant nature of a strong beating regime, it is quite reasonable to anticipate its formation in the neighborhood of the 1:1 resonance manifold. It is well known that the phenomenon of 1:1 resonance capture provides maximum energy transfer between weakly coupled identical oscillators. Thus, in order to depict analytically the regime of complete energy transfer, as well as to find necessary conditions for its existence, it is convenient to consider the system dynamics in a neighborhood of the 1:1 resonance manifold.

Assuming resonance interactions, we rewrite (4.1) as follows:

$$\frac{d^2 x_k}{dt^2} + \Omega^2 x_k = \varepsilon \mu [\chi (x_{3-k} - x_k)^n - x_k{}^n + \Omega^2 x_k], \ k = 1, 2, \tag{4.2}$$

where $\Omega$ denotes the resonance frequency depending on the system energy, $\varepsilon$ is a small parameter of the system, $\mu = 1/\varepsilon$. As in the previous sections, we assume that the sum in the square brackets is small but the expression $\mu [\chi (x_{3-k} - x_k)^n - x_k^n + \Omega^2 x_k]$ is of $O(1)$ in the vicinity of resonance.

We underline that the representation of equations of motion in the form (4.2) allows us to investigate the purely nonlinear system in the framework of the quasi-linear theory and to employ the earlier developed methods. In the first step, we introduce the new complex variables as follows:

$$x_k = \frac{1}{2t}(Y_k e^{i\Omega\tau_0} - Y_k^* e^{-i\Omega\tau_0}), \; v_k = \frac{dx_k}{dt} = \frac{\Omega}{2}(Y_k e^{i\Omega\tau_0} + Y_k^* e^{-i\Omega\tau_0}), \tag{4.3}$$

and then substitute (4.3) into (4.1) to obtain the equations for $Y_k$, $Y_k^*$ with the right hand sides of $O(\varepsilon)$ (see [178] for more details). In the next step, the complex amplitude $Y_k(t,\varepsilon)$ is sought in the form of the expansion $Y_k(t,\varepsilon) = \varphi_k(\tau_1) + \varepsilon\varphi_k^{(1)}(t,\tau_1) + \varepsilon^2\ldots$, where $\tau_1 = \varepsilon t$ is the leading-order slow time scale. Then, applying the multiple scales methodology, we derive the following equation for the leading-order slow amplitudes $\varphi_k^{(0)}(\tau_1)$ (detailed arguments are provided in [178]):

$$\frac{d\varphi_k}{d\tau_1} = i\mu \left(\frac{C_{(n-1)/2}^n}{(2\Omega)^n} |\varphi_k|^{n-1}\varphi_k - \frac{\Omega}{2}\varphi_k + \chi\frac{C_{(n-1)/2}^n}{(2\Omega)^n}|\varphi_{3-k} - \varphi_k|^{n-1}(\varphi_{3-k} - \varphi_k)\right), \; k=1,2, \tag{4.4}$$

where $C_{(n-1)/2}^n$ is the binomial coefficient. It is easy to conclude that system (4.4) possesses two integrals of motion

$$\begin{aligned} N^2 &= |\varphi_1|^2 + |\varphi_2|^2, \\ H &= |\varphi_1|^{n+1} + |\varphi_2|^{n+1} + \mu|\varphi_2 - \varphi_1|^{n+1}. \end{aligned} \tag{4.5}$$

The first integral of motion allows for a convenient change of coordinates

$$\varphi_1 = N\cos\theta e^{i\delta_1}, \; \varphi_2 = N\sin\theta e^{i\delta_2}. \tag{4.6}$$

Substituting (4.6) into (4.4), considering the relative phase $\delta = \delta_1 - \delta_2$ as a new variable, and rescaling the independent variable by law $\tau = [C_{(n-1)/2}^n N^{n-1}/(2\Omega)^n]\tau_1$, we reduce (4.4) to the real-valued system

$$\begin{aligned} \frac{d\delta}{d\tau} &= \mu[\cos^{n-1}\theta - \sin^{n-1}\theta + 2\chi(1 - \sin 2\theta\cos\delta)^{\frac{n-1}{2}}\cot 2\theta\cos\delta], \\ \frac{d\theta}{d\tau} &= \mu\chi(1 - \sin 2\theta\cos\delta)^{\frac{n-1}{2}}\sin\delta. \end{aligned} \tag{4.7}$$

It is important to note that system (4.7) does not involve the energy-dependent frequency $\Omega$ and governed only by the constant parameters $\chi$ and $n$, thus making the global dynamics of system (4.7) invariant to the slow amplitudes $|\varphi_1|$, $|\varphi_2|$. We show that variations of the governing parameters $\chi$ and $n$ lead to both local and global bifurcations.

### 4.2.1. Fixed points and NNMs in the neighborhood of resonance

The further study of the system dynamics is concentrated on the analysis of system (4.7). First, we find fixed points of (4.7). By setting $d\theta/d\tau = d\delta/d\tau = 0$, we obtain the following algebraic equations defining the fixed points of (4.7):

$$\cos^{n-1}\theta - \sin^{n-1}\theta + 2\chi(1 - \sin 2\theta \cos \delta)^{\frac{n-1}{2}} \cot 2\theta \cos \delta = 0,$$

$$(1 - \sin 2\theta \cos \delta)^{\frac{n-1}{2}} \sin \delta = 0. \tag{4.8}$$

By setting $\cos 2\theta = 0$, $\sin \delta = 0$, we obtain the following set of fixed points:

$$(\delta_1^{(1)}, \theta_1^{(1)}) = (0, \pi/4); (\delta_2^{(1)}, \theta_2^{(1)}) = (\pi, 3\pi/4); \tag{4.9}$$

$$(\delta_1^{(2)}, \theta_1^{(2)}) = (0, 3\pi/4); (\delta_2^{(2)}, \theta_2^{(2)}) = (\pi, \pi/4).$$

The first pair of fixed points corresponds to the in-phase nonlinear normal mode (NNM) of the original system (4.1), while the second one corresponds to the out-of-phase NNM.

Additional fixed points $(\delta_k^{(3)}, \theta_k^{(3)})$, $k = 1,2$, are defined by the conditions $\cos 2\theta \neq 0$, $\sin \delta = 0$, that is,

$$(\cos^{n-1}\theta - \sin^{n-1}\theta)\tan 2\theta - 2\chi(1 + \sin 2\theta)^{\frac{n-1}{2}} = 0, \delta = \pi, \tag{4.10}$$

$$(\cos^{n-1}\theta - \sin^{n-1}\theta)\tan 2\theta + 2\chi(1 - \sin 2\theta)^{\frac{n-1}{2}} = 0, \delta = 0. \tag{4.11}$$

As mentioned above, the global system dynamics is governed by the parameters $\chi$ and $n$. In Fig. 26 we plot the solutions of Eq. (4.10) corresponding to $\delta = \pi$ vs. variations of the parameter $\chi$ for four different values of $n$. It is easy to see that the branches of solutions bifurcate in the point ($\pi$, $\pi/4$). We will not illustrate solutions of (4.11) because they are similar to those of (4.10), but for the range of $\theta \in [\pi, 2\pi]$.

Figure 26 demonstrates a qualitative change of the solutions of (4.10) for $n > 5$. These topological changes represent the results of transition from a supercritical to a subcritical pitch-fork bifurcation undergone by the fixed points described in [211, 212]. Note that the appearance of the sub-critical bifurcation of the fixed points has a significant effect on the occurrence of strong beats as well as on the shape of the response.

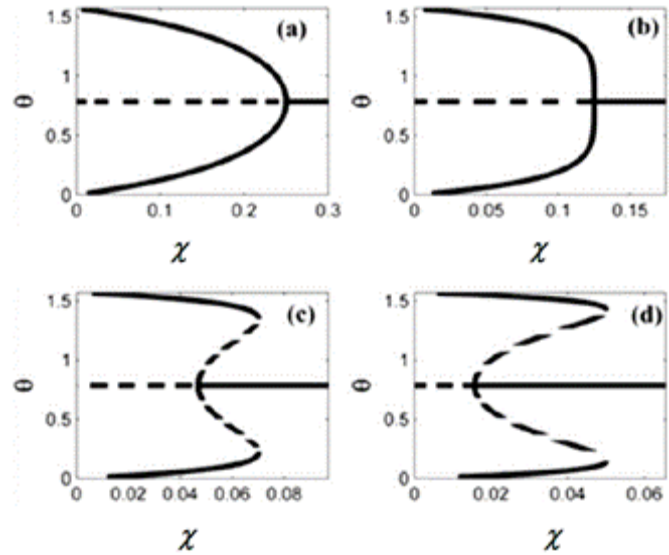

**Fig. 26.** Solutions of (4.10) vs. variation of coupling strength $\chi$: (a) $n = 3$; (b) $n = 5$; (c) $n = 5$; (d) $n = 9$. Stable branches of solution are denoted by bold solid lines, unstable branches of solutions are denoted by dashed lines. Horizontal axes $\theta = \pi/4$ correspond to the fixed point $\theta_2^{(2)}$.

Before proceeding with the analysis of strong beating response, we analytically prove the occurrence of a subcritical pitch-fork bifurcation at $n > 5$. In the first step, one can deduce that the branches of the solutions of (4.10) bifurcate in the fixed point $\theta_2^{(2)} = \pi/4$. To analyze the motion near $\theta = \pi/4$, Eq. (4.10) is rewritten as

$$\chi = \frac{(\cos^{n-1}\theta - \sin^{n-1}\theta)\tan 2\theta}{2(1 + \sin 2\theta)^{\frac{n-1}{2}}}.$$ (4.12)

It follows from (4.12) that $\lim_{\theta \to \pi/4} (\partial \chi / \partial \theta) = 0$. In the next step, we insert the expansion

$\theta = \pi/4 + \tilde{\theta}, \ (|\tilde{\theta}| \ll 1)$ into (4.12) to obtain

$$\chi = \frac{n-1}{2^n} + \frac{n(n-1)(n-2)}{3(2)^{n+1}} \tilde{\theta}^2 + O(\tilde{\theta}^3). \qquad (4.13)$$

The first term in the right-hand side of (4.13) represents the first bifurcation value $\chi_{cr}^{(1)}$ of coupling strength $\chi$

$$\chi_{cr}^{(1)} = 2^{-n}(n-1), \qquad (4.14)$$

at which the fixed point $(\delta_2^{(2)}, \theta_2^{(2)}) = (\pi, \pi/4)$ is transformed from the saddle point to the stable center for an arbitrary value of $n$. The transformation of the saddle point can be easily proved by performing a linear stability analysis [178].

It follows from expansion (4.13) that the coefficient of the quadratic term is always positive for $n > 5$. In the special case of $n = 5$ this coefficient equals zero. In this case, one can easily show that the coefficient of the higher-order term in the expansion (i.e., the forth-order term) is negative. Hence, expansion (4.13) proves the formation of a sub-critical pitchfork bifurcation for $n > 5$.

### 4.2.2. Limiting Phase Trajectories

Figures 27, 28 depict phase portraits of system (4.7) for $n = 3$ and $n = 7$. The choice of these values of the parameter $n$ is not arbitrary; it aims to show the global changes in the system dynamics caused by the transition from super-critical ($n < 5$) to sub-critical ($n > 5$) pitch-fork bifurcations.

In terms of the model (4.7), complete energy exchange between the oscillators is associated with the *Limiting Phase Trajectory* (LPT), which passes through zero point $\delta = \theta = 0$ and reaches the value $\theta = \pi/2$. We underline that, unlike the previous sections, there is no way to distinguish quasi-linear, moderately nonlinear and strongly nonlinear regimes, because the system does not exhibit the quasi-linear behavior.

Figures 27(a,b) demonstrate the special orbits (bold lines) satisfying the initial condition $\delta = \theta = 0$. However, motion along these orbits cannot lead to complete energy exchange between the oscillators, as the trajectory does not reach the value $\theta = \pi / 2$. Below we refer to this kind of the phase trajectory as the LPT of the first kind.

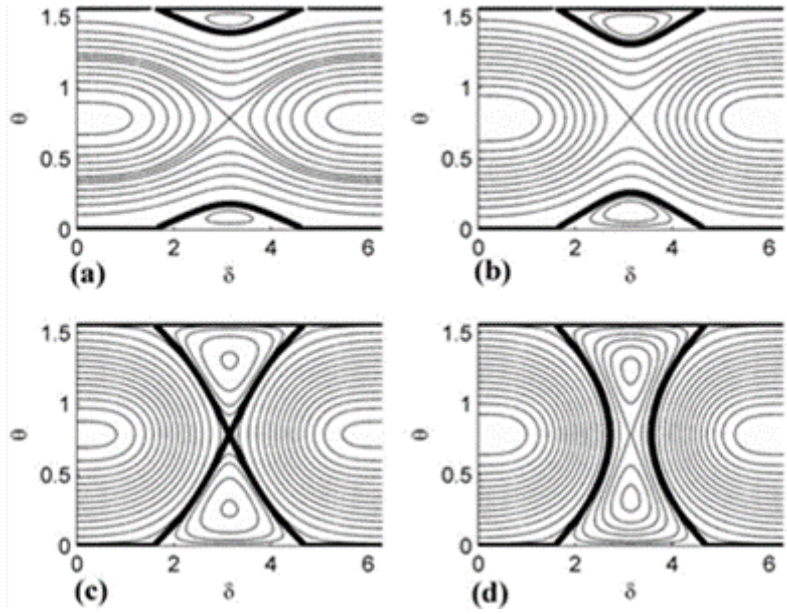

**Fig. 27.** Phase portrait of system (4.7) with $n = 3$ and different coupling strength: (a) $\chi = 0.075$, (b) $\chi = 0.1$; (c) $\chi = 0.1667$; (d) $\chi = 0.19$. Limiting Phase Trajectories (LPTs) are denoted by bold lines on each plot.

Increasing the value of $\chi$ up to a certain critical value $\chi_{cr}^{(2)}$, one can observe the coalescence of the LPT of the first kind with the separatrix (Fig. 27(c)). This coalescence leads to the global bifurcation resulting in the formation of an LPT of the qualitatively different type (Fig. 27(d)), that is, of the LPT of the second kind.

Phase portraits in Fig. 28(a) are qualitatively similar to those in Fig. 27(a). However, slightly increasing the strength of coupling $\chi$ (Fig. 28(b)), we observe the occurrence of an additional pair of unstable fixed points, and the transition of the fixed point ($\delta_2^{(2)}$, $\theta_2^{(2)}$) from the saddle point to the stable center. The latter observation is a result of the above described sub-critical pitch-fork bifurcation. Clearly, except the regular LPT starting at $\theta = 0$, there exists an additional branch of the LPT (we will refer to it as the "LPT bubble") encircling the center $(\delta_2^{(2)}, \theta_2^{(2)})$. This new type of LPT is illustrated in Fig. 28(c). At a certain critical value $\chi = \chi_{cr}^{(2)}$, the regular LPT collides with the 'bubble' LPT exactly at the saddle point, entailing the occurrence of the LPT of the second kind (Fig. 28(d)). The newborn LPT of the second kind has a topology different from that one in the case of $n = 3$ reported in previous sections.

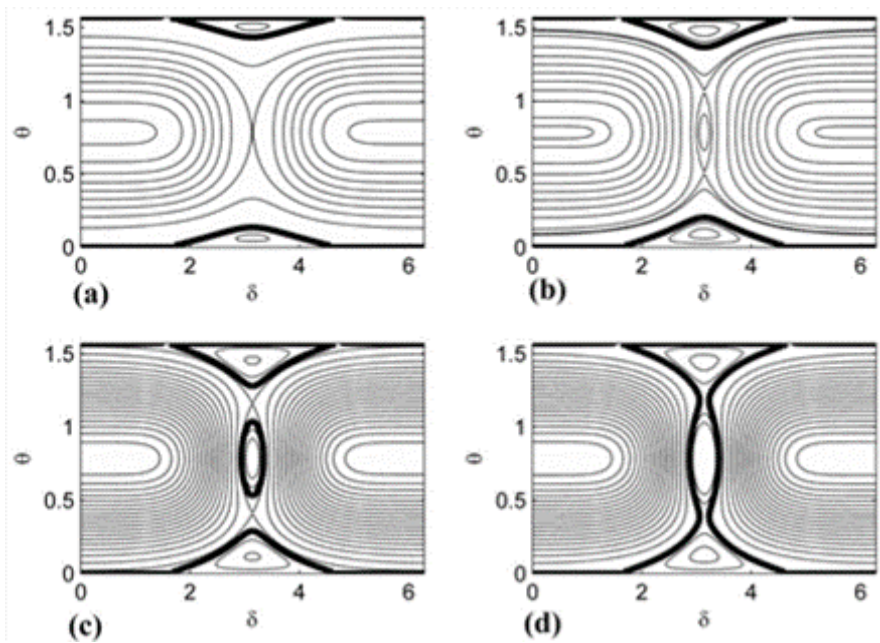

**Fig. 28.** Phase portrait of system (4.7) with $n = 7$ and different coupling strength: a) $\chi = 0.040$; b) $\chi = 0.055$; c) $\chi = 0.061$; d) $\chi = 0.063$. Limiting Phase Trajectories (LPTs) are denoted with the bold line on each plot.

It is seen in Fig. 28 that the LPT of the second kind in the case of $n > 5$ has a near rectangular shape instead of the triangle observed at $n = 3$ and $n = 5$.

In order to derive analytical conditions of the occurrence of the LPT of the second kind for any value of $n$, as well as to find a critical value $\chi_{cr}^{(2)}$, we consider the second integral of motion (4.5). Expressing $H$ in terms of $(\theta, \delta)$, one obtains

$$H(\delta, \theta, \chi) = \cos^{n+1}\theta + \sin^{n+1}\theta + \chi(1 - \sin 2\theta \cos \delta)^{(n+1)/2}. \tag{4.15}$$

Equation (4.15) depicts phase trajectories of (4.7). We recall that the LPT of any kind passes through zero point $\delta = \theta = 0$. Substituting $\theta = 0$ into (4.15) yields the following exact value of $H(\delta, \theta, \chi)$ corresponding to the LPT:

$$H_{LPT}(\delta, \theta, \chi) = 1 + \chi. \tag{4.16}$$

In the case of $n < 7$, an increase of the coupling parameter $\chi$ up to a critical value $\chi_{cr}^{(2)}$ corresponds to the transition from the LPT of the first kind to the LPT of the second kind. Since the LPT passes through the saddle point $(\delta_2^{(2)}, \theta_2^{(2)}) = (\pi, \pi/4)$, the critical value $\chi = \chi_{cr}^{(2)}$ can be found from the equality

$$H_{LPT}(\pi, \pi/4, \chi_{cr}^{(2)}) = 1 + \chi_{cr}^{(2)}. \tag{4.17}$$

It follows from (4.15), (4.17) that

$$\chi_{cr}^{(2)} = (1 - 2^{-\frac{(n-1)}{2}})(2^{\frac{(n-1)}{2}} - 1). \tag{4.18}$$

We recall that the critical value $\chi_{cr}^{(2)}$ is associated with the transition from the LPT of the first kind to the LPT of the second kind only for $n = 3$ and $n = 5$.

If $n > 5$, then, requiring the fixed point $(\delta_2^{(2)}, \theta_2^{(2)})$ to lie on the LPT, we find a critical value $\chi_{cr}^{(3)}$ corresponding to the occurrence of the LPT 'bubble'. To this end, we require the branch of the LPT to cross a saddle point branching out from $(\delta_2^{(2)}, \theta_2^{(2)})$ through the pitchfork bifurcation. This branch of the solutions is given by Eq. (4.10). Thus, solving the following nonlinear system:

$$\begin{aligned}
(\cos^{n-1}\theta - \sin^{n-1}\theta)\tan 2\theta - 2\chi_{cr}^{(3)}(1 + \sin 2\theta)^{\frac{n-1}{2}} &= 0, \\
\cos^{n+1}\theta + \sin^{n+1}\theta + \chi_{cr}^{(3)}(1 + \sin 2\theta)^{\frac{n+1}{2}} &= 1 + \chi_{cr}^{(3)},
\end{aligned} \tag{4.19}$$

one can derive a new criterion for the transition from localization to complete energy transfer for the case of $n > 5$. In Table 1 we summarize critical values of coupling strength $\chi$ corresponding to different types of dynamical transitions.

| | Pitchfork bifurcation value | Formation of the LPT 'bubble' | Transition from localization to a complete energy transfer |
|---|---|---|---|
| $n < 7$ | $\chi_{cr}^{(1)} = 2^{-n}(n-1)$ | - | $\chi_{cr}^{(2)} = (1 - 2^{-\frac{(n-1)}{2}})(2^{\frac{(n-1)}{2}} - 1)$ |
| $n \geq 7$ | $\chi_{cr}^{(1)} = 2^{-n}(n-1)$ | $\chi_{cr}^{(2)} = (1 - 2^{-\frac{(n-1)}{2}})(2^{\frac{(n-1)}{2}} - 1)$ | $\chi_{cr}^{(3)}$ - solution of (25) |

**Table 1.** Critical values of coupling strength $\chi$ vs. $n$ for different types of dynamical transitions

### 4.3. Numerical analysis of the fundamental model

We perform numerical verifications of the theoretical model suggested in the previous section. We compare the response of the original system (4.1) with initial conditions $x_1(0) = 1$, $x_2(0) = 0$; $v_1(0) = v_2(0) = 0$ with the slow envelope (4.4) satisfying the corresponding initial conditions $\varphi_1(0) = i$, $\varphi_2(0) = 0$. To confirm the validity of the theoretical models for two different topologies of the LPT, we choose two representative values of $n$, namely $n = 3$ and $n = 7$.

From the results in Fig. 29(a) ($n = 3$), it is clear that the choice of coupling strength $\chi$ below the predicted threshold ($\chi < \chi_{cr}^{(2)} = 0.167$) leads to energy localization on the first oscillator. However, if the value of $\chi$ is increasing above the threshold ($\chi > \chi_{cr}^{(2)}$), we clearly observe the occurrence of a strong beating response, i.e., complete energy exchanges between the oscillators (Fig. 29(b)). Moreover, the slow flow envelope given by (4.4) is in a very good agreement with the full model (4.1). This means that the analytical model clearly recovers the mechanism of the transition from localization to recurrent energy transfer observed in the full model.

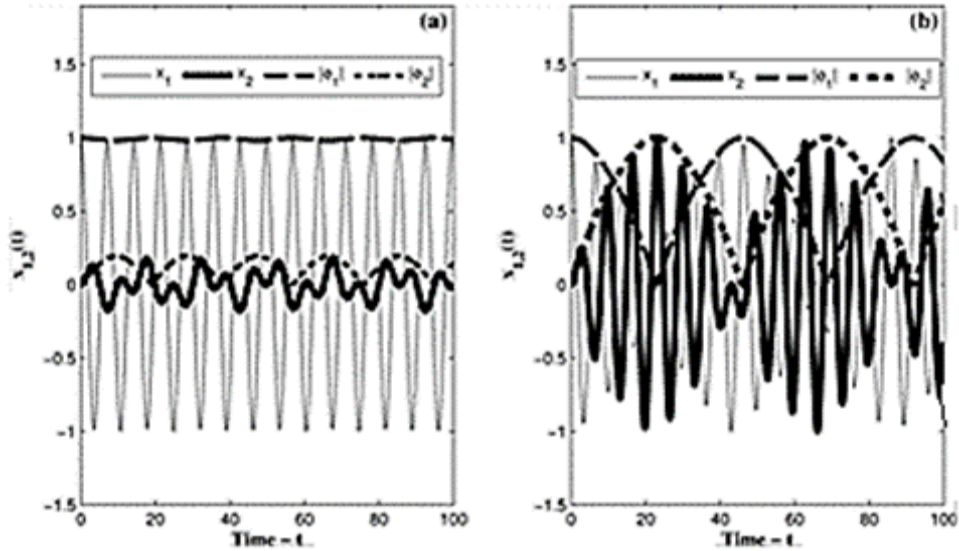

**Fig. 29.** Superposition of the slow envelope (4.4) on the full response (4.1) for $n = 3$ and different coupling strength: (a) $\chi = 0.08 < \chi_{cr}^{(2}$; (b) $\chi = 0.19 > \chi_{cr}^{(2}$. Initial conditions: $x_1(0) = 1$, $x_2(0) = 0$; $v_1(0) = v_2(0) = 0$ for the full system; $\varphi_1(0) = i$, $\varphi_2(0) = 0$ for the envelope.

To better illustrate the effect of the transition from localization to a complete transport of energy, we construct the Poincaré maps corresponding to system (4.1). To this end, we first restrict the system dynamics to an isoenergetic manifold. It is easy to show that, due to homogeneity of (4.1), the total system energy can be normalized to unity ($E = 1$). Thus, fixing the total energy to a constant level, we restrict the dynamical flow to the three-dimensional iso-energetic manifold $E(x_1, \dot{x}_1, x_2, \dot{x}_2) = 1$. Transversely intersecting the three-dimensional iso-energetic manifold by the two-dimensional cut plane $T : \{x_1 = 0\}$, one obtains the Poincaré map $P : \Sigma \rightarrow \Sigma$, where the Poincaré section is defined as $\Sigma = \{x_1 = 0, \dot{x}_1 > 0\} \cap \{E(x_1, \dot{x}_1, x_2, \dot{x}_2) = 1\}$.

Fundamental time-periodic solutions of a basic period $T$ correspond to the period 1 equilibrium points in the Poincaré map. Additional sub-harmonic solutions of periods $nT$ correspond to the period $n$ equilibrium points of the Poincaré map, i.e., to the orbits that pierce the cut section $n$ times before repeating themselves. Clearly, the construction of the Poincaré map ($P : \Sigma \rightarrow \Sigma$) effectively reduces the global system dynamics to the plane $(x, v)$.

The Poincaré section in Fig. 30a corresponds to the case of $\chi < \chi_{cr}^{(2)}$ (energy localization) while Fig. 30b corresponds to the case of $\chi > \chi_{cr}^{(2)}$ (strong beating response).

From the close observation of the results in Fig. 30 one can identify the formation of special orbits (marked with bold dots) passing through the origin. It is clear that this the invariant set emanating from the origin constitutes a special orbit leading to complete energy exchange between the oscillators (strong beating response). However, the results in Fig. 30a suggest that a similar orbit emanating from the origin does not lead to complete energy exchange. These special orbits of the Poincaré sections can be directly correlated to the LPT of the reduced model. This special orbit illustrated in Fig. 30a corresponds to energy localization (LPT of the first kind) while that one in Fig. 30b corresponds to complete energy exchange (LPT of the second kind).

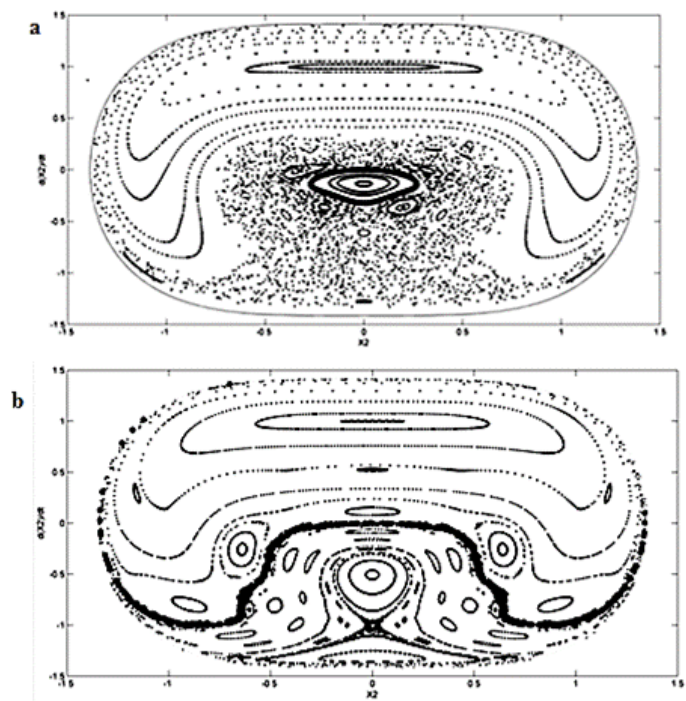

**Fig. 30.** Poincaré map for $n = 3$; $\chi = 0.08 < \chi_{cr}^{(2)}$ (a), $\chi = 0.2 > \chi_{cr}^{(2)}$ (b); LPT is marked with bold dots.

Figures 31, 32 present computational results for $n = 7$. Figure 31 depicts the time-response for $n = 7$. Note that the response of the reduced-order model (4.4) agrees fairly well

with the response of the full model (4.1) despite the relatively high power of nonlinearity. The theoretical prediction of the threshold value $\chi_{cr}^{(2)} = 0.0625$ is confirmed in Fig. 30.

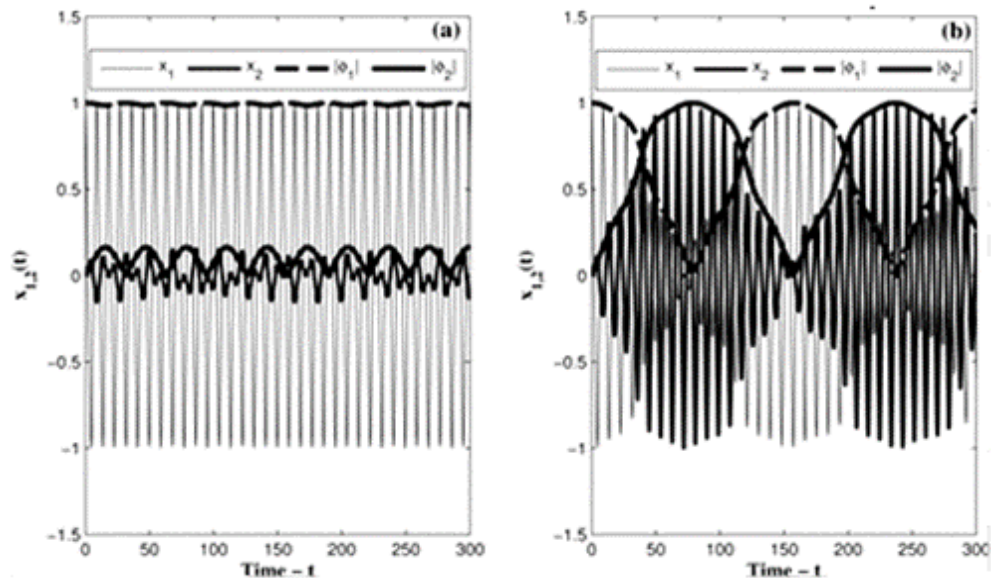

**Fig. 31.** Superposition of the slow envelope (4.4) on the full response (4.1) for $n = 7$ and different coupling strength: (*a*) $\chi = 0.05 <\chi_{cr}^{(2)}$; (*b*) $\chi = 0.05 <\chi_{cr}^{(2)}$. Initial conditions are indicated in Fig. 29.

Figure 32 demonstrates that, despite the prevalence of the 'chaotic sea' region, which covers almost the entire map, one can still observe the preservation of the special orbits corresponding to the LPTs of the first and the second types.

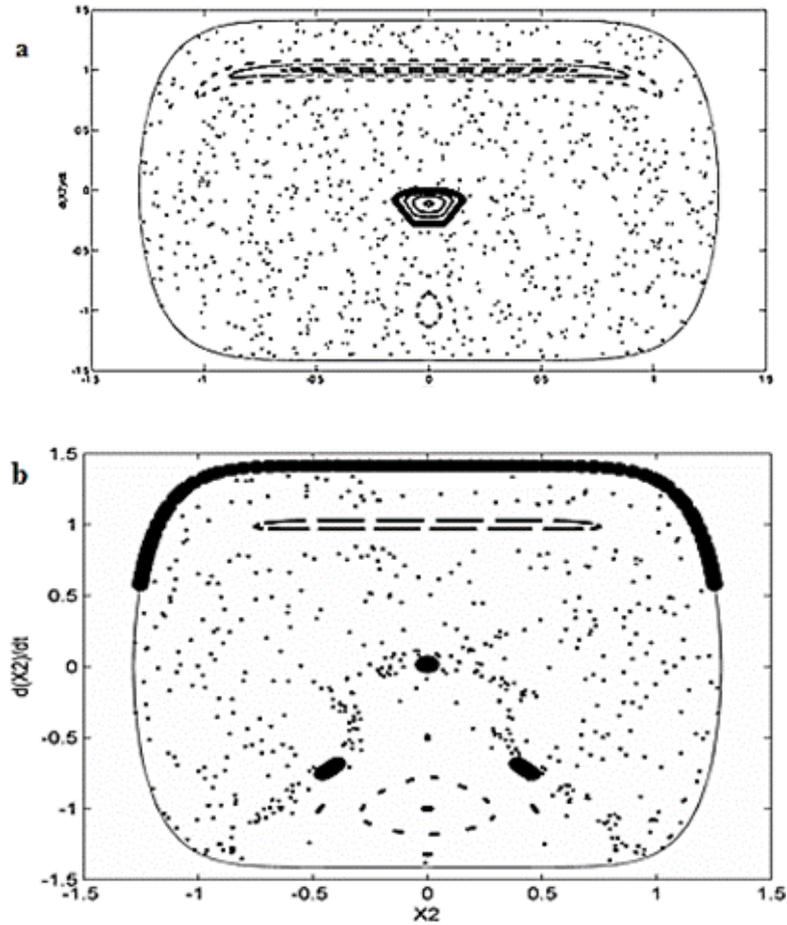

**Fig. 32.** Poincaré map for $n = 7$; a: $\chi = 0.05 < \chi_{cr}^{(2)}$; b: $\chi = 0.07 > \chi_{cr}^{(2)}$. LPT is marked with bold dots.

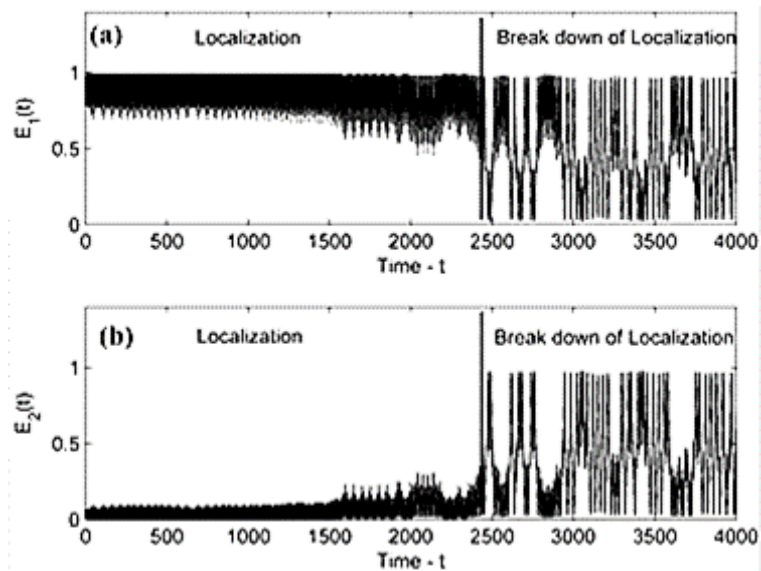

**Fig. 33.** Instantaneous energy corresponding to the mixed type of responses exhibiting temporal localization and sudden bursts of the beating-like behavior ($\chi_b = 0.09$, $n = 3$): (a) energy of the first oscillator; (b) energy of the second oscillator. Initial conditions: $x_1(0) = x_2(0) = 0$; $v_1(0) = \sqrt{2}$, $v_2(0) = 0$.

Numerical simulations show that, with an increase of the coupling parameter $\chi$, the occurrence of the beating response is usually preceded by the regimes of a 'mixed' type. Such regimes exhibit a temporal energy localization followed by distinct irregular transitions into a beating-like response (Fig. 33) alternating with subsequent localizations.

In Tables 2, 3 we compare the numerically obtained critical values of coupling $\chi$ with the theoretically derived values $\chi_{cr}^{(2)}$, $\chi_{cr}^{(3)}$ for different values of $n$. In numerical simulations, the total system energy is normalized to unity $E = 1$. Table 2 corresponds to initial excitations $x_1(0) = 0$, $v_1(0) = \sqrt{2}$ , Table 3 - to $x_1(0) = [(n+1)(1+\chi)^{-1}]^{1/(n+1)}$, $v_1(0) = 0$. We consider the following critical values: $\chi_b$, corresponding to the breakdown of localization and the occurrence of the response of a 'mixed' type, and $\chi_{SB}$, related to pure beatings (Fig. 33). This "mixed" type of response has been observed in the interval $\chi_b < \chi < \chi_{SB}$, or, in other words, in the interval between energy localization ($\chi < \chi_b$) and beating, associated with complete energy exchange ($\chi > \chi_{SB}$).

| | $n = 3$ | $n = 5$ | $n = 7$ | $n = 9$ | $n = 11$ | $n = 13$ |
|---|---|---|---|---|---|---|
| $\chi_{cr}^{(2)}$ ($n \leq 5$) $\chi_{cr}^{(3)}$ ($n \geq 7$) | 0.1667 | 0.1071 | 0.0625 | 0.0445 | 0.0347 | 0.0285 |
| $\chi_{SB}$ | 0.17 | 0.112 | 0.069 | - | - | - |
| $\chi_b$ | 0.09 | 0.0875 | 0.054 | 0.038 | 0.03 | 0.027 |

**Table 2.** Comparison of the threshold values $\chi_{cr}^{(2)}$, $\chi_{cr}^{(3)}$ obtained from the theoretical analysis with the critical parameters $\chi_b$, $\chi_{SB}$ obtained by numerical calculation of the response of system (4.1) subject to impulse excitation $x_1(0) = x_2(0) = 0$; $v_1(0) = \sqrt{2}$ , $v_2(0) = 0$.

| | $n = 3$ | $n = 5$ | $n = 7$ | $n = 9$ | $n = 11$ | $n = 13$ |
|---|---|---|---|---|---|---|
| $\chi_{\text{cr}}^{(2)}$ | 0.1667 | 0.1071 | 0.0625 | 0.0445 | 0.0347 | 0.0285 |
| $\chi_{SB}$ | 0.173 | 0.11 | 0.0637 | - | - | - |
| $\chi_b$ | 0.165 | 0.108 | 0.0635 | 0.044 | 0.035 | 0.029 |

**Table 3.** Comparison of the threshold values $\chi_{\text{cr}}^{(2)}$, $\chi_{\text{cr}}^{(3)}$ obtained from the theoretical analysis with the critical parameters $\chi_b$, $\chi_{SB}$ obtained by numerical calculation of the response of system (4.1) subject to initial displacement $x_1(0) = [(n + 1)(1 + \chi)^{-1}]^{1/(n+1)}$, $x_2(0) = 0$; $v_1(0) = v_2(0) = 0$.

An insignificant difference in the critical values $\chi_b$ and $\chi_{SB}$ in Tables 2 and 3 can be explained by high sensitivity of system (4.1) to the change of the initial conditions. However, despite these deviations, one can note a good agreement between the numerical and analytical critical values in Tables 2, 3.

It is important to note that the formation of regular beatings for $n \geq 9$ is problematic. However, even in the absence of a regular response, one can observe a transition from localization to complete energy exchange at $\chi = \chi_{\text{cr}}^{(2)}$ provided the latter exhibits highly irregular motion, i.e., the chaotic-like behavior. It follows from the results in Tables 2, 3 that, despite the absence of a regular beating response for $n \geq 9$, the analytical model predicts fairly well a critical value $\chi_b$ corresponding to the breakdown of localization followed by irregular energy transfer between the oscillators.

## 5. Non-conventional synchronization of weakly coupled active oscillators

In this section we describe a new type of self-sustained oscillations associated with the phenomenon of synchronization. Previous studies in the model of two weakly coupled Van der Pol oscillators considered their synchronization in the regimes close to nonlinear normal modes (NNMs). In [87, 114] it was shown for the first time that in the important case of the threshold excitation an alternative synchronization mechanism can emerge even if the conventional synchronization becomes impossible. We identify this mechanism as an appearance of the dynamic attractor with the complete periodic energy exchange between the oscillators and then show that it can be interpreted as a dissipative analogue of highly intensive beats in a conservative system. This type of motion is therefore opposite to the NNM-type synchronization in which, by definition, no energy exchange can occur. The analytical study is based on the LPT concept but, in contrast to the conservative systems, in the present case the LPT can be regarded as an attractor. Finally, it is shown that within the LPT approach, the localized mode represents an attractor in the range of model parameters wherein the LPT as well as the in-phase and out-of-phase NNMs become unstable.

The chain of coupled dissipative oscillators described by the Van der Pol (VdP) or Van der Pol-Duffing (VdP-D) equations represents one of the fundamental models of nonlinear dynamics [148, 207] with numerous applications in different fields of physics and biophysics [26,89,148,141,155]. In the continuum limit, it can be reduced to the complex Ginsburg-Landau equation [2,101,133], which admits periodic or localized solutions in several significant cases [89,101,133,148]. The simplest discrete model that combines two nonlinear dissipative oscillators was considered earlier. As it was mentioned above, the main attention was paid to synchronization of the oscillators in the regimes close to NNMs [26,89,123,148,141,155,202]. We have demonstrated that, in the conservative case, another type of motion characterized by complete energy exchange between oscillators and defined as LPT can play the fundamental role. This dynamical regime is opposite to NNMs by its physical meaning. The present section analytically demonstrates that the LPT corresponding to complete energy exchange between weakly coupled oscillators appears to be an attractor which is alternative to that of the NNM's type. Revealing a new class of attractors may serve as the starting point for deeper understanding and further applications of the synchronization phenomenon.

## 5.1. Lienard equations

We begin with a system of two linearly coupled Lienard oscillators, separately described by the equation

$$\frac{d^2u}{dt^2} + f(u)\frac{du}{dt} + g(u) = 0 \qquad (5.1)$$

It is known that such an oscillator possesses stable periodic attractors (limit cycles), whenever functions $f$ and $g$ satisfy the conditions of the well-known Poincare-Bendixon theorem (see, e.g. [89]). Usually, $f$ is an even whereas $g$ is an odd analytical function represented in the form of the truncated power series

$$f(u) = a_0 + a_2u^2 + a_4u^4 + O(u^6), \; g(u) = b_1u + b_3u^3 + O(u^5). \qquad (5.2)$$

The signs of the coefficients are responsible for the qualitative dynamical behavior of the model. If $a_0 < 0$, $a_2 > 0$, $a_4 = 0$, and $b_1 > 0$, the oscillator takes the VdP-D form. In this case, the equilibrium point ($u = 0$, $du/dt = 0$) represents an unstable focus. As a result, the energy flows into the system near the equilibrium for any small magnitude of initial perturbations. However, it would be physically reasonable to assume the existence of a threshold of the initial excitation above which the global dynamical process can be triggered. Within the above expansion for $f(u)$, the only way to introducing the threshold is to choose the coefficient as follows: $a_0 > 0$, $a_2 < 0$, $a_4 > 0$. This choice provides local stability of the equilibrium point so that the energy in-flow into the system is triggered only above certain non-zero level of excitation, when $f(u)$ becomes negative [89].

## 5.2. Coupled active oscillators

Our principal goal is to demonstrate that the presence of the energy threshold allows revealing a new type of synchronization in the system of two weakly coupled active oscillators, We denote the numerical and scaling factors as follows: $a_0 = 2\varepsilon\gamma$, $a_2 = -8\varepsilon g$, $a_4 = 16\varepsilon d$, $b_1 = 1$, $b_2 = 8\varepsilon\alpha$, and the strength of coupling $2\varepsilon\beta$, where $0 < \varepsilon << 1$ is a small parameter of the system. As a result, the model under consideration takes the form

$$\frac{d^2u_1}{d\tau_0^2} + u_1 + 2\beta\varepsilon(u_1 - u_2) + 8\varepsilon\alpha u_1^3 + 2\varepsilon(\gamma - 4gu_1^2 + 8du_1^4)\frac{du_1}{d\tau_0} = 0,$$  (5.3)

$$\frac{d^2u_2}{d\tau_0^2} + u_2 + 2\beta\varepsilon(u_2 - u_1) + 8\varepsilon\alpha u_2^3 + 2\varepsilon(\gamma - 4gu_2^2 + 8du_2^4)\frac{du_2}{d\tau_0} = 0.$$

As in Section 3.1, we introduce the change of variables $Y_r = (v_r + iu_r)e^{-i\tau_0}$, $r = 1, 2$, such that $Y_r(\tau_0, \tau_1, \varepsilon) = \varphi_r^{(0)}(\tau_1) + \varepsilon\varphi_r^{(0)}(\tau_0, \tau_1) + O(\varepsilon^2)$, $\tau_1 = \varepsilon t$ and then conclude that the leading order slow terms $\varphi_r^{(0)}(\tau_1)$ can be presented in the form (1.4): $\varphi_1^{(0)}(\tau_1) = a(\tau_1)\exp(i\beta\tau_1)$, $\varphi_2^{(0)}(\tau_1) = b(\tau_1)\exp(i\beta\tau_1)$, with the slow complex amplitudes $a$, $b$ given by

$$\frac{da}{d\tau_1} - 3ia\,|\,a\,|^2\,a + (\gamma - g\,|\,a\,|^2 + d\,|\,a\,|^4)a + i\beta b = 0,$$

$$\frac{db}{d\tau_1} - 3ia\,|\,b\,|^2\,b + (\gamma - g\,|\,b\,|^2 + d\,|\,b\,|^4)b + i\beta a = 0.$$  (5.4)

Finally, rewriting $a$, $b$ in the polar form

$$a = R_1\exp(i\delta_1),\ b = R_2\exp(i\delta_2)$$  (5.5)

and considering the phase difference $\Delta = \delta_1 - \delta_2$ as a new variable, we derive the following system of real-valued equations:

$$\frac{dR_1}{d\tau_1} + \gamma R_1 - g R_1^3 + d R_1^5 + \beta R_2 \sin \Delta = 0,$$

$$\frac{dR_2}{d\tau_1} + \gamma R_2 - g R_2^3 + d R_2^5 - \beta R_1 \sin \Delta = 0, \qquad (5.6)$$

$$R_1 R_2 \frac{d\Delta}{d\tau_1} + 3\alpha R_1 R_2 (R_2^2 - R_1^2) + \beta (R_2^2 - R_1^2) \cos \Delta = 0.$$

A principal advantage of transforming of the complex equations (5.4) into the real form (5.6) is that this allows us to reveal more easily the conditions providing existence of a symmetry, and therefore, to consider both types of dynamical synchronization (NNMs and LPTs) from the unified standpoint, as discussed below.

### 5.3. NNMs and LPTs

System (5.6) is non-integrable but it possesses the discrete symmetry, that is, it preserves its form under the following coordinate replacement:

(a) $R_1 \rightarrow R_2; R_2 \rightarrow R_2; \Delta \rightarrow -\Delta;$

(b) $R_1 \rightarrow -R_2; R_2 \rightarrow -R_1; \Delta \rightarrow \Delta.$ \qquad (5.7)

The symmetries (a) and (b) provide the existence of the in-phase ($R_1 = R_2, \Delta = 0$) and the out-of-phase ($R_1 = R_2, \Delta = \pi$) NNMs, respectively. In a general case, Eqs. (5.6) do not explicitly reveal any other symmetry, discrete or continuous, except of the temporal translation. However, if any non-trivial continuous symmetry exists under certain conditions, it can be found in the framework of the Lie group theory [144], by manipulating the infinitesimal differential operator of the dynamical system (5.6), $X = X_0 + X_1$, where

$$X_0 = \xi(R_1, R_2, \Delta) \frac{\partial}{\partial \tau_1} + \eta(R_1, R_2, \Delta) \frac{\partial}{\partial R_1} + \zeta(R_1, R_2, \Delta) \frac{\partial}{\partial R_2} + \varsigma(R_1, R_2, \Delta) \frac{\partial}{\partial \Delta},$$

and $X_1$ is the first continuation of the operator $X_0$, whose components are given by time derivatives in system (5.6). Following the technique of [144] and considering the partial differential equations for the components of operator $X$, we reveal the existence of the rotation group in the plane $(R_1, R_2)$ with the invariant $I \equiv N = R_1^2 + R_2^2$. For the rotational symmetry to take place, the parameters of system (5.6) must satisfy the relation $g^2 = 9\gamma d/2$, while the initial conditions provide a certain excitation level given by the constant $N = 2g/3d$. In this case, introducing the coordinate transformation $R_1 = \sqrt{N}\cos\theta$ and $R_2 = \sqrt{N}\sin\theta$, we obtain the system

$$\frac{d\theta}{d\tau_2} = \frac{1}{2}\left(\sin\varDelta - \lambda\sin 4\theta\right),$$

$$\sin 2\theta \frac{d\varDelta}{d\tau_2} = \cos 2\theta\cos\varDelta + 2k\sin 4\theta,$$

$$(5.8)$$

where $\tau_2 = \beta\tau_1$, and the parameters $k = 3\alpha N/2\beta$ and $\lambda = N^2 d/8\beta$ characterize nonlinearity and dissipation relative to coupling of the generators, respectively.

### 5.4. Analysis of the phase plane and analytical solutions

Figure 34 depicts the system behavior for different combinations of the nonlinearity and dissipation parameters. First, we consider the case $\lambda = 0$, when system (5.8) is conservative and the NNMs are stable (Fig. 34(a)). Two branches of the LPT are associated with complete energy exchange between the generators. In order to avoid the conservative type bifurcation of the NNMs occurring at $k = ½$, we choose the parameter of nonlinearity between 0 and 1/2. When the parameter $\lambda$ is relatively small, the system has two unstable focuses corresponding to unstable NNMs of the original system (5.1) (Fig. 34, (b), (c), (d)). The focuses transform into unstable nodes when $\lambda^2 > 1 - 2k$ (Fig. 34, (e) and (f)). If the dissipative parameter $\lambda$ does not exceed the value $0.5(1 + \sqrt{1 - 4k^2})$, the only attractor of the system is the LPT with intensive energy exchange between the oscillators.

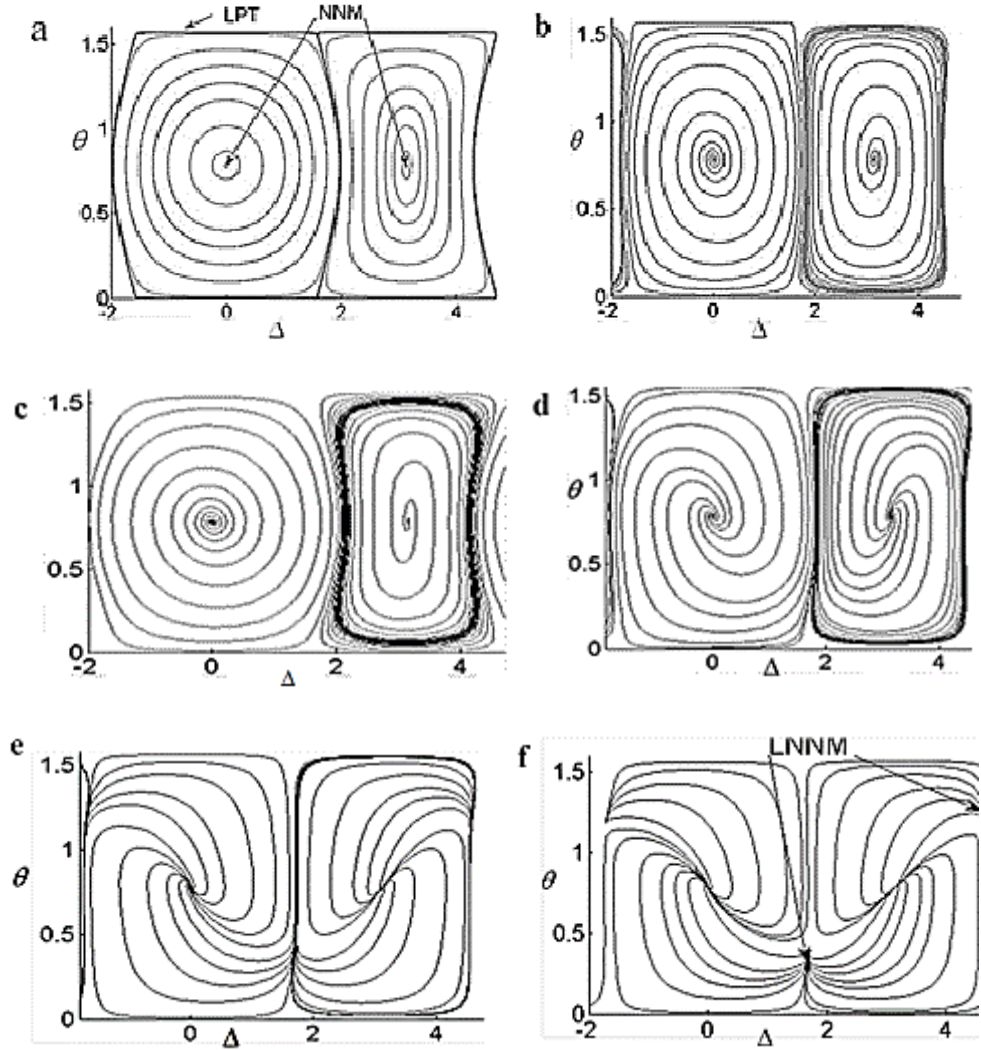

**Fig. 34.** Phase planes of system (5) in terms of the variables $\theta$ and $\Delta$: (a) $k=0.2$, $\lambda=0$ (conservative system), b) $k=0.2, \lambda=0.1$; (c) $k=0.44, \lambda=0.1$ (d) $k=0.2, \lambda=0.5$; (e) $k=0.1$, $\lambda=0.9$; (f) $k=0.1$, $\lambda=0.99$. The results are in a good agreement with simulations of the original system (1).

In Fig. 33, one can see that the phase shift between the oscillators on the LPT remains near $\pm\pi/2$ (Fig. 33 b), while the sign changes almost instantly as the system approaches the LTP attractor. Therefore, the oscillators become synchronized in a non-conventional way that can be qualified as the 'LPT-type synchronization.' Furthermore, if $\lambda \geq 0.5(1+\sqrt{1-4k^2})$, the LPT becomes unstable, and the attractor is a stationary point corresponding to the localized NNM with energy predominantly trapped on one of the two oscillators (Fig. 34 (f)).

It is important to note that the evolution of the LPT leading to the transition from energy exchange to energy localization, turns out to be independent on the evolution of the stationary points and occurs 'later' than the transformation from an unstable focus to an unstable node (Fig. 34(e).)

In the range of intensive energy exchange, one can obtain an analytical solution of system (5.8) by using the saw-tooth transformations [149]:

$$\theta = A\tau + \frac{\lambda}{4}[\cos(4A\tau) - 1]e + ..., \quad \Delta = \pi - \left[\frac{\pi}{2} - 2k\sin(2A\tau)\right]e + ..., \quad (5.9)$$

where the basis functions $\tau$ and $e$ are defined by formulas (2.20) and depicted in Fig. 7. Note that numerical solutions shown in Fig. 35(a, b) appear to be in good agreement with analytical solutions (5.9) demonstrated in Fig. 35(c, d).

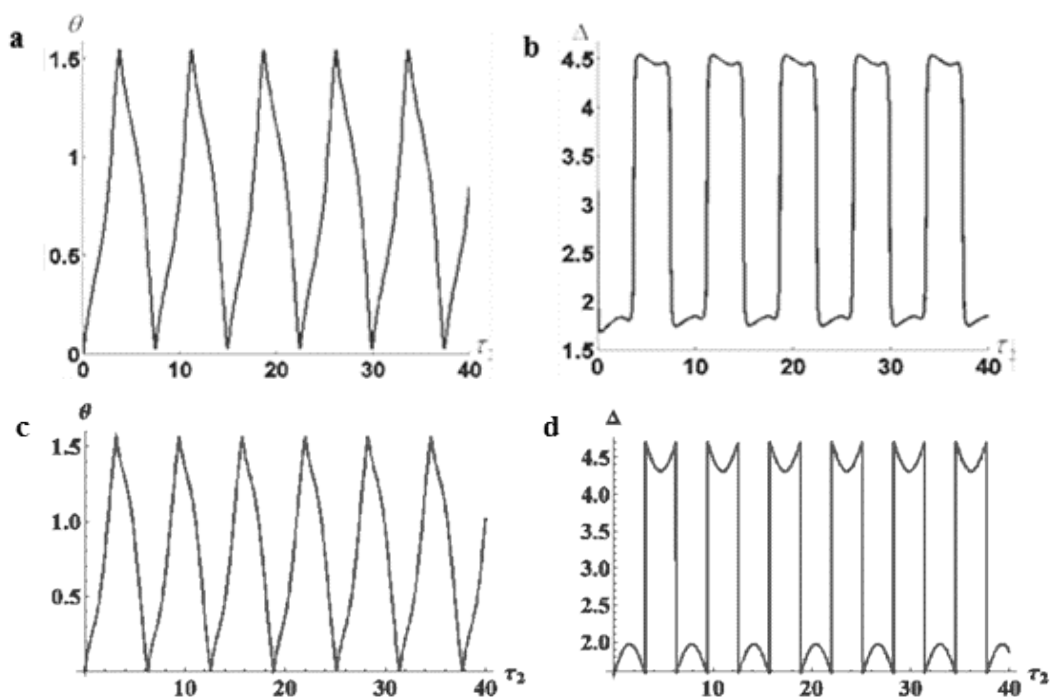

**Fig. 35.** Numerical (exact) solutions $\theta$ (a) and $\Delta$ (b) and their non-smooth approximations (c) and (d)

The behavior of the system before and after the transition from non-conventional synchronization on the LPT to synchronization on the localized NNM is illustrated in Fig. 36(a) and 36(b), in terms of the original variables. The amplitudes $R_j = \sqrt{u_j^2 + v_j^2}, j = 1,2$, presented in Fig. 36 are obtained by numerical integration of the original system (5.3). Non-conventional synchronization on the LPT far from the localization threshold is shown in Fig.

36(c). Also, we present for comparison the plot demonstrating the well-studied conventional synchronization, realized on the out-of-phase NNM (Fig. 36(d)). In the latter case, the set of parameters is taken far enough from the accepted symmetry conditions.

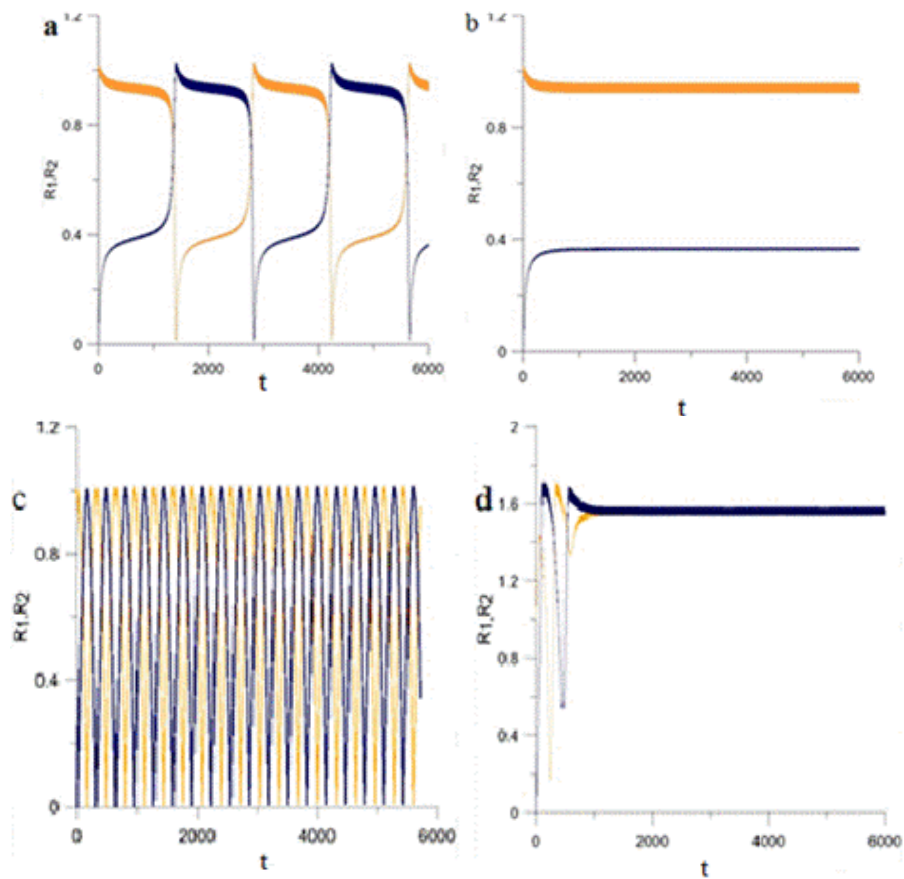

**Fig. 36.** Transitions from energy exchange to energy localization in system (5.1) of two coupled generators. The amplitudes $R_j = \sqrt{u_j^2 + v_j^2}, j = 1,2$ are obtained by numerical integration of the original system (5.3) ( $R_1$ - orange line, $R_2$ - blue line) (a) energy exchange just before the transition to localization for the parameters $k$ = 0.1, $\lambda$ = 0.98; (b) the behavior just over the localization threshold, in which most part of energy is localized on one of the oscillators for the parameters $k$ = 0.1, $\lambda$ = 0.99; (c) nonconventional synchronization corresponding to the parameter set of Fig. 34(b), far from the localization threshold; and (d) conventional synchronization on out-of-phase NNM when the parameters of generators do not satisfy the accepted symmetry conditions.

It is worth noticing that the presented scenario seems to be quite general in the area of model parameters wherein the conventional synchronization is impossible (it is confirmed by the analytical estimates and numerical simulations presented in [87, 114]. However, the considered case allows the detailed analytical description and reveals a new important type of synchronization. Besides, one can see how complete energy exchange between different parts of the systems on the LPT goes over to the energy localization. By the adiabatic change of the system parameters, the LPT-type synchronization can be realized between the out-of-phase and the in-phase NNM-type synchronizations. Now, after our prediction of the domain of the dissipative parameters wherein the LPT-type synchronization exists, such type of synchronization can be found experimentally in the physical, chemical, and biological systems modelled by a pair of coupled generators.

# 6. Limiting phase trajectories in passage through resonance in time-variant systems

In this section we analyze passage through resonance and associated effects (autoresonance and tunneling) in oscillator arrays with slowly time-varying parameters. We show that there exists an initial interval of motion, wherein the effect of slow change of parameters is negligible, and the dynamics of the system corresponds to motion along the LPT of a similar time-invariant system. When the time increases, the effect of slowly-changing parameters becomes visible and results in a drastic change of the system dynamics.

We begin with the consideration of a nonlinear (Duffing) oscillator with a slowly varying linear stiffness subjected to a periodic excitation (Section 6.1). It is important to underline that properties of this system do not directly correspond to the properties of a forced oscillator with constant parameters (Section 2.1). In the system under consideration, the amplitude of oscillations either converges to a certain stationary level or the motion can be considered as small oscillations near a slowly increasing mean value (Figs. 44, 45). The phenomenon of the permanent growth of energy in a classical anharmonic oscillator subject to slow variations of forcing and/or resonant frequencies is referred to as *autoresonance* (AR), or nonstationary resonance.

Section 6.2 investigates energy transfer from an excited nonlinear actuator (the Duffing oscillator) to a coupled passive linear oscillator. Two classes of problems are studied analytically and numerically: (1) a periodic force with constant frequency is applied to the nonlinear (Duffing) oscillator with slowly time-decreasing linear stiffness; (2) the time-independent nonlinear oscillator is excited by a force with slowly increasing frequency. In both cases, stiffness of the attached linear oscillator and linear coupling remain constant, and the system is initially engaged in resonance. It is shown that in the system of the first type autoresonance in the nonlinear oscillator may entail a high-energy regime in the coupled oscillator as well. On the contrary, in the system of the second type the most part of energy remains localized on the excited oscillator, whereas the coupled oscillator exhibits bounded low-energy oscillations.

Sections 6.3 and 6.4 are devoted to analytical and numerical investigations of tunneling in coupled oscillators.

## 6.1. Autoresonance in a SDOF nonlinear oscillator

The occurrence of autoresonance may be informally explained by an example of capture into resonance of the Duffing oscillator. Let us assume that the oscillator is initially at rest and subjected to an external force of small amplitude $\varepsilon$ and constant frequency $\omega_0$ equal to the frequency of linear oscillations. It is known that an increase of the natural frequency due to continuous growth of the amplitude of nonlinear oscillations results in breakup of resonance in the oscillator with constant parameters. However, if slowly decreasing stiffness counterbalances an effect of increasing amplitudes and sustains the value of the natural frequency near its initial value $\omega_0$, the oscillator remains captured into resonance with the external force. After first studies for the purposes of particle acceleration [129, 206], AR has become a very active field of research. Theoretical approaches, experimental evidence and applications of AR in different fields of natural science, from plasmas to planetary dynamics, are reported in numerous papers [46]; additional theoretical and computational results can be found in [11,17,25,68]; recent advances in this field are discussed, e.g., in [3,46,137, 140, 170].

In this section, we study the dynamics of the Duffing oscillator with slowly varying linear stiffness subjected to a periodic excitation with constant frequency. The equation of motion is given by

$$\frac{d^2u}{d\tau_0^2} + (1 - \varepsilon\zeta(\tau))u + 8\varepsilon\alpha u^3 = 2\varepsilon F\cos\tau_0, \tag{6.1}$$

where $\varepsilon > 0$ is a small parameter of the system, $\zeta(\tau) = s + b\tau$, $\tau = \varepsilon\tau_0$ is the slow time scale. As in the previous sections, initial conditions $u = 0$, $v = du/d\tau_0 = 0$ at $\tau_0 = 0$ determine the LPT of system (6.1) corresponding to the maximum possible energy transfer from the source of energy to the oscillator. We consider the case of $s > 0$, $\alpha > 0$.

Asymptotic solutions of Eq. (6.1) for small $\varepsilon$ can be obtained in the same way as in Section 2.1. First, we introduce the change of variables $Y = (v + iu)e^{-i\tau_0}$, where the amplitude $Y$ is constructed in the form of the multiple-scales expansion with a slow main term: $Y(\tau_0, \tau, \varepsilon) = \varphi^{(0)}(\tau) + \varepsilon\varphi^{(1)}(t, \tau) + O(\varepsilon^2)$. Using rescaling

$$\tau_1 = s\tau, \ \varphi = \Lambda^{-1}\varphi^{(0)}, \ \Lambda = (s/3\alpha)^{1/2}, f = F/s\Lambda = F\sqrt{3\alpha/s^3}, \ \beta = b/s^2, \qquad (6.2)$$

and repeating the reasoning of Sec. 2.1, a two-parameter equation similar to (2.8) is derived:

$$\frac{d\varphi}{d\tau_1} + i(1 + \beta\tau_1 - |\varphi|^2)\varphi = -if, \varphi(0) = 0. \qquad (6.3)$$

The polar representation $\varphi = ae^{i\Delta}$ transforms Eq. (6.3) into the system

$$\frac{da}{d\tau_1} = -f\sin\Delta,$$

$$\frac{d\Delta}{d\tau_1} = -(1 + \beta\tau_1) + a^2 - a^{-1}f\cos\Delta. \qquad (6.4)$$

with initial conditions $a(0) = 0$, $\Delta(0) = -\pi/2$. It now follows from (6.2) - (6.4) that

$$u(\tau_0, \varepsilon) = \Lambda a(\tau_1)\sin(\tau_0 + \Delta(\tau_1) + \tau_1) + O(\varepsilon), \qquad (6.5)$$

$$v(\tau_0, \varepsilon) = \Lambda a(\tau_1)\cos(\tau_0 + \Delta(\tau_1) + \tau_1) + O(\varepsilon).$$

### 6.1.1. Critical parameters

For better understanding of the occurrence of unbounded modes, we first consider the underlying *time-invariant* system (2.11), namely,

$$\frac{da}{d\tau_1} = -f \sin \Delta,$$

$$\frac{d\Delta}{d\tau_1} = -1 + a^2 - a^{-1} f \cos \Delta. \qquad (6.6)$$

with initial conditions $a(0) = 0$, $\Delta(0) = -\pi/2$ corresponding to the LPT. It was proved in Section 2.1 that there exist two critical relationships

$$f_1 = \sqrt{2/27} \approx 0.2721, f_2 = 2/\sqrt{27} \approx 0.3849 \qquad (6.7)$$

which define the boundaries between different types of the dynamical behavior. By definition given in Section 2.1, conditions $f < f_1$, $f_1 < f < f_2$, and $f > f_2$ characterize *quasi-linear*, *moderately nonlinear*, and *strongly nonlinear* dynamics, respectively.

Now we extend the above-mentioned results to system (6.4). Figure 37 present the results of numerical simulations for system (6.4) with detuning $\beta\tau_1$ and the following values of the system parameters: $\beta = \pm 0.07$, $f = 0.34$ ($f_1 < f < f_2$).

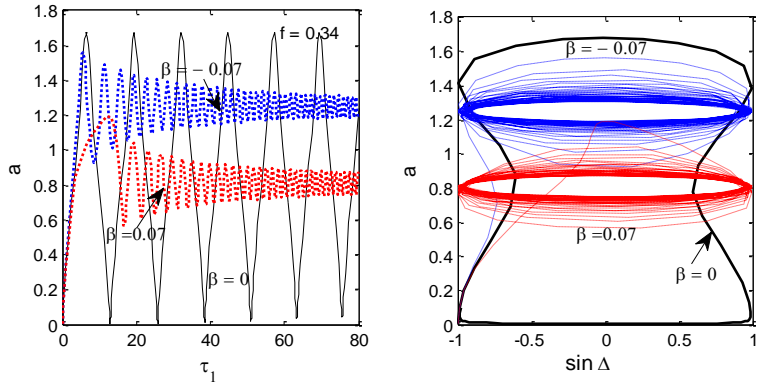

**Fig. 37**. Plots of $a(\tau_1)$ and phase portraits of system (6.4) with parameters $\beta = \pm 0.07$, $f = 0.34$. Plots of the LPTs for the corresponding time-invarinat systems ($\beta = 0$, black solid lines) are shown for comparison.

It is seen in Fig. 37 that the solution of Eq. (6.4) is very close to the LPT of the time-independent system during the first half-cycle of oscillations. If $\beta > 0$, detuning $\beta\tau_1$ increases with an increase of $\tau_1$, thereby shifting the system to the domain of small oscillations; if $\beta < 0$, detuning decreases with an increase of $\tau_1$; in the latter case, the system passes through resonance with large amplitude of oscillations. This implies that the LPT of the time-invariant system ($\beta = 0$) determines the maximum achievable energy level.

The projection of the trajectory $a(\tau_1)$ onto the phase plane represents the spiral orbit with an attracting focus ($a = a_0$, $\sin\Delta = 0$), where $a_0 = \lim a(\tau_1)$ as $\tau_1 \to \infty$.

Figure 38 depicts the emergence of AR from stable bounded oscillations under the change of rate $\beta > 0$. As seen in Fig. 38, under very slow sweep the transition from bounded oscillations to AR takes place if the parameter $f$ is close to the critical value $f_1$. In particular, this means that the peak of the LPT in the time-invariant system ($\beta = 0$) determines a minimum energy level achievable in the process of capture into resonance.

It is shown in Fig. 38 that at $f = 0.274$ the transition occurs at $\beta \approx 0.001$; the difference between $f$ and $f_1 = 0.2721$ is less than 1%; at $f \approx 0.28$ the transition takes place at $\beta \approx 0.006$; the difference between $f$ and $f_1$ is less than 2.7%. On the other hand, for $f = 0.34$, the critical rate $\beta \approx 0.061$; the difference between $f$ and $f_1$ is about 20%. This implies that the inequality $f > f_1$ can be interpreted as *the necessary condition* of the emergence of AR.

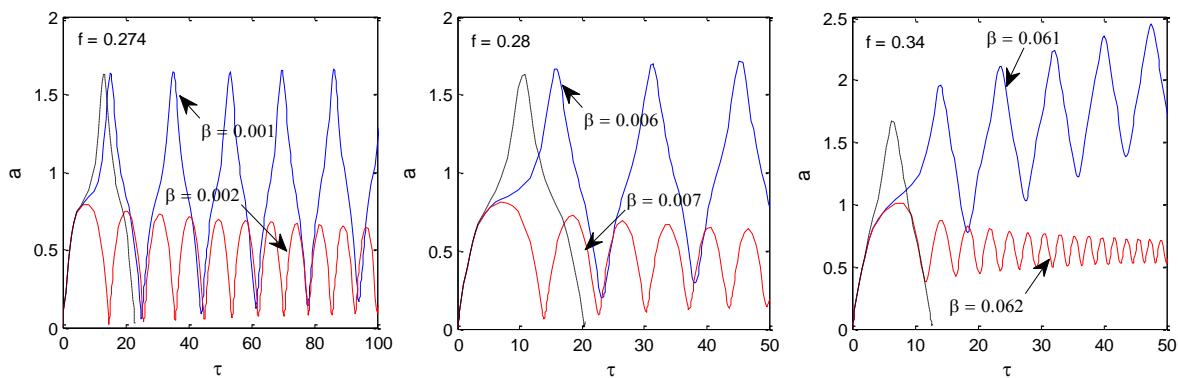

**Fig. 38**. Transitions to AR in system (6.4) with different parameters $f$ and $\beta$; the cycle of oscillations in the time-independent system (black line) is demonstrated for comparison.

Figure 39 elucidates the nature of AR oscillations. It is seen that in the first half-cycle of oscillations the amplitude $a(\tau_1)$ is close to the LPT of a moderately nonlinear mode of motion (cf. Fig. 5(b)). Then the shape of the trajectory changes, and it turns into quasi-linear oscillations near an upward quasi-steady center $(\bar{a}(\tau_1),\ 0)$ with a slowly increasing value of $\bar{a}(\tau_1)$. The the quasi-stationary amplitude $\bar{a}(\tau_1)$ is calculated below.

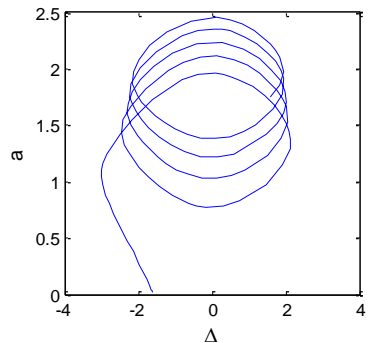

**Fig. 39**. Transition to AR in system (6.4) with parameters $f = 0.34$, $\beta = 0.061$.

The obtained numerical results motivate the derivation of an analytical threshold between bounded and unbounded oscillations. In order to evaluate the critical rate $\beta$ corresponding to the transition from bounded to unbounded oscillations, we employ the fact that for sufficiently small $\tau_1$ the solution $a(\tau_1)$ of system (6.4) is very close to the LPT of the time-invariant system (6.6). We recall that the LPT of the moderately nonlinear $(f_1 < f < f_2)$ time-invariant systems has a distinctive inflection at an instant $\tau_1 = T^*$ (Fig. 38). We introduce the parameter $\tilde{f}(\tau_1) = f/(1 + \beta\tau_1)^{3/2}$ such that $\tilde{f}(0) = f > f_1$. Numerical results in Figs. 37, 38 indicate that an adiabatically varying system in which $\tilde{f}(0) > f_1$ gets captured into the domain of small oscillations provided $\tilde{f}(T^*) < f_1$. Under this assumption, the critical rate is given by

$$\beta^* = (T^*)^{-1}[(f/f_1)^{2/3} - 1]. \tag{6.8}$$

If $\beta < \beta^*$, the system admits AR. In order to check the correctness of equality (6.8), we calculate the critical rate $\beta^*$ in the system with linear-in-time detuning ($n = 1$). First, the instant $T^*$ will be found from the obtained numerical results. In the next step, the analytical estimate of $T^*$ and the respective value $\beta^*$ will be derived.

We recall that the point of inflection is determined by the conditions $da/d\tau_1 \neq 0$, $d^2a/d\tau_1{}^2 = 0$. It follows from (6.6) that the latter condition corresponds to $d\Delta/d\tau_1 = 0$, i.e., the envelope $a(\tau_1)$ achieves the point of inflection at the minimal value of the phase $\varDelta$ (Fig. 40).

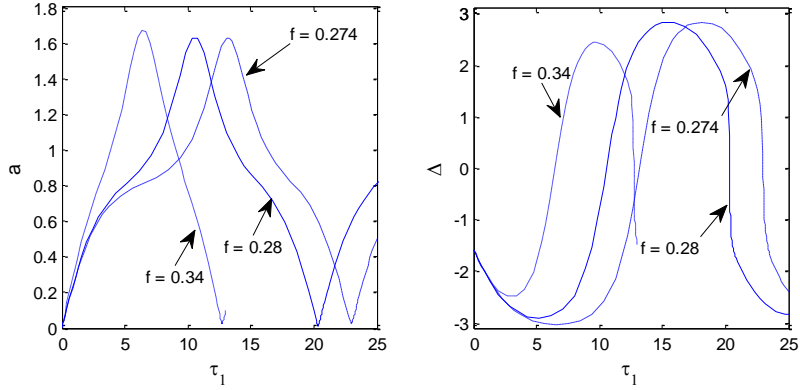

**Fig. 40**. Envelopes $a(\tau_1)$ and phases $\varDelta(\tau_1)$ for different values of the parameter $f$

As seen in Fig. 40, $T^* \approx 6.5$ for $f = 0.274$; this yields $\beta^* \approx 0.00075$, while computational results give $0.001 < \beta < 0.002$. Then, $T^* \approx 5$ and $\beta^* \approx 0.004$ for $f = 0.28$, while the numerical simulation gives $0.006 < \beta < 0.007$. Note that for $f = 0.28$ the threshold parameter $\mu_{th} = |\beta_{th}|^{-3/4} f = 0.41$ yields $\beta_{th} = (f/\mu_{th})^{4/3} = 0.467$, which is vastly larger than the real threshold rate. In a similar way, we find that for $f = 0.34$ the critical rate $\beta^* = 0.053$, while the numerical simulation gives $0.061 < \beta < 0.062$. Note that at $f = 0.34$ the inflection of the curve $a(\tau_1)$ is practically indistinguishable but the phase has the distinct minimum at $T^* \approx 3$ (Fig. 40).

It is important to note that formula (6.8) defines the critical rate for systems with both linear and nonlinear-in-time detuning laws. For example, in the case of quadratic detuning $\beta\tau_1^2$ and $f = 0.28$, we find $\beta^* = 0.0008$; at the same time, the numerical simulation gives $0.001 < \beta < 0.002$ (Fig. 40).

An analytical derivation of both an instant $T^*$ and a point of inflection $a^*$, $\varDelta^*$ is built upon the results obtained in Section 2.1. We recall that the amplitude $a(\tau_1)$ corresponding to the LPT of conservative system (6.6) satisfies the second-order equations (2.17), namely,

$$\frac{d^2a}{d\tau_1^2} + \frac{dU}{da} = 0 \qquad (6.9)$$

with potential (2.18). The inflection point is defined by the condition $dU/da = 0$ at $a = a^*$, or

$$a^* = \sqrt{2/3} = 0.8165. \tag{6.10}$$

The value of the phase $\varDelta^*$ at inflection is defined by the equality $d\varDelta/d\tau_1 = 0$. Using (6.6), (6.10), it is easy to derive that $\cos\varDelta^* = a^*[(a^*)^2 - 1]/f$, where $a^*[(a^*)^2 - 1] = -f_1$, and thus,

$$\cos\varDelta^* = -f_1/f. \tag{6.11}$$

Since the maximum of $U(a)$ is also defined by the condition $dU/da = 0$, the potential barrier passes through the point of inflection $a = a^*$. It follows then that the time up to inflection $T^*$ equals the time $\tau_1^*$ needed to reach the potential barrier. Using the same arguments as in Section 2.1, one can find that

$$T^* = \tau_1^* \approx 3\ln\frac{(f^2 - f_1^2)^{1/2}}{f - f_1} = \frac{3}{2}\ln\frac{f + f_1}{f - f_1} . \tag{6.12}$$

For example, in the case of $f = 0.274$ and $f = 0.28$ we obtain the approximations $\tau^* \approx 7.48$ and $\tau_1^* \approx 5.9$, which exceed the corresponding numerical values $T^*$ (Fig. 40) for 15%. It follows then that the substitution of $\tau_1^*$ for $T^*$ into (6.8) gives the rate $\beta_1^* < \beta^*$. Therefore, detuning rate $\beta < \beta_1^* < \beta^*$ allows the occurrence of autoresonance.

### 6.1.2. Numerical study of capture into resonance

The obtained numerical results allow for the representation of the complex amplitude $\varphi(\tau_1)$ as $\varphi(\tau_1) = \overline{\varphi}(\tau_1) + \tilde{\varphi}(\tau_1)$, where the terms $\overline{\varphi}(\tau_1)$ and $\tilde{\varphi}(\tau_1)$ denote a quasi-stationary value of $\varphi(\tau_1)$ and rather small fast fluctuations near $\overline{\varphi}$, respectively. The state $\overline{\varphi}$ can be approximately calculated as a stationary point of Eq. (6.3) with "frozen" detuning $\zeta_0$. We thus obtain

$$(\zeta_0 - |\overline{\varphi}|^2)\overline{\varphi} = -f, \overline{\varphi} \approx \pm\sqrt{\zeta_0} + f/2\zeta_0. \tag{6.13}$$

$$\overline{\varphi} \approx \pm\sqrt{\zeta_0} , \ \overline{a} = |\overline{\varphi}| \approx \sqrt{\zeta_0} \ \text{ if } |f/2\zeta_0| << 1.$$

The quasi-stationary amplitude $\bar{a} \approx \sqrt{\zeta_0}$ corresponds to the backbone curve [3,139] and expresses a relationship between the amplitude and the frequency of free oscillations (Fig. 41). The rapidly oscillating component $\tilde{\varphi}(\tau_1)$ can be approximately computed from the linearized equation

$$\frac{d\tilde{\varphi}}{d\tau_1} + 2i\zeta_0(\tau_1)(\mathrm{Re}\,\tilde{\varphi}) = \mp \frac{\beta}{2\sqrt{\zeta_0(\tau_1)}}. \tag{6.14}$$

Capture into resonance in the system with parameters $\beta = 0.05$, $f = 0.34$ is depicted in Fig. 41. Numerical results demonstrate the convergence of the amplitude $a(\tau_1)$ to a monotonically increasing backbone curve and phase-locking at $\tau_1 \to \infty$.

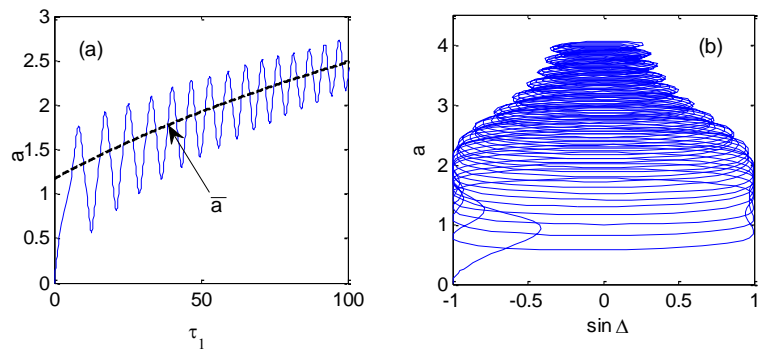

**Fig. 41.** Capture into resonance: (a) convergence of $a(\tau_1)$ to a monotonically increasing backbone $\bar{a}(\tau_1)$; (b) phase locking at $\tau \to \infty$.

## 6.2. Autoresonance versus localization in weakly coupled oscillators

In this section we extend the notion of the resonance energy transfer along the LPT on a 2DOF system consisting of a passive linear oscillator weakly coupled with a nonlinear (Duffing) actuator excited by an external force.

We recall that the analysis of passage through resonance was first concentrated on the basic nonlinear oscillator but then the developed methods and approaches were extended to two- or three-dimensional systems. Examples in this category are excitations of continuously phase-locked plasma waves [10], particle transport in a weak external field with slowly changing frequency [38, 209], control of nanoparticles [72], etc. Some particular results [10, 46] demonstrated that external forcing with a slowly-varying frequency applied to a pair of coupled nonlinear oscillators generates AR in both oscillators. We show that this conclusion is not universal; in particular, it does not hold for a pair of weakly coupled linear and nonlinear (Duffing) oscillators considered in this section.

In this section we examine passage through resonance in two classes of systems. The first class includes the systems, in which a periodic force with constant frequency acts on the Duffing oscillator with slowly time-decreasing linear stiffness; in the systems of the second class, the time-invariant nonlinear oscillator is excited by a force with slowly increasing frequency. In both cases, stiffness of the linear oscillator and linear coupling remain constant, and the entire system is initially captured into resonance.

We demonstrate that AR in the nonlinear actuator may entail oscillations with growing amplitude in the coupled oscillator only in the system of the first class with constant excitation frequency, whereas in the system of the second class the most part of energy remains localized on the excited oscillator, and a portion of energy transferred to the linear oscillator is insufficient to provide growing oscillations. This means that the systems that seem to be almost identical exhibit different dynamical behavior caused by their different resonance properties. It is important to note that the change of frequency of the forcing field is usually considered as an effective tool for producing the desired resonance dynamics, and the failure of this approach in a multi-degrees-of freedom system has not been discussed thus far in the literature.

### 6.2.1. Energy transfer in a system with constant excitation frequency

The equations of motion of two coupled oscillators are given by

$$
\begin{aligned}
&m_1 \frac{d^2 u_1}{dt^2} + c_1 u_1 + c_{10}(u_1 - u_0) = 0, \\
&m_0 \frac{d^2 u_0}{dt^2} + C(t)u + ku^3 + c_{10}(u_0 - u_1) = A\cos \omega t,
\end{aligned}
$$

(6.15)

where $u_0$ and $u_1$ denote absolute displacements of the nonlinear and linear oscillators, respectively; $m_0$ and $m_1$ are their masses; $c_1$ denotes stiffness of the linear oscillator; $c_{10}$ is the linear coupling coefficient; $k$ is the coefficients of cubic nonlinearity; $C(t) = c_0 - (k_1 + k_2 t)$, $k_{1,2} > 0$; $A$ and $\omega$ are the amplitude and the frequency of the periodic force. The system is initially at rest, that is, $u_k = 0$, $v_k = du_k/dt = 0$ at $t = 0$, $k = 0, 1$. We recall that these initial conditions determine the LPT of system (6.15) associated with maximum possible energy transfer from the source of energy to the oscillator.

As in the previous sections, we define the small parameter $2\varepsilon = c_{10}/c_1 << 1$. Considering weak nonlinearity and taking into account resonance properties of the system, we denote

$$c_1/m_1 = c_0/m_0 = \omega^2, \; A = \varepsilon m \omega^2 F, \qquad (6.16)$$

$$k_1/c_0 = 2\varepsilon s, \; k_2/c_0 = 2\varepsilon^2 b\omega, \; k/c_0 = 8\varepsilon\alpha, \; c_{10}/c_1 = 2\varepsilon\lambda_1, \; c_{10}/c_0 = 2\varepsilon\lambda_0.$$

and then reduce the equations of motion to the form:

$$\frac{d^2 u_1}{d\tau_0^2} + u_1 + 2\varepsilon\lambda_1(u_1 - u_0) = 0,$$

$$\frac{d^2 u_0}{d\tau_0^2} + (1 - 2\varepsilon\zeta(\tau))u_0 + 2\varepsilon\lambda_0(u_0 - u_1) + 8\varepsilon\alpha u_0^3 = 2\varepsilon F \sin\tau_0, \qquad (6.17)$$

where $\tau_0 = \omega t$ is the fast time scale, and $\tau = \varepsilon\tau_0$ is the leading-order slow time scale; $\zeta(\tau) = s + b\tau$. The asymptotic analysis of system (6.17) is analogous to the analysis performed in Section 3.1 and Section 6.1. As in the previous sections, we introduce the following transformations:

$$v_k + iu_k = Y_k e^{i\tau_0}, \; Y_k(\tau_0, \tau, \varepsilon) = \varphi_k^{(0)}(\tau) + \varepsilon\varphi_k^{(0)}(\tau_0, \tau) + \varepsilon^2..., \qquad (6.18)$$

$$\tau_1 = s\tau, \; \varphi_k = \Lambda^{-1}\varphi_k^{(0)}, \; \Lambda = (s/3\alpha)^{1/2}, \; k = 0, 1,$$

and then perform the separation of the fast and slow time-scales. As a result, we obtain the following system for the leading-order slow complex amplitudes $\varphi_0(\tau_1)$ and $\varphi_1(\tau_1)$:

$$\frac{d\varphi_1}{d\tau_1} - i\mu_1(\varphi_1 - \varphi_0) = 0, \varphi_1(0) = 0,$$

$$\frac{d\varphi_0}{d\tau_1} - \iota\mu_0(\varphi_0 - \varphi_1) + i(\zeta_0(\tau_1) - |\varphi_0|^2)\varphi_0 = -if, \varphi_0(0) = 0. \tag{6.19}$$

with coefficients $f = F/s\Lambda$, $\beta = b/s^2$, $\mu_k = \lambda_k/s$ and detuning $\zeta_0(\tau_1) = 1 + \beta\tau_1$. For more details concerning the asymptotic analysis of similar systems, one can refer to [79-81, 85].

The study of tunneling in weakly coupled oscillator [81, 85] proved that an asymptotic solution of the nonlinear equation can be obtained separately in the following cases: either coupling is weak enough to provide the condition $\mu_1\mu_2 \ll \beta$, or the mass of the attached oscillator is much less than the mass of the actuator, $m_1 \ll m_0$. For brevity, only the first case is discussed below. Under this assumption, the term proportional to $\mu_0$ can be removed from (6.19) in the main approximation. The resulting truncated system

$$\frac{d\varphi_1}{d\tau_1} - i\mu_1\varphi_1 = -i\mu_1\varphi_0, \varphi_1(0) = 0,$$

$$\frac{d\varphi_0}{d\tau_1} + i(\zeta_0(\tau_1) - |\varphi_0|^2)\varphi_0 = -if, \varphi_0(0) = 0. \tag{6.20}$$

includes an independent nonlinear equation for the complex amplitude $\varphi_0$ and a linear equation for $\varphi_1$, in which $\varphi_0$ plays the role an external excitation. The effect of weak coupling on motion of the nonlinear oscillator may be considered in subsequent iterations [80, 81].

The response $\varphi_1(\tau)$ can be defined from (6.19) or (6.20) by the expression:

$$\varphi_1(\tau_1) = -i\mu_1 \int_0^{\tau_1} e^{i\mu_1(\tau-s)}\varphi_0(s)ds. \tag{6.21}$$

As in Section 6.1, the solution of the nonlinear equation is represented as $\varphi(\tau_1) = \bar{\varphi}(\tau_1) + \tilde{\varphi}(\tau_1)$, where $\bar{\varphi}(\tau_1)$ and $\tilde{\varphi}(\tau_1)$ denote a quasi-stationary value of $\varphi(\tau_1)$ and fast fluctuations near $\bar{\varphi}(\tau_1)$, respectively. Since the contribution from fast fluctuations $\tilde{\varphi}_0(\tau)$ to integral (6.21) is relatively small compared to the contribution from the slowly-varying component $\bar{\varphi}_0$, we employ approximation (6.13) to obtain:

$$\varphi_1(\tau_1) \approx -i\mu_1 e^{-i\mu_1\tau_1} J(\tau_1), \; J(\tau_1) = \int_0^{\tau_1} e^{i\mu_1 s} \overline{\varphi}_0(s)\,ds, \qquad (6.22)$$

where $\overline{\varphi}_0(\tau_1) = \sqrt{1 + \beta\tau_1}$ . Integration by parts gives

$$J(\tau_1) = -i\mu_1^{-1}[e^{i\mu_1\tau_1}\overline{\varphi}_0(\tau_1) - \overline{\varphi}_0(0)] - \Phi(\tau_1),$$

$$\Phi(\tau_1) = \frac{\beta}{2}\int_0^{\tau_1}\frac{e^{i\mu_1 s}}{\sqrt{1 + \beta s}}\,ds. \qquad (6.23)$$

It follows from (6.23) that $\Phi(\tau_1) = \sqrt{\beta}F(\tau_1)$, where $F(\tau_1)$ is a Fresnel-type integral bounded for any $\tau_1 > 0$. Hence, $\varphi_1(\tau_1) = \overline{\varphi}_1(\tau_1) + \widetilde{\varphi}_1(\tau_1) + O(\sqrt{\beta})$, where

$$\overline{\varphi}_1(\tau_1) = \overline{\varphi}_0(\tau_1), \; \widetilde{\varphi}_1(\tau_1) \approx -\overline{\varphi}(0)e^{i\mu_1\tau_1}. \qquad (6.24)$$

Although equality $\overline{\varphi}_1(\tau_1) = \overline{\varphi}_0(\tau_1)$ can be directly obtained from (6.19), transformations (6.21) – (6.24) formally demonstrate the equality of the averaged amplitudes for both oscillators, as well as the occurrence of growing oscillations in the linear oscillator.

Let $a_k$ and $a_k^{tr}$ ($k = 0, 1$) denote real-valued amplitudes oscillations in the full system (6.19) and the truncated system (6.20), respectively. Plots of $a_k(\tau_1)$ and $a_k^{tr}(\tau_1)$ in the systems with parameters

$$\beta = 0.05, \mu_0 = 0.02, \mu_1 = 0.25, f = 0.34 \qquad (6.25)$$

are presented in Fig. 42. It is important to note that a single oscillator with parameters $\mu_0 = 0$, $\beta^* \approx 0.06, f = 0.34$ admits AR (see Section 6.1).

It is seen that the amplitudes $a_k$ (solid lines) and $a_k^{tr}$ (dotted lines) calculated for the full system (6.19) and the truncated system (6.20), respectively, are close to each other. Dashed lines depict the identical backbone curves. Furthermore, Fig. 42(a) shows that, as in the case of a SDOF oscillator, in the first half-period of oscillations the amplitudes $a_0$ and $a_0^{tr}$ are close to the LPT of the time-invariant system but then motion turns into small fast oscillations near the backbone curve $\overline{a}_0 \approx \sqrt{\zeta_0}$ .

Figure 42($b$) demonstrates that irregular oscillations of the linear oscillator at the early stage are then transformed into regular oscillations near the backbone curve. In the system with coupling $\mu_1 = 0.15$ we obtain the period of oscillations $T_1 = 2\pi/\mu_1 \approx 25.12$, and the amplitude of fluctuations $|\tilde{\varphi}_1(\tau_1)| = |\overline{\varphi}(0)| = 1$; both these values are close to the parameters in Fig. 42($b$).

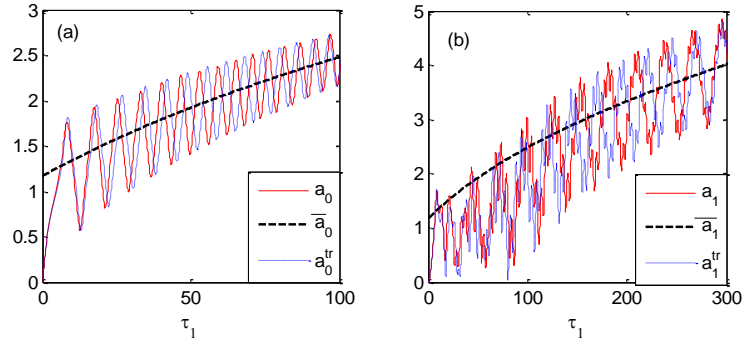

**Fig. 42.** Amplitudes and phases of oscillations in the full and truncated systems with parameters (6.25): ($a$) amplitudes of the nonlinear oscillators; ($b$) amplitudes of the linear oscillators; dashed lines represent quasi-stationary amplitudes.

Although both oscillators possess gradually increasing amplitudes, the nature of the emerged processes is different. Figure 42(a) depicts AR in the nonlinear oscillator with the trajectory corresponding to the LPT of large oscillations. At the same time, Fig. 42(b) demonstrates forced oscillations, in which the response of the nonlinear oscillator acts as an external force.

### 6.2.2. Energy localization and transport in a system with a slowly-varying forcing frequency

In this section we briefly analyse energy transport in coupled oscillators with constant parameters and a slowly changing forcing frequency. The equations of motion are reduced the form similar to (6.17)

$$\frac{d^2u_1}{d\tau_0^2} + u_1 + 2\varepsilon\lambda_1(u_1 - u_0) = 0,$$

$$\frac{d^2u_0}{d\tau_0^2} + u + 2\varepsilon\lambda(u_0 - u_1) + 8\varepsilon\alpha u_0^3 = 2\varepsilon F\sin(\tau_0 + \theta(\tau)),$$

$$\frac{d\theta}{d\tau} = s + b\tau.$$

(6.26)

Transformations (6.18) together with the change of variables $\tau_1 = s\tau$, $\phi_j = \varphi_j e^{-i\theta}$, yield the following dimensionless equations for the slow complex amplitudes $\phi_j(\tau_1)$:

$$\frac{d\phi_1}{d\tau_1} + i\zeta_0(\tau_1)\phi_1 - i\mu_1(\phi_1 - \phi_0) = 0, \phi_1(0) = 0,$$

$$\frac{d\phi_0}{d\tau_1} - \iota\mu_0(\phi_0 - \phi_1) + i(\zeta_0(\tau) - |\phi_0|^2)\phi = -if, \phi_0(0) = 0. \tag{6.27}$$

where $\zeta_0(\tau_1) = 1 + \beta\tau_1$. Note that Eqs. (6.27) are similar to (6.19) but the time-dependent coefficient $\zeta_0(\tau_1)$ is now involved in both equations. Details of transformations are provided in [79, 80].

As in the previous section, the sought amplitude of the nonlinear oscillator is presented as $\phi_0 = \bar{\phi}_0 + \tilde{\phi}_0$, where $\bar{\phi}_0(\tau_1)$ and $\tilde{\phi}_0(\tau_1)$ denote the quasi-stationary amplitude and small fast fluctuations near $\bar{\phi}_0(\tau_1)$, respectively. It is easy to show that the state $\bar{\phi}_0(\tau_1)$ and the backbone curve $\bar{a}_0 = |\bar{\phi}_0|$ are defined by relations (6.13).

After calculating the nonlinear response $\phi_0(\tau_1)$, the solution $\phi_1(\tau_1)$ can be directly found from the first equation (6.27). Ignoring the effect of small fast fluctuations, we obtain

$$\phi_1(\tau_1) \approx -i\frac{\mu_1}{2\beta}e^{-iS(\tau_1)/2\beta}K(\tau_1), S(\tau_1) = (1 + \beta\tau_1)^2,$$

$$K(\tau_1) = K_0(\tau_1) - K_0(1), \ K_0(\tau_1) = \int_0^{S(\tau_1)} e^{iz/2\beta}z^{-1/4}dz. \tag{6.28}$$

Although the expression for $K_0(\tau_1)$ cannot be found in closed form, the limiting value $K_0(\infty)$ can be explicitly evaluated, and equals $K_0(\infty) = (2\beta)^{4/3}\Gamma(\tfrac{3}{4})e^{3i\pi/8}$, where $\Gamma$ is the gamma function [56]. Hence, $a_1(\tau_1) \to a_{1\infty} = \mu_1(2\beta)^{1/3}\Gamma(\tfrac{3}{4})$ as $\tau_1 \to \infty$. This result indicates that AR in the nonlinear actuator is unable to generate oscillations with permanently growing energy in the attached oscillator but the transferred energy suffices to sustain linear oscillations with bounded amplitude.

Figure 43 depicts the amplitudes of oscillations $a_0(\tau_1) = |\phi_0(\tau_1)|$ and $a_1(\tau_1) = |\phi_1(\tau_1)|$ calculated from Eqs. (6.27). Figure 43(a) shows that the amplitude of nonlinear oscillations is very close to its analogue presented in Fig. 42(a) but the amplitude of linear oscillations in Fig. 43(b) drastically differs from the amplitude of oscillations with growing energy in Fig. 42(b). The shape of the amplitude $a_1(\tau_1)$ is similar to the resonance curve with a noticeable peak in the initial stage of motion, but then motion turns into small oscillations with near the limiting amplitude close to the theoretically predicted value $a_{1\infty} \approx 0.1$.

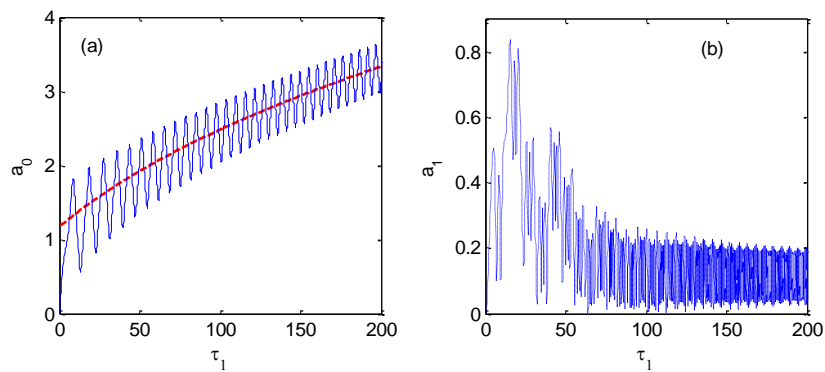

**Fig. 43.** Amplitudes of oscillations: (a) nonlinear oscillations; dashed line denotes the backbone curve $\bar{a}_0$; (b) linear oscillations.

### 6.2.3. Energy transfer in a system with slowly-varying natural and excitation frequencies

A key conclusion from the obtained results is that in the system of with slowly changing excitation frequency the energy transferred from the nonlinear oscillator is insufficient to produce oscillations with growing energy in the attached linear oscillator. The different dynamical behavior can be considered as a consequence of different resonance properties of the systems. In the system with a constant frequency of external forcing but slowly-varying parameters of the nonlinear actuator both oscillators are captured in resonance. If the forcing frequency slowly increases and the parameters of the system remain constant, resonance in the nonlinear oscillator is still sustained by increasing amplitude, while the frequency of the linear oscillator falls into the domain beyond the resonance. This implies that the slow change of the linear stiffness of the actuator can be considered as a parameter controlling the occurrence of high-energy oscillations with growing amplitude in the linear oscillator.

As an illustrating example, we consider a system with slowly changing linear stiffness of the actuator. The system dynamics is described by the equations

$$\frac{d^2 u_1}{d\tau_0^2} + u_1 + 2\varepsilon\lambda_1(u_1 - u_0) = 0,$$

$$\frac{d^2 u_0}{d\tau_0^2} + (1 - 2\varepsilon\xi(\tau))u_0 + 2\varepsilon\lambda_0(u_0 - u_1) + 8\varepsilon\alpha u_0^3 = 2\varepsilon F\sin(\tau_0 + \theta(\tau)), \qquad (6.29)$$

$$\frac{d\theta}{d\tau} = \zeta(\tau),$$

where $\zeta(\tau) = s + b_1\tau$, $\xi(\tau) = b_3\tau^3$, $\tau = \varepsilon\tau_0$; all other coefficients are defined by relations (6.16). We recall that if $\xi(\tau) = 0$, then AR may appear only the nonlinear oscillator. We will show that slow changes in both natural and excitation frequencies of the actuator may sustain growing oscillations in the coupled linear oscillator.

Transformations (6.18), together with the change of variables $\tau_1 = s\tau$, $\phi_j = \varphi_j e^{-i\theta}$, result in the following equations for the slow complex amplitudes $\phi_j(\tau_1)$:

$$\frac{d\phi_1}{d\tau_1} + i\zeta_1(\tau_1)\phi_1 - i\mu_1(\phi_1 - \phi_0) = 0, \phi_1(0) = 0,$$

$$\frac{d\phi_0}{d\tau_1} + i[\zeta_0(\tau_1) - |\phi_0|^2]\phi_0 - i\mu_0(\phi_0 - \phi_1) = -if, \phi_0(0) = 0, \qquad (6.30)$$

where

$$\tau_1 = s\tau, \zeta_0(\tau) = \zeta_1(\tau) + \xi_1(\tau), \zeta_1(\tau_1) = 1 + \beta_1\tau_1, \xi_1(\tau_1) = \beta_3\tau_1^3,$$

$$\beta_1 = b_1/s^2, \beta_3 = b_3/s^4;$$

other parameters are defined in (6.19) and (6.27). Note that details of converting the full system (6.29) into the equations for slow complex amplitudes (6.30) can be found in [79].

It is easy to obtain from (6.30) that the quasi-steady states $\bar{\phi}_j$ can be evaluated as

$$|\bar{\phi}_0(\tau_1)| \approx [\zeta_1(\tau_1) + \xi_1(\tau_1)]^{1/2} \sim O(\tau^{3/2}),$$

$$|\bar{\phi}_1(\tau_1)| \approx \mu_1[\zeta_1(\tau_1) + \xi_1(\tau_1)]^{1/2}/(\zeta_1(\tau_1) - \mu_1) \sim O(\tau^{1/2}). \qquad (6.31)$$

Expressions (6.31) imply simultaneous (but not equal) growth of backbone curves and thus suggest growing energy of oscillations of both oscillators provided linear stiffness of the actuator varies with rate $\xi_1(\tau)$ exceeding the rate $\zeta_1(\tau_1)$ of the change of the forcing frequency.

We illustrate these conclusions by the results of numerical simulations for the system with parameter $\beta_1 = 10^{-3}$, $\beta_3 = 10^{-5}$, $f = 0.34$, $\mu_0 = 0.01$, $\mu_1 = 0.15$ (Fig. 44).

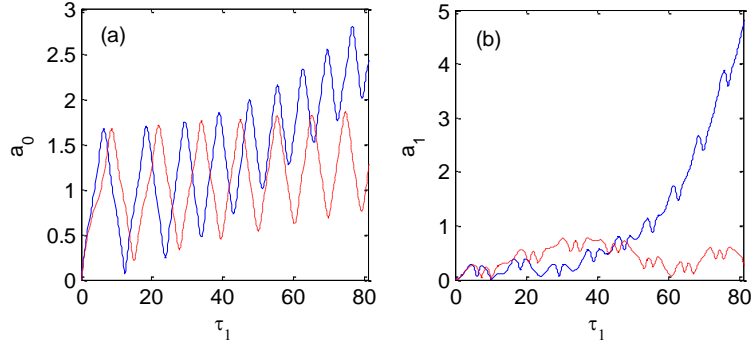

**Fig**. **44**. Amplitudes of oscillations of the actuator (*a*) and the linear oscillator (*b*); solid lines corresponds to system (6.30); dashed lines correspond to the time-independent actuator ($\beta_3 = 0$).

It is seen that an additional slow change of the actuator stiffness entails an increase of the nonlinear response, thus enhancing energy transfer and making it sufficient to sustain growing oscillations of the linear oscillator. A thorough theoretical analysis of this model is omitted but the obtained results motivate further analytical investigation of feasible energy transfer in the multi-dimensional arrays (see, e.g. [79]).

### 6.3. Classical and quantum tunneling in coupled linear oscillators

In this part of the report we develop an analytical framework to investigate irreversible energy transfer in a system of two *unforced* weakly-coupled oscillators with slowly time-varying frequencies. In the system under consideration one of the oscillators is initially excited by an initial impulse, while the second one is initially at rest. As shown in the previous sections, these initial conditions provide motion along the LPT with a maximum possible energy transfer from the excited oscillator to the second one.

The analysis developed in this section, gives special attention to an analogy between energy transfer in a system of classical oscillators and the quantum Landau-Zener tunnelling. Due to its generality, the Landau-Zener scenario has been applied to numerous problems in various contexts, such as, e.g., laser physics [162], semiconductor super-lattices [156], tunnelling of optical [192] or acoustic waves [92, 164] and quantum information processing [163], to name just a few examples. Although a passage between two energy levels is an intrinsic feature of all above-mentioned processes, manifestation of a direct analogy between energy transfer in a classical system with time-varying parameters and quantum Landau-Zener tunnelling is a recent development [115]. It was shown that the equations of the adiabatic passage through resonance in a system of two weakly coupled linear oscillators with a slowly-varying frequency detuning are identical to the equations of the Landau-Zener tunnelling problem. This conclusion may be treated as an extension of a previously found analogy between *adiabatic* quantum tunnelling and energy exchange in weakly coupled oscillators with constant parameters demonstrated in Section III.

It is well known that an analytical solution for the problem of transient tunnelling is prohibitively difficult even in the linear case [210]. The purpose of this section is to derive an explicit asymptotic approximation, which depicts irreversible energy transfer on the LPT. In addition, the obtained asymptotic solution provides a simple and accurate prediction of tunnelling over a finite time-interval, unlike a common approach considering fixed initial conditions at $t \to -\infty$ and a stationary solution at $t \to \infty$.

Note that an approximate solution of the quasi-linear problem can be found with the help of an iteration procedure, wherein the linear solution is chosen as an initial approximation. We skip the discussion of this problem, because it does not provide qualitatively new results. Details of the asymptotic analysis can be found in the authors' work [81].

*6.3.1. Main equations*

We study energy transport in a system of two weakly coupled linear oscillators of equal mass $m$. The first oscillator with linear stiffness $c$ is excited by an initial impulse $V$; the coupled oscillator with time-dependent linear stiffness $C_2(t) = c - (k_1 - k_2 t)$, $k_{1,2} > 0$ is initially at rest; the oscillators are connected by a linear coupling of stiffness $c_{12}$. The absolute displacements and velocities of the oscillators are denoted by $u_i$ and $V_i = du_i/dt$, $i = 1, 2$. We will demonstrate that the second oscillator with time-dependent frequency acts as an energy sink and ensures a visible reduction of oscillations of the excited mass.

Assuming weak coupling, we define the small parameter of the problem $2\varepsilon = c_{12}/c << 1$. Then, considering rescaled dimensionless parameters

$$c_{12}/c = 2\varepsilon\lambda, \ \lambda = 1; \ k_1/c = 2\varepsilon\sigma, \ k_2/(c\omega) = \varepsilon^2\beta, \ \omega^2 = c/m$$

and introducing the dimensionless time-scales $\tau_0 = \omega t$, $\tau_1 = \varepsilon\tau_0$, we obtain the following dimensionless equations of motion

$$\frac{d^2 u_1}{d\tau_0^2} + u_1 + 2\varepsilon\lambda(u_1 - u_2) = 0, \tag{6.32}$$

$$\frac{d^2 u_2}{d\tau_0^2} + (1 - 2\varepsilon\zeta(\tau_1))u_2 + 2\varepsilon\lambda(u_2 - u_1) = 0,$$

where $\zeta(\tau_1) = \sigma - \beta\tau_1$; initial conditions at $\tau_0 = 0$ are given by $u_1 = u_2 = 0$; $v_1 = V/\omega = V_0$, $v_2 = 0$, $v_i = du_i/d\tau_0$. It is important to note that system (6.32) may be considered as resonant only in a finite time interval wherein $|\zeta(\tau_1)| \sim 1$. In this case, the value of $\varepsilon\zeta(\tau_1)$ is small and instant frequencies of the overall system remain close.

In analogy to the previous sections, solutions of system (6.32) are sought with the help of the multiple-scale techniques. First, we define new complex-valued amplitudes $\eta$ and $\varphi$ by formulas $\eta(\tau_0, \varepsilon) = (v_1 + iu_1)e^{-i\omega_\varepsilon\tau_0}$, $\varphi(\tau_0, \varepsilon) = (v_2 + iu_2)e^{-i\omega_\varepsilon\tau_0}$, where $\omega_\varepsilon = 1 + \varepsilon\lambda + O(\varepsilon^2)$. In the next step, the functions $\eta$ and $\varphi$ are constructed in the form of the asymptotic series

$$\eta(\tau_0, \varepsilon) = \eta_0(\tau_1) + \varepsilon\eta_1(\tau_0, \tau_1) + \varepsilon^2 \ldots, \quad \varphi(\tau_0, \varepsilon) = \varphi_0(\tau_1) + \varepsilon\varphi_1(\tau_0, \tau_1) + \varepsilon^2 \ldots$$

Reproducing standard arguments, we derive the following first-order equations for the slow complex amplitudes $\eta_0$ and $\varphi_0$

$$\frac{d\eta_0}{d\tau_1} = -i\lambda\varphi_0(\tau_1), \quad \eta_0(0) = V_0, \tag{6.33}$$

$$\frac{d\varphi_0}{d\tau_1} - i\Omega(\tau_1)\varphi_0 = -i\lambda_2\eta_0.$$

where $\Omega(\tau_1) = -\sigma + \beta\tau_1$ (see [78, 85, 115] for more details). Once the envelopes $\eta_0(\tau_1)$, $\varphi_0(\tau_1)$ are found, the leading-order approximations to the solutions $u_1$, $u_2$ are given by

$$u_{10}(\tau_0, \varepsilon) = |\eta_0(\tau_1)|\sin(\omega_\varepsilon\tau_0 + \delta(\tau_1)), \quad \delta(\tau_1) = \arg(\eta_0(\tau_1)), \tag{6.34}$$

$$u_{20}(\tau_0, \tau_1) = |\varphi_0(\tau_1)|\sin(\omega_\varepsilon\tau_0 + \alpha(\tau_1)), \quad \alpha(\tau_1) = \arg\varphi_0(\tau_1).$$

Partial energy of the oscillators on the LPT is approximately calculated as

$$e_{10}(\tau_1) = \tfrac{1}{2}(<u_{10}^2> + <v_{10}^2>) = \tfrac{1}{2}|\eta_0(\tau_1)|^2. \tag{6.35}$$

$$e_{20}(\tau_1) = \tfrac{1}{2}(<u_{20}^2> + <v_{20}^2>) = \tfrac{1}{2}|\varphi_0(\tau_1)|^2,$$

where $<\cdot>$ denotes the averaging over the "fast" period $T = 2\pi/\omega_\varepsilon$. Note that expressions (6.35) ignore the residual terms of $O(\varepsilon)$ associated with potential energy of weak coupling and small slow change of linear stiffness.

We obtain from (6.33), (6.35) that for small $\tau_1$ the following approximations are valid:

$$\eta_0(\tau_1) = V_0(1 - \tfrac{1}{2}\lambda^2\tau_1^2), \quad e_{10}(\tau_1) = \tfrac{1}{2}V_0^2(1 - \lambda^2\tau_1^2). \tag{6.36}$$

$$\varphi_0(\tau_1) \approx -i\lambda V_0\tau_1, \quad e_{20}(\tau_1) \approx \tfrac{1}{2}(\lambda V_0\tau_1)^2.$$

It now follows from (6.36) that in the initial time-interval the energy of the excited oscillator is decreasing while the energy of the second oscillator is increasing. It follows from (6.36) that an instant $\tau_1^*$ at which $e_{10}(\tau_1^*) = e_{20}(\tau_1^*)$ is defined by the equality $\tau_1^* = 1/(\sqrt{2}\lambda)$. It is obvious that $\tau_1^*$ decreases with an increase of coupling. This conclusion agrees with experimental results [78,115].

Finally, we note that Eqs. (6.33) are equivalent to the second-order differential equation

$$\frac{d^2\varphi_0}{d\tau_1^2} - i\Omega(\tau_1)\frac{d\varphi_0}{d\tau_1} + (\lambda^2 - i\beta)\varphi_0 = 0, \qquad (6.37)$$

with initial conditions $\varphi_0 = 0$, $d\varphi_0/d\tau_1 = -i\lambda V_0$ at $\tau_1 = 0$. The equivalence of Eq. (6.37) to the equation of the Landau-Zener transient tunnelling problem [90, 210] is discussed in [85].

### 6.3.2. Approximate analysis of energy transfer in the linear system

It is well known that an analytical solution for the Landau-Zener problem of transient tunnelling dynamics is prohibitively complicated even in the linear case [210]. However, It follows from (6.37) that asymptotic solutions can be greatly simplified if $\beta_2 >> \lambda^2$ [85]. Under these assumptions, the term $\lambda^2$ can be ignored in Eq. (6.37). For convenience, we denote $\beta = 2\beta_1^2$. We thus have an approximate solution $\tilde{\varphi}_0(\tau_1)$

$$\tilde{\varphi}_0(\tau_1) = -i\lambda V_0 I(\tau_1), I(\tau_1) = \frac{1}{\beta_1} e^{iB(\tau_1)} F(\tau_1,\theta_0) e^{i\theta_0^2}, \qquad (6.38)$$

where $B(s) = (\beta_1 s + \theta_0)^2$, $\theta_0 = -\sigma/2\beta_1$, and

$$F(\tau_1,\theta_0) = \int_{\theta_0}^{\beta_1\tau_1+\theta_0} e^{-ih^2} dh = [C(\beta_1\tau_1 + \theta_0) - C(\theta_0)] - i[S(\beta_1\tau_1 + \theta_0) - S(\theta_0)],$$

$C(x)$ and $S(x)$ are the cos- and sin-Fresnel integrals. Once the envelope $\tilde{\varphi}_0(\tau_1)$ is found, the envelope $\tilde{\eta}_0(\tau_1)$ is expressed as

$$\tilde{\eta}_0(\tau_1) = V_0 - i\lambda \int_0^{\tau_1} \tilde{\varphi}_0(s) ds. \qquad (6.39)$$

An approximate solution of system (6.32) is now given by

$$\tilde{u}_1(\tau_0,\varepsilon) = |\tilde{\eta}_0(\tau_1)| \sin(\omega_\varepsilon\tau_0 + \tilde{\alpha}(\tau_1)), \tilde{\alpha}(\tau_1) = \arg(\tilde{\eta}_0(\tau_1)), \qquad (6.40)$$

$$\tilde{u}_2(\tau_0,\varepsilon) = |\tilde{\varphi}_0(\tau_1)| \sin(\omega_\varepsilon\tau_0 + \tilde{\alpha}(\tau_1)), \tilde{\alpha}(\tau_1) = \arg\tilde{\varphi}_0(\tau_1).$$

We evaluate the amplitudes of oscillations in two limiting cases:

1. If $\beta_1 \tau_1 << \sqrt{2}$ , then it follows from (6.38) that

$$|\varphi_0(\tau_1)| \approx \lambda V_0 \tau_1 \qquad (6.41)$$

2. If $\beta_1 \tau_1 >> \sqrt{2}$ , then it follows from the properties of the Fresnel integrals [56] that

$$\tilde{\varphi}_0(\tau_1) \rightarrow \bar{\varphi_0} = -i \frac{\lambda V_0}{\beta_1} \{ [\sqrt{\tfrac{\pi}{8}} - C(\theta_0)] - i[\sqrt{\tfrac{\pi}{8}} - S(\theta_0)] \}, \qquad (6.42)$$

$$|\bar{\varphi_0}| = \frac{\lambda V_0}{\beta_1} \{ [\sqrt{\tfrac{\pi}{8}} - C(\theta_0)]^2 + [\sqrt{\tfrac{\pi}{8}} - S(\theta_0)]^2 \}^{1/2}, \text{ as } \tau \rightarrow \infty.$$

Expressions (6.42) imply that at large times the second oscillator (the energy sink) exhibits quasi-stationary oscillations with constant amplitude $|\bar{\varphi_0}|$ and energy $\bar{e}_{20} = \frac{1}{2}|\bar{\varphi_0}|^2$. This illustrates almost irreversible energy transfer from the initially excited oscillator to the sink being initially at rest.

We now compare exact (numerical) solutions of the initial systems (6.32) with approximations (6.40). The parameters of the system are given by $\varepsilon = 0.136$; $V_0 = 1$; $\sigma = 0.83$; $\beta = 2.7$; $\lambda = 1$. In Fig. 45 one can observe close proximity of the exact solutions and their approximations for each oscillator separately (Figs. 45(a, b)). Almost irreversible energy transfer is manifested as a decrease of energy of the first oscillator with a simultaneously growing energy of the second oscillator is shown in Fig. 45(c). One can conclude that the first half-period of oscillations is similar to motion along the LPT of a corresponding time-invariant system but at later times the system dynamics changes and the amplitude of oscillations tend to certain limiting values.

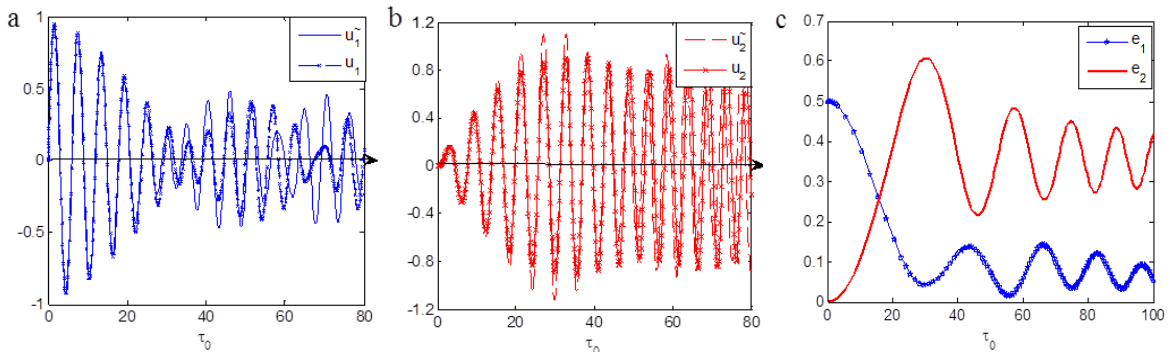

**Fig. 45.** Exact (numerical) and approximate solutions $u_1$ and $\tilde{u}_1$ (a); $u_2$ and $\tilde{u}_2$ (b); energy of the oscillator $e_1$ and the sink $e_2$ (c).

## 6.4. Moderately and strongly nonlinear adiabatic tunneling

In this section we examine irreversible energy transfer along the LPTs for moderately and strongly nonlinear regimes in a system with slowly time-varying parameters. We demonstrate the equivalence of the equations for the slow passage through resonance in the classical system and the equations of nonlinear LZ tunneling. It is noted here that the well-known LZ tunneling is typical for moderately nonlinear regimes, while the system with strongly nonlinear behavior can exhibit both the transition from the energy localization to intense energy exchange (with a large inductive period) as well as the rapid passage through the separatrix. It is important to underline that the revealed mathematical equivalence suggests a unified approach to the study of such physically different processes as energy transfer in classical oscillatory systems under slow driving and nonlinear quantum LZ tunneling.

### 6.4.1. Moderately nonlinear regimes

In this part of the paper, we consider adiabatic tunneling in a system of two coupled oscillators similar to (3.3), namely,

$$\frac{d^2 u_1}{d\tau_0^2} + u_1 + 2\varepsilon(u_1 - u_2) + 8\varepsilon\alpha u_1^3 = 0,$$

$$\frac{d^2 u_2}{d\tau_0^2} + (1 + 2\varepsilon g(\tau_2))u_2 + 2\varepsilon(u_2 - u_1) + 8\varepsilon\alpha u_2^3 = 0,$$

(6.43)

where $g(\tau_2) = g_0 + g_1\tau_2$, $\tau_2 = \varepsilon^2\tau_0$, $\tau_1 = \varepsilon\tau_0$. It is well-known that energy transfer between coupled oscillators with adiabatically changed parameters is divided into two stages, with each characterizing by adiabatic invariance but separated by an abrupt jump at a moment of tunneling. Breaking of adiabaticity under slow driving has been intensively studied over last decades using various approximations (see, e.g., [63-65, 71,162]). We investigate slow transient processes before and after tunneling. The asymptotic analysis accounts for the fact that in the first interval of motion most part of energy is localized on the excited oscillator while the residual energy of the coupled oscillator is small enough. After tunneling, localization of energy takes place on the coupled oscillator but energy of the initially excited oscillator becomes small. An introduction of the small parameter characterizing a relative energy level allows an explicit asymptotic solution.

As in Section 3.1.1, we introduce the change of variables $v_j + iu_j = Y_j e^{i\tau_0}$, $j = 0,1$, such that $Y_j = \varphi_j^{(0)} + \varepsilon \varphi_j^{(0)} + O(\varepsilon^2)$, with the leading order components $\varphi_1^{(0)} = ae^{i\tau_1}$, $\varphi_2^{(0)} = be^{i\tau_1}$. The equations for the slow complex envelopes $a$ and $b$ are given by

$$\frac{da}{d\tau_1} + ib - 3i\alpha|a|^2 a = 0, \tag{6.44}$$

$$\frac{db}{d\tau_1} + ia - 3i\alpha|b|^2 b - 2ig(\tau_2)b = 0,$$

with initial conditions $a(0) = 1$, $b(0) = 0$. We recall that the chosen initial conditions correspond to the initial unit impulse applied to the first oscillators (cf. (3.4)). It is easy to prove that the conservation law $|a|^2 + |b|^2 = 1$ holds true for system (6.44) despite its non-stationarity. Also, we note that Eqs. (6.44) are identical to the nonlinear analog of the LZ equations of quantum tunneling (see, e.g., [94,191,209,213] for further details).

The change of variables $a = \cos\theta e^{i\delta_1}$, $b = \sin\theta e^{i\delta_2}$, $\Delta = \delta_1 - \delta_1$ reduces (6.44) to the form similar to (3.5)

$$\frac{d\theta}{d\tau_1} = \sin\Delta, \tag{6.45}$$

$$\sin 2\theta \frac{d\Delta}{d\tau_1} = 2(\cos\Delta + 2k\sin 2\theta)\cos 2\theta - 2g(\tau_2)\sin 2\theta,$$

with initial conditions $\theta(0) = 0$, $\Delta(0) = \pi/2$. Figure 47 depicts the plots of the energy $2e_1 = |a|^2$, $2e_2 = |b|^2$ in the interval $0 \le \tau_1 \le 1000$ for system (6.45) with parameters

$$k = 0.65, \ g_0 = -0.5, \ \varepsilon g_1 = 0.001.$$

Figures 46, 47 depict energy of the coupled oscillators during adiabatic tunneling. Figure 46 shows that the dynamics of each oscillator is close to motion along the LPT of the time invariant system in the initial interval of motion wherein the change of detuning is negligible.

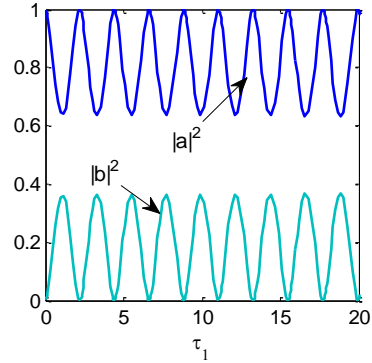

**Fig**. **46**. Plots of $|a|^2$ and $|b|^2$ in the initial interval of motion correspond to motion along the unperturbed LPT with maximum energy exchange between the oscillators.

At later times the change of energy and the occurrence of tunneling at $T^* \approx 585$ become evident (Fig. 47). The detuning value $g(T^*) = g_0 + \varepsilon g_1 T^* = 0.085$ is close to critical detuning $g^* \approx 0.083$ for a similar time-invariant system with $g_1 = 0$ (Section 3.1). Figure 47(c) represents the phase portrait in the plane ($\theta$, $V = \sin\Delta$), which also demonstrates two stage of motion: oscillations along the LPT of time-invariant systems in the initial interval turn into relatively small oscillations converging to a slowly moving center with a final jump to another energy level.

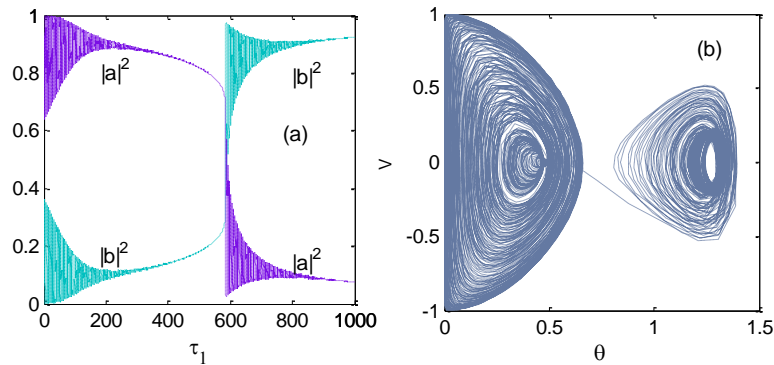

**Fig. 47**. (a) - plots of $|a|^2$ and $|b|^2$ with instant jumps between the energy levels; (b) - phase portrait of system (6.45).

Now we briefly analyze the dynamics in the interval $S_1$: $0 \leq \tau_1 < T^*$ before tunneling. To this end, we employ the change of variables

$$\Psi = |b|e^{i\Delta} = |\sin\theta|e^{i\Delta} \qquad (6.46)$$

which reduces (6.45) to the following complex-valued equation:

$$\frac{d\Psi}{d\tau_1} = 4i\left[\frac{1 - \frac{1}{2}(\Psi^2 + 2|\Psi|^2)}{\sqrt{1 - |\Psi|^2}} + 2\omega(\tau_2)\Psi - 8k\Psi|\Psi|^2\right], \; \Psi(0) = 0, \qquad (6.47)$$

where $\omega(\tau_2) = 2k - g(\tau_2) = \omega_0 - g_1\tau_2$, $\omega_0 = 2k - g_0$. It is important to note that the frequency $\omega(\tau_2)$ directly depends on the coefficient $k$, thereby reflecting the effect of nonlinearity even in the linear approximation of Eq. (6.47).

Since $|b| = |\Psi| << 1$ in $S_1$, we introduce the rescaled variable $\psi = \varepsilon^{-1/2}\Psi$ and the new time scale $s = \varepsilon^{-1/2}\tau_1$, as well as rescaled coefficients $8k = \varepsilon^{-1/2}\kappa$, $2\omega_0 = \varepsilon^{-1/2}w_0$, where $\kappa$ and $w_0$ are of $O(1)$. Finally, we denote $\varepsilon^{3/2}g_1 = \beta_1^2/4$ (keeping in mind that $\beta_1 << 1$). When we substitute the rescaled variables and coefficients into (6.47) and ignore the terms of orders higher than $\varepsilon$, we obtain the equation

$$\frac{d\psi}{ds} = 4i[w(s)\psi + 1 - \frac{1}{2}\varepsilon(\psi^2 + |\psi|^2) - \kappa\varepsilon\psi|\psi|^2], \; \psi(0) = 0, \qquad (6.48)$$

where $w(s) = w_0 - \beta_1^2 s/2$. Thus we get a quasi-linear equation, and the earlier developed iteration procedure can be employed to construct an approximate solution. The initial iteration $\psi_0(s)$ is chosen as a solution of the linear equation

$$\frac{d\psi_0}{ds} = 4i[w(s)\psi_0 + 1], \; \psi_0(0) = 0, \qquad (6.49)$$

$$\psi_0(s) = 4i\beta^{-1}e^{i(\phi_0(s) - \alpha^2)}\Phi_0(s),$$

where $\phi_0(s) = w_0 s - (\beta_1 s)^2$, $\alpha = w_0/\beta_1$, and

$$\Phi_0(s) = \int_{-\alpha}^{\beta_1 s - \alpha} e^{ih^2}\,dh = [C(\beta_1 s - \alpha) + C(\alpha)] + i[S(\beta_1 s - \alpha) + S(\alpha)],$$

with $C(h)$ and $S(h)$ being the cos- and sin-Fresnel integrals, respectively. Once $\psi_0$ is found, the main approximation to the function $|b|$ is calculated as $|b_0| = \varepsilon^{1/2}|\psi_0|$. Note that the parameter $w_0$ depends on the coefficient of nonlinearity $k$, and thus, the behavior of the solution $\psi_0(s)$ is conditioned by nonlinearity.

In the next step, the first iteration $\psi_1$ is found from the linearized equation

$$\frac{d\psi_1}{ds} = 4i\{[w(s) - \varepsilon\varkappa|\psi_0(s)|^2]\psi_1 - \varepsilon(\tfrac{1}{2}\psi_0^2(s) + |\psi_0(s)|^2) + 1\},\ \psi_1(0) = 0, \qquad (6.50)$$

$$\psi_1(s) = \psi_0(s) + 4i\,e^{i\phi_1(s)}\,\Phi_1(s),$$

where

$$\phi_1(s) = \phi_0(s) - \varepsilon\varkappa\int_0^s |\psi_0(z)|^2\,dz\,,\ \ \Phi_1(s) = \int_0^s e^{-i\phi_1(z)}dz - \varepsilon\int_0^s e^{-i\phi_1(z)}R(z)dz,$$

$$R(z) = \tfrac{1}{2}\psi_0^2(z) + |\psi_0(z)|^2.$$

Once the solution $\psi_1$ is derived, the first iteration to the function $|b|$ is calculated as $|b_1| = \varepsilon^{1/2}|\psi_1|$. The exact solution $|b(\tau_1)|^2$ and iterations $|b_0(\tau_1)|^2$ and $|b_1(\tau_1)|^2$ are presented in Fig. 48.

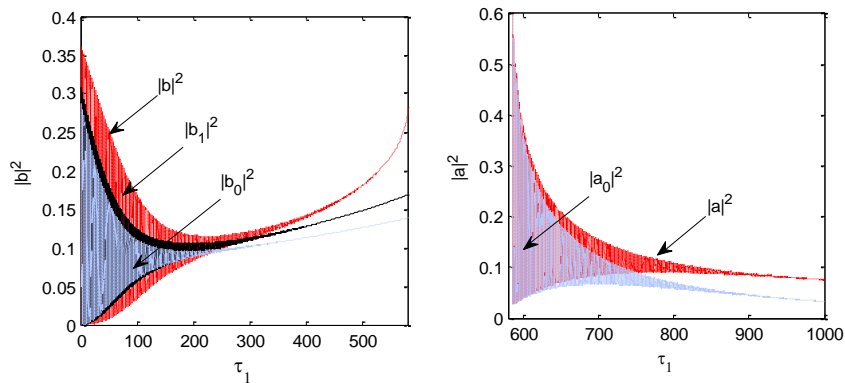

**Fig. 48**. Exact and approximate solutions before and after tunneling: plots of the exact solution $|b|^2$ and the iterations $|b_0|^2$ and $|b_1|^2$ in the interval $S_1$ (left); plots of the exact solution $|a|^2$ and the iterations $|a_0|^2$ in the interval $S_2$ (right)

It is seen that in the first half of the interval $S_1$ the maximal divergence between the exact solution and its linear approximation is less than 15%. The divergence in the second part of $S_1$ is due to the different behavior of exact and approximate solutions near the point of transition: while the exact solution demonstrates a sudden increase at an instant of tunneling, the linear approximation tends to a certain limit.

The dynamical behavior in the interval $S_2$: $\tau_1 > T$ after tunneling is studied in a similar way. Since $|a| << 1$, the variable analogous to (6.46) are introduced

$$Z = -|a|e^{i\Delta} = -|\cos\theta|e^{i\Delta} \qquad (6.51)$$

Transformations of the same sort that led to Eqs. (6.47) yield the following equation:

$$\frac{dZ}{d\tau_1} = 4i[2\omega_1(\tau_2)Z - 1 + \tfrac{1}{2}(Z^2 + |Z|^2) - 8\kappa Z|Z|^2], \ Z(T) = p_0, \qquad (6.52)$$

where $\omega_1(\tau_2) = 2k + g(\tau_2) = \omega_{10} + g_1\tau_2$, $\omega_{10} = 2k + g_0$. As shown in Section 3.1, initial condition for Eq. (6.52) at $\tau_1 = T$ should be defined from the condition of the coalescence of the stable and unstable states. This means that the quantity $\theta(T) = \theta_T$ (see (3.8)) determines the initial condition $|a(T)| = |\cos\theta_T| = p_0$.

Given $|Z| << 1$ in $S_2$, we introduce the following transformations of the variables and the parameters: $Z = \varepsilon^{1/2}z$, $\tau_1 = \varepsilon^{1/2}s$, $8k = \varepsilon^{-1/2}\kappa$, $2\omega_{10} = \varepsilon^{-1/2}w_{10}$. Then we denote $\varepsilon^{3/2}g_1 = \beta_1^2/4$, $\varepsilon^{-1/2}T = s_0$, $\varepsilon^{1/2}p_0 = p_{10}$. Substituting the rescaled quantities into (6.52) and ignoring the higher-order terms, we obtain the quasi-linear equation similar to (6.48)

$$\frac{dz}{ds} = 4i[w_1(s)z - 1 + \tfrac{1}{2}\varepsilon(z^2 + 2|z|^2) - \varepsilon\kappa z|z|^2], \ z(s_0) = p_{10}, \qquad (6.53)$$

with the adiabatically increasing parameter $w_1(s) = w_{10} + \beta_1^2 s/2$. The initial iteration $z_0(s)$ is given by the following expressions

$$\frac{dz_0}{ds} = 4i[w_1(s)z_0 - 1], \ z_0(s_0) = p_{10},$$
$$z_0(s) = (p_1 - 4i\Theta(s))e^{i\delta(s)}, \qquad (6.54)$$

where $\delta(s) = 4[w_{10}s + (\beta_1 s)^2]$, $\Theta(s) = e^{i\alpha_1^2}F_1(s)/2\beta_1$, and

$$F_1(s) = \int_{\alpha_1}^{2\beta_1 s + \alpha_1} e^{-ih^2}dh = [C(2\beta_1 s + \alpha_1) - C(\alpha_1)] - i[S(2\beta_1 s + \alpha_1) - S(\alpha_1)].$$

Applying inverse rescaling, we get the initial iteration $|a_0| = \varepsilon^{1/2}|z_0|$. As seen in Fig. 48(b), the plot of $|a_0(\tau_1)|^2$ is close to $|a(\tau_1)|^2$. Finally, calculating the stationary state $|\bar{a}_0| = \lim|a_0(\tau_1)|$ as $\tau_1 \to \infty$, we obtain

$$|\bar{a}_0| = \varepsilon^{1/2}\bar{z}_0 = |p_0 + 1/\omega_{10}|. \qquad (6.55)$$

The resulting non-zero value of $|\bar{a}_0|$ implies the existence of the residual energy depending on the initial condition $p_0$. Given $k = 0.65$, $g_0 = -0.5$, we obtain $|\bar{a}_0| = 0.445$. It follows from (6.55) that in the interval $S_2$: $\tau_1 > T$ the quantity $e_2$ tends to the limiting value $\bar{e}_2 = |\bar{b}_0|^2/2 = (1 - |\bar{a}_0|^2)/2 = 0.401$ as $\tau_1 \to \infty$. The quantity $\bar{e}_2$ characterizes the amount of energy transferred from the excited oscillator with initial energy $e_{10} = 1/2$ to the coupled oscillator being initially at rest.

### 6.4.2. Strongly nonlinear regimes

The analysis of strongly nonlinear dynamics of a time-invariant system (Section 3) demonstrates that the change of detuning $g$ may entail a transition from weak to strong energy exchange. In this section we show that this conclusion remains valid for adiabatic strongly nonlinear tunneling.

Figure 49 demonstrates the phase portrait and the transient evolution of the angle $\theta$ for system (6.45) with parameters $k = 0.9$, $g_0 = -0.25$; $\varepsilon g_1 = 0.001$.

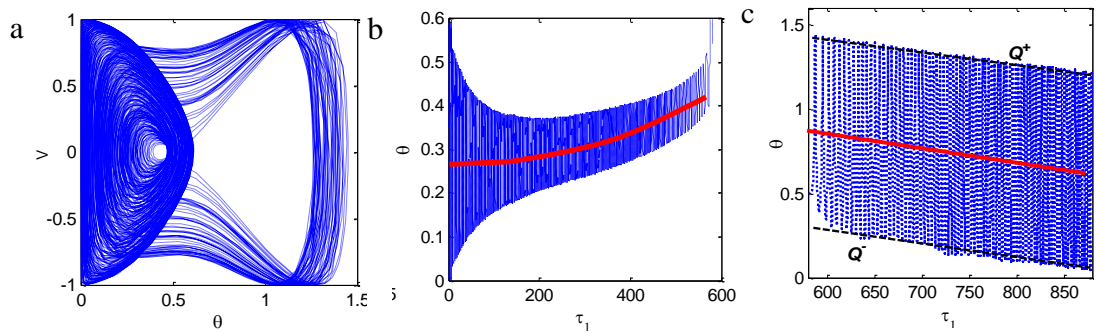

**Fig. 49.** Transient processes in the interval $0 \le \tau_1 \le 900$: (a) phase portrait of the system; (b) plot of $\theta(\tau_1)$ before tunneling; (c) plot of $\theta(\tau_1)$ after tunneling

The phenomenon of energy localization with relatively small oscillations near the slowly varying quasi-steady state is observed in the initial interval $S_1$: $0 \leq \tau_1 < T^*$; then, in the interval $S_2$: $\tau_1 > T^*$, energy localization changes to intense exchange. Note that tunneling occurs at $T^* = 580$ corresponding to $g = 0.33$. We recall this value of $g$ coincides with the result obtained for the stationary system in Section 3.1.

The adiabatic convergence to the transition point at the first stage of motion can be examined in the same way as in Section 6.4.1. We briefly analyze the dynamical behavior in the interval $S_2$, where the variations of $\theta(\tau_1)$ are large enough, and the asymptotic approach of Section 3 is inapplicable. We note that the local minima $\theta^-(\tau_1)$ and maxima $\theta^+(\tau_1)$ of $\theta(\tau_1)$ lie on the slowly-varying envelops $Q^-(\tau_2)$ and $Q^+(\tau_2)$, respectively (Fig. 49(c)). Assuming that initial values $|Q^-|$ and $|\pi/2 - Q^+|$ are small enough (Fig. 49(c)), the initial approximation can be chosen as $\theta_0^- = Q_0^- = 0$, $\theta_0^+ = Q_0^+ = \pi/2$. This yields the first approximation

$$Q_1^\pm(\tau_1) = \theta_0^\pm - \varepsilon g_1(\tau_1 - T). \tag{6.56}$$

(see [110] for more details). Figure 49(c) demonstrates a good agreement of approximations (6.77) with the precise (numerical) solution in the interval $S_2$.

# 7. Energy exchange, localization and transfer in finite oscillatory chains. Brief review

In this section we discuss an extension of the LPT concept to systems with more than two degrees of freedom. First, a chain of three identical weakly coupled Duffing oscillators (3DOF) is considered. This model helps us establish to which extent the LPT theory developed for the 2DOF systems (Section 3) is applicable, and to clarify the mechanism of the LPTs breaking. Possible extensions of the LPT concept to high-dimensional oscillator chains is discussed in the end of this section

## 7.1. Bifurcations of LPTs and routes to chaos in an anharmonic 3DOF chain

### 7.1.1. Equations of the slow flow model in a 3DOF chain

Assuming weak nonlinearity and weak linear coupling, we reduce the equations of motion to the following non-dimensional form:

$$\frac{d^2 x_k}{d\tau_0^2} + x_k + \varepsilon s x_k^3 = \varepsilon s \chi [(1 - \delta_{1k})(x_{k-1} - x_k) - (1 - \delta_{3k})(x_k - x_{k+1})], \; 1 \leq k \leq 3, \tag{7.1}$$

where $\delta_{ij}$ denotes the Kronecker delta, the small parameter $\varepsilon$ $(0 < \varepsilon << 1)$ defines weak nonlinearity, weak coupling stiffness is denoted by $\varepsilon s \chi$, $s = 3/8$ (the coefficient $s$ is introduced for convenience). We also note that $x_0 = x_4 = 0$. The selected initial conditions, defining the LPT of system (7.1), correspond to an impulse imposed to the first oscillator ($k = 1$) with the system being initially at rest, i.e.,

$$x_k(0) = 0, \; v_1 = 1, \; v_2 = v_3 = 0 \text{ at } t = 0; \; v_k = dx_k/dt. \tag{7.2}$$

As in the previous sections, an analysis of system (7.1) is performed with the help of the multiple time scales method. First, we introduce new complex variables $Y_k$, $Y_k^*$ by formulas

$$Y_k = (v_k + iu_k)e^{-i\tau_0}, Y_k^* = (v_k - iu_k)e^{i\tau_0}, \; k = 1,2,3, \tag{7.3}$$

As shown previously, the substitution of (7.3) into (7.1) results in the equations for $Y_k$, $Y_k^*$ with the right-hand sides $O(\varepsilon)$. In the next step, the complex amplitude $Y_k(\tau_0, \varepsilon)$ is constructed in the form of the multiple scale expansion

$$Y_k(\tau_0, \tau_1, \varepsilon) = \varphi_k(\tau_1) + \varepsilon \varphi_k^{(1)}(\tau_0, \tau_1) + \varepsilon^2 ...,$$

where $\tau_1 = \varepsilon \tau_0$ is the slow time scale. Applying the multiple scales methodology, we derive the following equation for the slow envelopes $\varphi_k$

$$\frac{d\varphi_k}{d\tau_1} = i |\varphi_k|^2 \varphi_k + i\chi[(1 - \delta_{1k})(\varphi_k - \varphi_{k-1}) + (1 - \delta_{3k})(\varphi_k - \varphi_{k+1})] \;,\; 1 \le k \le 3, \qquad (7.4)$$

$$\varphi_1(0) = i, \; \varphi_2(0) = \varphi_3(0) = 0,$$

It is important to underline that Eqs. (7.4) involve the only parameter $\chi$. System (7.4) possesses two integrals of motion

$$\sum_{k=1}^{3} |\varphi_k|^2 = 1,$$

$$H = \frac{i}{2} \sum_{k=1}^{3} |\varphi_k|^4 + i\chi[|\varphi_1 - \varphi_2|^2 + |\varphi_2 - \varphi_3|^2] = const. \qquad (7.5)$$

If we express the variables $\varphi_k$ through spherical coordinates in 3D

$$\varphi_1 = e^{i\delta_1} \cos\theta\cos\phi, \varphi_2 = e^{i\delta_2} \sin\theta, \varphi_3 = Ne^{i\delta_3} \sin\theta\sin\phi . \qquad (7.6)$$

and then insert (7.6) into (7.4), we obtain the following real-valued equations for the coordinates $\theta$, $\phi$, and the relative phases $\Delta_{12} = \delta_1 - \delta_2$, $\Delta_{23} = \delta_2 - \delta_3$:

$$\frac{d\theta}{d\tau_1} = \chi(\cos\phi\sin\Delta_{12} - \sin\phi\sin\Delta_{23}),$$

$$\frac{d\phi}{d\tau_1} = \chi\tan\theta\sin\phi\sin\Delta_{12} + \cos\phi\sin\Delta_{23},$$

$$\frac{d\Delta_{12}}{d\tau_1} = \cos^2\theta\cos^2\phi - \sin^2\theta - \chi[1 - \cot\theta\sin\phi\cos\Delta_{23}$$

$$+ \frac{2}{\sin 2\theta\cos\phi}(\sin^2\theta - \cos^2\theta\cos^2\phi)\cos\Delta_{12}], \qquad (7.7)$$

$$\frac{d\Delta_{23}}{d\tau_1} = \sin^2\theta - \cos^2\theta\sin^2\phi + \chi[1 - \cot\theta\cos\phi\cos\Delta_{12}$$

$$-\frac{2}{\sin 2\theta\sin\phi}(\cos^2\theta\sin^2\phi - \sin^2\theta)\cos\Delta_{23}].$$

The second integral of motion can also be expressed in terms of the new coordinates

$$H_{4D} = \frac{i}{2}[\sin^4\theta + \cos^4\theta(\cos^4\phi + \sin^4\phi)] - \tag{7.8}$$

$$-i\chi\sin 2\theta(\cos\phi\cos\Delta_{12} + \sin\phi\cos\Delta_{23}) - i\frac{\chi}{2}\cos 2\theta.$$

In contrast to its 2DOF counterpart, the slow-flow model (7.7) of the 3DOF system is non-integrable, and thus, they suggest chaotic response. As shown below, there exist the values of coupling $\chi$, for which regular regimes coexist with non-regular ones. We recall that regular stationary regimes are associated with NNMs, while regular highly non-stationary regimes correspond to strongly modulated resonant responses, i.e., LPTs.

### 7.1.2. Nonlinear Normal Modes

As in the 2DOF systems, NNMs in the 3DOF system represents stationary points in the phase space of the system in the slow time scale. The set of solutions beyond the stationary points correspond to non-stationary processes, among which one can distinguish both the LPTs with maximum possible energy exchange as well as chaotic motion.

We begin with the bifurcation analysis of NNMs. Figures 50 and 51 depict the slow amplitudes $|\varphi_k|$ for each oscillator ($1 \le k \le 3$), corresponding to the out-of-phase modes ($\Delta_{12} = \Delta_{23} = \pi$) and the mixed modes ($\Delta_{12} = 0$, $\Delta_{23} = \pi$), respectively. The in-phase modes ($\Delta_{12} = \Delta_{23} = 0$) are not presented in this work, as they corresponds to a rather trivial solution $|\varphi_k| = 1/\sqrt{3}$, $k = 1, 2, 3$.

Figures 50 and 51 depict NNMs with energy localization either on the first oscillator, or, alternatively, on both the first and second oscillators. It is important to underline that these NNMs are similar to NNMs in the above-considered 2DOF models. The stable and the unstable localized modes bifurcate through the saddle-center bifurcation, while an additional mode, which is localized on the central oscillator, is stable and persists for any value of coupling $\chi$.

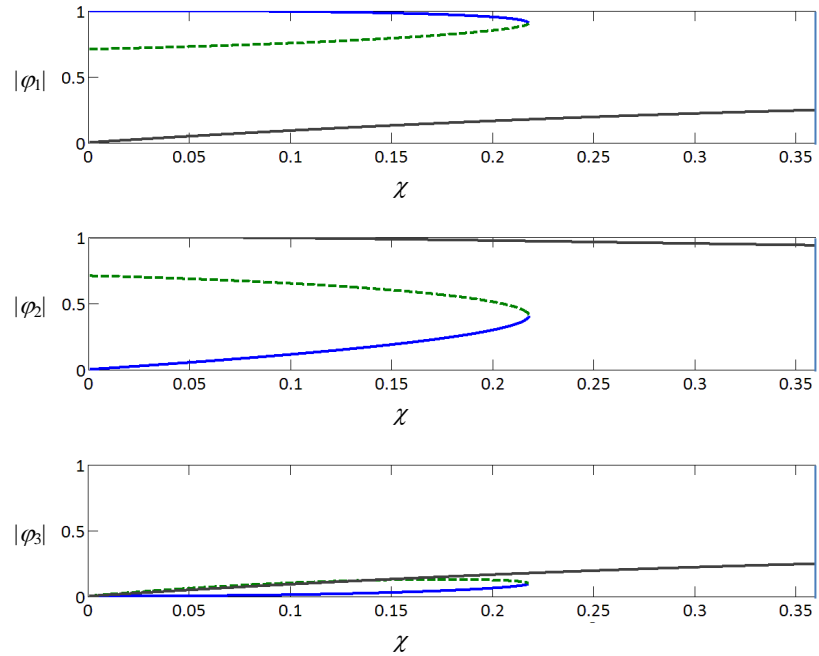

**Fig. 50.** Bifurcation diagram of NNMs ($\Delta_{12} = \Delta_{23} = \pi$). Each mode is indicated with different color. Unstable modes are denoted with dashed lines.

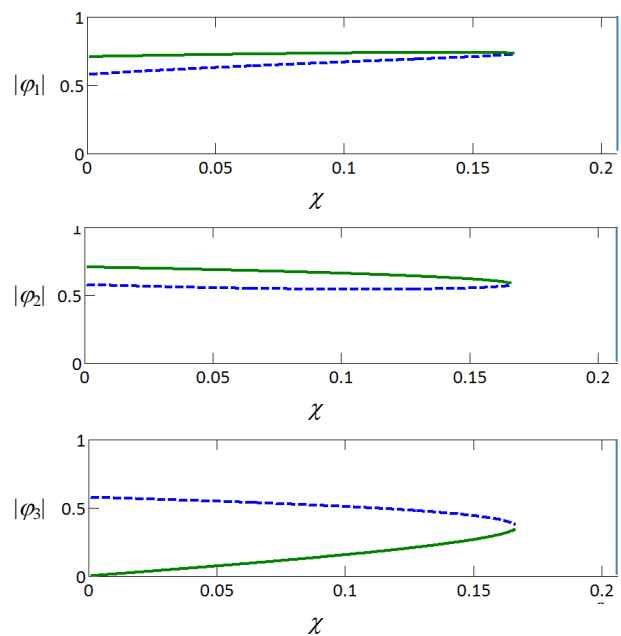

**Fig. 51.** Bifurcation diagram of NNMs ($\Delta_{12} = 0$, $\Delta_{23} = \pi$). Each mode is indicated with different color. Unstable modes are denoted with dashed lines.

*7.1.3. Emergence and bifurcations of LPTs in the chain of three coupled oscillators*

In the earlier work [82], the properties of LPTs in an $n$-dimensional array of linear oscillators with a strongly nonlinear energy sink were investigated. In this paragraph, we explore LPTs in the nonlinear system (7.1). We will show that LPTs in a multi-particle nonlinear chain can also play a dominant role in predicting the emergence and bifurcations of a special class of nonstationary regimes characterized by weak and strong beating between the first and the second oscillators but with insignificant energy transfer to the third oscillator. On the other hand, we will show that the failure of the above-mentioned regimes yields significant energy transport to the third oscillator, which results in chaotic motion with the equipartition of energy between all oscillators.

One can distinguish three types of motion in the system under consideration. The first type of motion is characterized by strong energy localization on the first oscillator, with moderate energy exchanges between the first oscillator and the rest of the chain. This type of the response is analogous to a weakly modulated regime in the system of two oscillators (Section 3). In analogy with the 2DOF system, this type of response with small energy pulsation can be attributed to the quasi-periodic LPT of the first type. The second type of nonstationary regimes is characterized by intense energy exchange between the first and second oscillators, with negligibly small energy transfer to the third oscillator. This type of response is analogous to the strongly modulated regime in the 2DOF system, and thus, it can be attributed to the quasi-periodic LPT of the second type. The results of numerical simulations demonstrating the peculiarities of the LPTs of both types will be given below.

The third type of motion discussed in the present section corresponds to the breakdown of the LPT of the second type and the emergence of irregular energy transport between all oscillators in the chain. It is obvious that this type of motion has no analogy in the 2DOF system. Note that for some particular choice of initial conditions (close to a simple initial impulse in (7.1)) one can also find a periodic LPT similar to that in the 2DOF system. The periodic LPTs will be illustrated below.

For understanding the phenomenon of consecutive transitions between different types of the LPTs as well as for predicting the formation of a fully delocalized response, we construct a reduced-order model based on the "master-slave" decomposition. Assuming that the third oscillator is weakly excited in comparison to the first and second oscillators and using the representation $\varphi_2 = e^{i\delta_2}\sin\theta$ (see (7.6)), we reduce system (7.4) to the form:

$$
\begin{aligned}
\frac{d\varphi_1}{d\tau_1} &= i\,|\varphi_1|^2\,\varphi_1 + i\chi(\varphi_1 - \varphi_2), \\
\frac{d\varphi_2}{d\tau_1} &= i\,|\varphi_2|^2\,\varphi_2 + i\chi(2\varphi_1 - \varphi_1), \\
\frac{d\varphi_3}{d\tau_1} &= i\chi\varphi_2 = i\chi\sin\theta e^{i\delta_2}.
\end{aligned}
\tag{7.9}
$$

The system of the first two equations in (7.9) possesses two integrals of motion

$$
\begin{aligned}
\sum_{k=1}^{2}|\varphi_k|^2 &= 1, \\
H_{6D} = \frac{i}{2}\sum_{k=1}^{2}|\varphi_k|^4 &+ i\chi[|\varphi_1 - \varphi_2|^2 + |\varphi_2|^2] = const.
\end{aligned}
\tag{7.10}
$$

A proper change of variables $\varphi_1 = e^{i\delta_1}\cos\theta, \varphi_2 = e^{i\delta_2}\sin\theta$ reduces the first two equations in (7.9) to a planar system with two angular variables $\theta$ and $\Delta = \delta_1 - \delta_2$

$$
\begin{aligned}
\frac{d\theta}{d\tau_1} &= \chi\sin\Delta, \\
\frac{d\Delta}{d\tau_1} &= \cos^2\theta - \sin^2\theta + \chi[2\cot 2\theta\cos\Delta - 1].
\end{aligned}
\tag{7.11}
$$

It is easy to prove that system (7.11) is Hamiltonian, with the Hamiltonian

$$
H = \tfrac{1}{2}(\cos^4\theta + \sin^4\theta) - \chi(\sin 2\theta\cos\Delta - \sin^2\Delta).
$$

Phase portraits of system (7.11) for different values of the parameter $\chi$ are given in Fig. 52(a, b). One can observe localization of energy on the first oscillator at $\chi = 0.1667$ (Fig.

52(a)), and formation of intense beating at the higher value $\chi = 0.18$ (Fig. 52(b)). The LPTs are indicated by the bold solid lines.

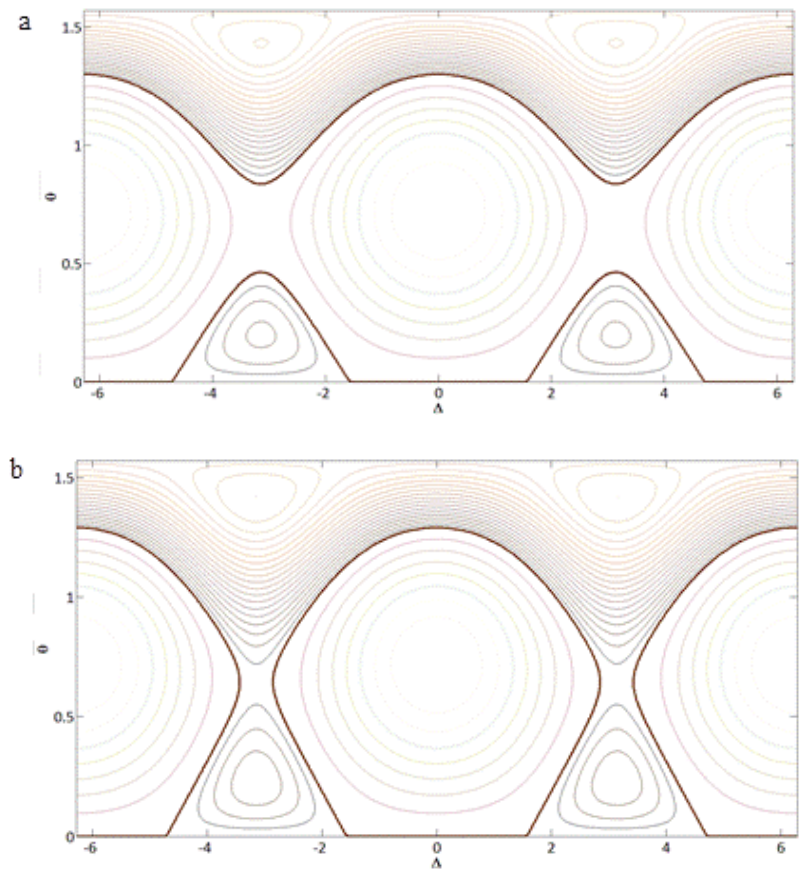

Fig. 52. Phase portraits of the reduced system with $\chi = 0.1667$ (a) and $\chi = 0.18$ (b).

It was shown in the previous sections that the transition from the LPT of the first type (moderate energy exchange) to the LPT of the second type (strong energy exchange) corresponds to the passage of the LPT of the first type through the separatrix. Using the same arguments as in Sections 2, 3, we derive the critical parameter corresponding to the passage through separatrix as $\chi = \chi_1 = 0.1746$.

We now consider the occurrence and the disappearance of the LPTs of the second type. We show that, at certain coupling strength, the LPT vanishes, yielding significant energy transfer to the third oscillator and generating chaotic energy transport in the entire system.

We restrict our consideration by the truncated system (7.9). Taking into account the resonant interaction between the LPT of the first pair of oscillators and the third oscillator, we formulate a robust analytical criterion for the annihilation of strong beating between the first and second oscillators.

It follows from Eqs. (7.9) that the unbounded growth of the response $\varphi_3$ can be attained under the condition of resonance, i.e., when the spectrum of $\cos\theta$ coalesces with that of $\exp(i\delta_2)$, thus yielding the secular growth of $\varphi_3$. First, we note that the phases $\delta_1(\tau)$ and $\delta_2(\tau)$ represent relatively small oscillations near a slowly increasing drift, namely,

$$\delta_j(\tau) = \omega\tau + \vartheta_j(\tau), j = 1, 2, \tag{7.12}$$

where $\omega$ is the average rate of rotation corresponding to the LPT, $\vartheta_j(\tau)$ is an oscillating function. Next, $\theta(\tau)$ is assumed to be a periodic function with the period $T_{mod} = 2\pi/\Omega$ corresponding to the period of modulation of strong resonant beating between the first and second oscillators. It now follows that the slow envelope $\varphi_3$ exhibits the secular growth if the angular frequency $\omega$ coincides with the modulation frequency $\Omega$, that is, $\omega = \Omega$. This condition allows one to determine the critical value $\chi_2$ from the equality

$$\omega(\chi_2) = \Omega(\chi_2). \tag{7.13}$$

The derivation of analytical expressions for $\omega$ and $\Omega$ as well as the analytical calculation of $\chi_2$ are beyond the scope of this report. To illustrate the criterion for the second transition phase we used Fast Fourier Transform to calculate both the frequency of modulation $\Omega$ and the angular frequency $\omega$ for the two distinct values of coupling $\chi = 0.3 < \chi_2$ and $\chi = 0.365 = \chi_2$. The results are summarized in Table 1.

| $\chi$ | $\Omega$ | $\omega$ |
|--------|----------|----------|
| 0.3 | 0.67 | 0.83 |
| 0.365 | 0.835 | 0.835 |

**Table 1**. Dependence of the frequency of modulations $\Omega$ and angular frequency $\omega$ on coupling strength $\chi$

### 7.1.4. Numerical results

Numerical results of this paragraph illustrate the occurrence of the aforementioned quasi-periodic LPTs due to gradual increase of the coupling parameter $\chi$. In addition to quasi-periodic LPTs, we demonstrate complete spontaneous energy delocalization for the high values of coupling. The numerical results have been obtained for system (7.1) with initial conditions (7.2).

1. *Quasi-periodic LPTs of the first type*. Figure 53 depicts instantaneous energy of each oscillator for weak coupling $\chi < \chi_1$. One can observe significant energy localization on the first oscillator together with weak energy exchange between the oscillators (weak modulation). In analogy to the 2DOF system, the response with significant energy localization on the first oscillator is further referred to as the quasi-periodic LPT of the first type.

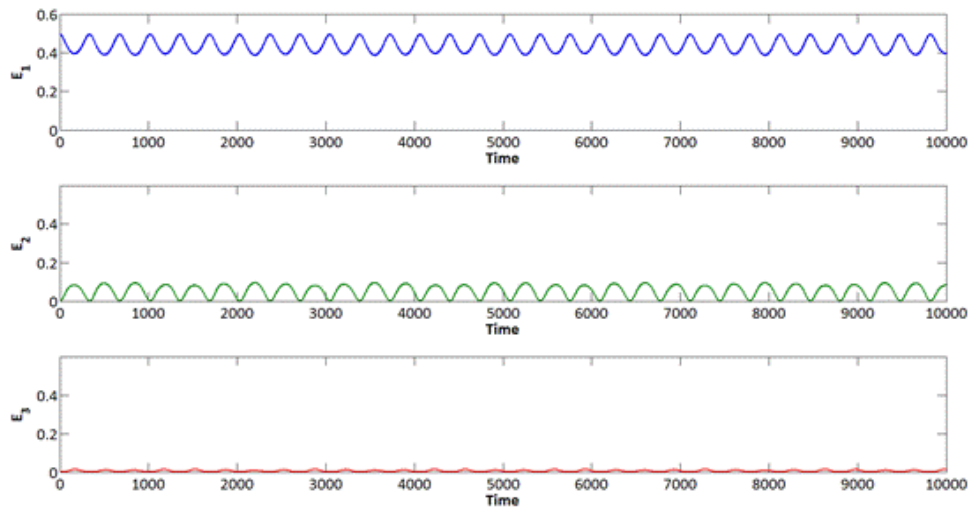

**Fig. 53.** Instantaneous energy recorded on each of the oscillators with coupling strength $\chi = 0.11 < \chi_1 = 0.175$.

2. *Quasi-periodic LPTs of the second type*. Given that the coupling strength $\chi > \chi_1$, we reveal a global change of the response (Fig. 54). Instead of energy localization on the first oscillator or alternative energy delocalization (i.e., energy exchange through the whole chain), the formation of an intermediate state characterized by regular strongly modulated beating between the first and the second oscillators is observed.

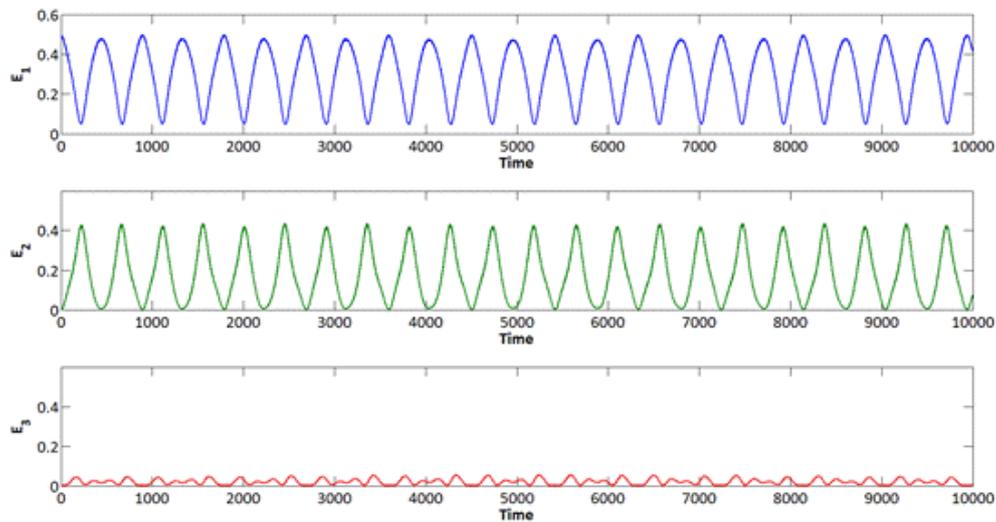

**Fig. 54.** Instantaneous energy recorded on each oscillator provided the coupling strength $\chi = 0.18$ ($\chi_1 < \chi < \chi_2$) is above the first threshold $\chi_1 = 0.175$ and below the second threshold $\chi_2 = 0.36$.

It is noted in Fig. 54 that an insignificant amount of energy is transported to the third oscillator. This type of highly modulated response with significant energy localization on the first and the second oscillators is referred to as a quasi-periodic LPT of the second type.

*3. Complete energy delocalization.* Further increase of the coupling parameter entails an additional global change of the system response. In this case, regimes of regular local pulsations (quasi-periodic LPTs) are completely destroyed, yielding irregular chaotic behavior in the chain (Fig. 55).

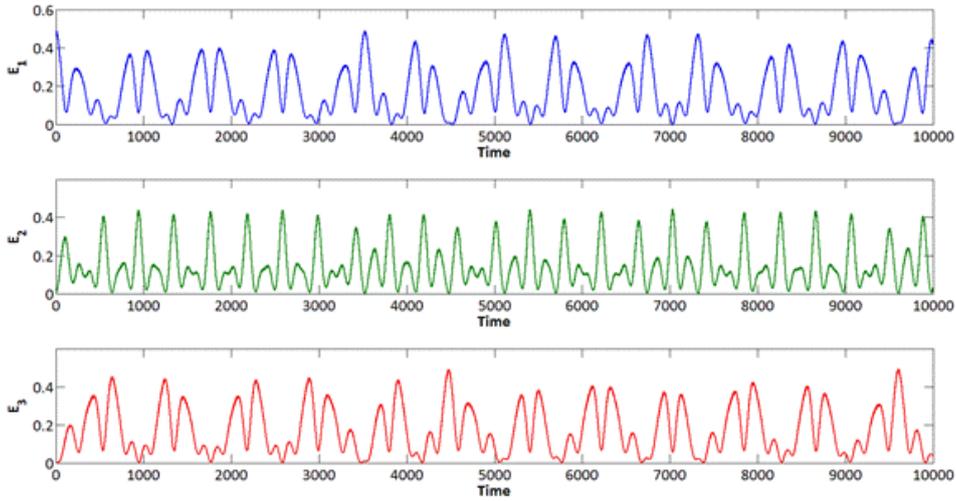

**Fig. 55.** Instantaneous energy recorded on each of the oscillators at $\chi = 0.33 < \chi_2 = 0.36$

The diagrams in Fig. 56 demonstrate consecutive transitions between the locally pulsating regimes as well as the emergence of highly delocalized ones. At $\chi = \chi_1 = 0.175$ we observe the first transition from energy localization on the first oscillator to the intermediate state of energy localization on the first two oscillators with intense energy exchange between them. The second transition at $\chi \approx \chi_2 = 0.365$ is from the moderately localized state (strongly pulsating regimes) to complete energy delocalization, i.e., maximal amplitudes of the response recorded on all the three oscillators are nearly equal.

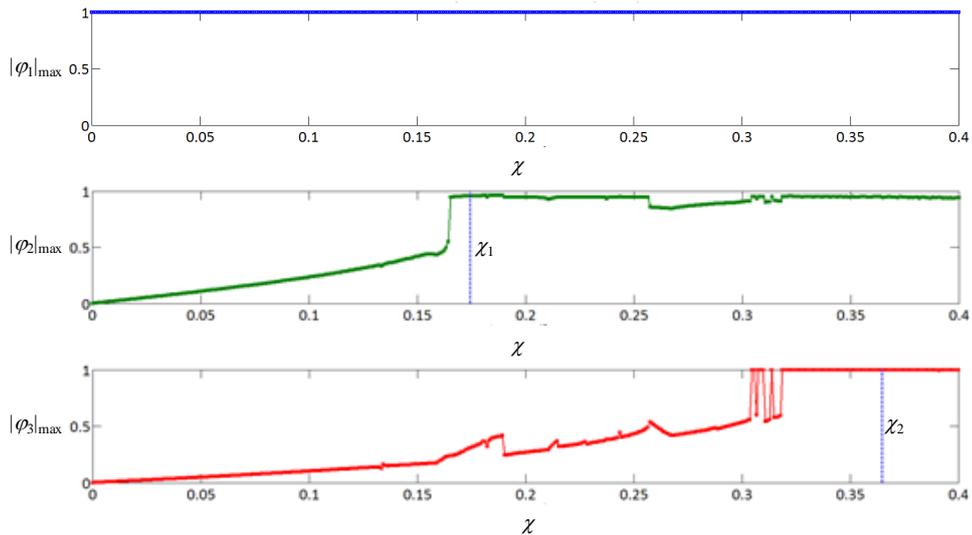

**Fig. 56.** Transition diagrams

It is obvious that the first transition is well approximated in the framework of the LPT concept for the simplified 2DOF model, wherein the effect of the third weakly excited oscillator introduces asymmetry similar to that considered in the 2DOF oscillators (Section 3). As a result, the numerical critical value for the 3DOF system $\chi_{1\text{num}} = 0.165$ is slightly less than the critical value $\chi_{1\text{LPT}} = 0.175$ defined for the 2DOF system (the difference is of the order of 6%). The point of the second transition from the localized to entirely delocalized states is also in a fairly good agreement with the transition diagram, namely, the numerical critical point $\chi_{2\text{num}} = 0.31$ is close to the theoretically found value $\chi_{2\text{th}} = 0.36$ (with the difference ~ 14%).

### 7.1.5. Spatially localized pulsating regimes

To assess the global dynamics of regular regimes as well as transition to chaos, we construct the Poincaré sections for the slow-flow model (7.7) at fixed energy levels.

As shown above, the global system dynamics can be reduced to the three-dimensional (3D) manifold. To this end, we fix the value of the Hamiltonian $H$ to a constant $h$, thus restricting the flow of (7.7) to the 3D isoenergetic manifold.

$$H_{4D}(\theta, \phi, \Delta_{12}, \Delta_{23}) = h. \tag{7.14}$$

Transversely intersecting the 3D manifold by a two-dimensional (2D) cut plane $S$: $\{\theta = \theta_0\}$, we construct the 2D Poincaré map $P: \Sigma \rightarrow \Sigma$, where the Poincaré section is defined as

$$\Sigma = \{\theta = \theta_0, d\theta/d\tau_1 > 0\} \cap \{H_{4D}(\theta, \phi, \Delta_{12}, \Delta_{23}) = h\}, \tag{7.15}$$

the restriction $d\theta/d\tau_1 > 0$ is imposed to indicate the orientation of the Poincaré map.

Despite apparent limitations, the constructed Poincaré maps faithfully reveal periodic orbits. The fundamental time-periodic solutions of a basic period $T$ corresponds to the period-1 equilibrium points of the Poincaré map, i.e., to orbits of system (7.7), which recurrently pierce the cut section at the same point. Additional subharmonic solutions of periods $nT$ may exist corresponding to the period-$n$ equilibrium points of the Poincaré map, i.e., to orbits that pierce the cut section $n$ times before repeating themselves. In the present study, we construct the Poincaré map $P : \Sigma \rightarrow \Sigma$ such that the global system dynamics is mapped onto the $(\Delta_{12}, \phi)$ plane.

To study consequent transitions from the highly localized to delocalized states, we set the value of the integral of motion corresponding to the initial energy localization on the first oscillators. This restriction corresponds to the following choice of the initial conditions for the reduced slow flow model: $\theta(0) = \pi/2$, $\phi(0) = \Delta_{12}(0) = \Delta_{23}(0) = 0$. This choice of initial conditions yields the following value of the second integral of motion

$$H_{4D}(\pi/2,0,0,0) = H_{\chi} = \frac{i}{2}(1 - \frac{\chi}{2}). \tag{7.16}$$

The Poincaré sections illustrated in Figs. 57 - 61 correspond to different values of $\chi$ and $H_{\chi}$. On each Poincaré diagram, the points of the quasi-periodic LPT are denoted with red color.

Starting from coupling $\chi < \chi_1$, we obtain the orbits encircling the center (Fig. 57). The center corresponds to the periodic LPT characterized by weak energy pulsations and mostly localized on the first oscillator. This response, referred to as the periodic LPT of the first type, allows us to determine the transition from intensive energy exchange to energy localization in the reduced 2DOF model. The boundary of the resonant island corresponding to this type of the LPTs is characterized by stronger modulation, which, however, does not describe the maximal possible energy transport between the oscillators (quasi-periodic LPT of the first type). Note that this conclusion is in the perfect agreement with the results obtained for the 2DOF systems.

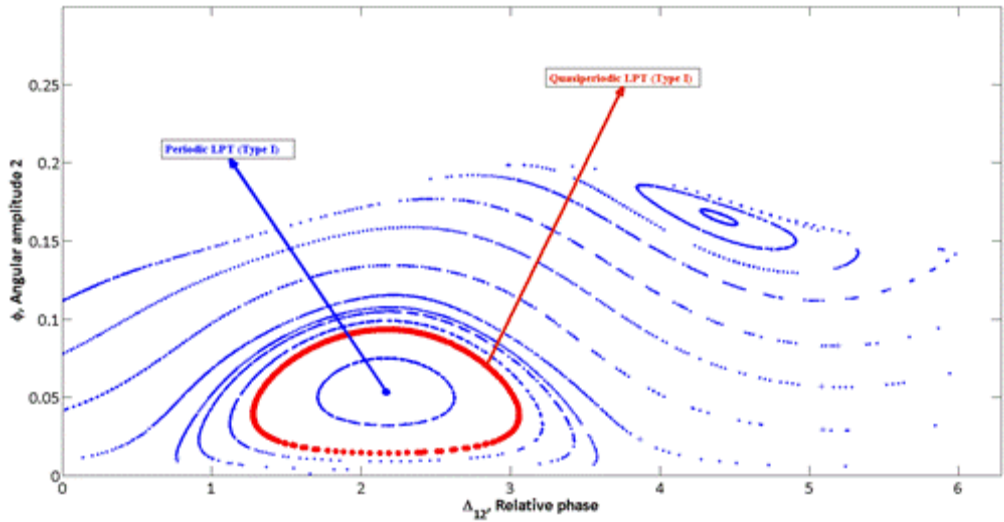

**Fig. 57.** Poincaré section: system parameters: $\theta_0 = 1.4$, $\chi = 0.15$.

An increase of coupling strength $\chi$ entails the growth of the quasi-periodic LPT (Fig.58), which then destructed while passing the chaotic region (Fig. 59). The annihilation of this regular orbit corresponds to the first transition from energy localization on the first oscillator to recurrent strong energy exchange between the first and second oscillators. It is worth noting the intermittent nature of highly pulsating regime emerging right after the breakdown of the quasi-periodic LPT. This intermittency is emphasized by its chaotic-like behavior. Naturally, this peculiarity cannot be observed in the framework of the reduced model.

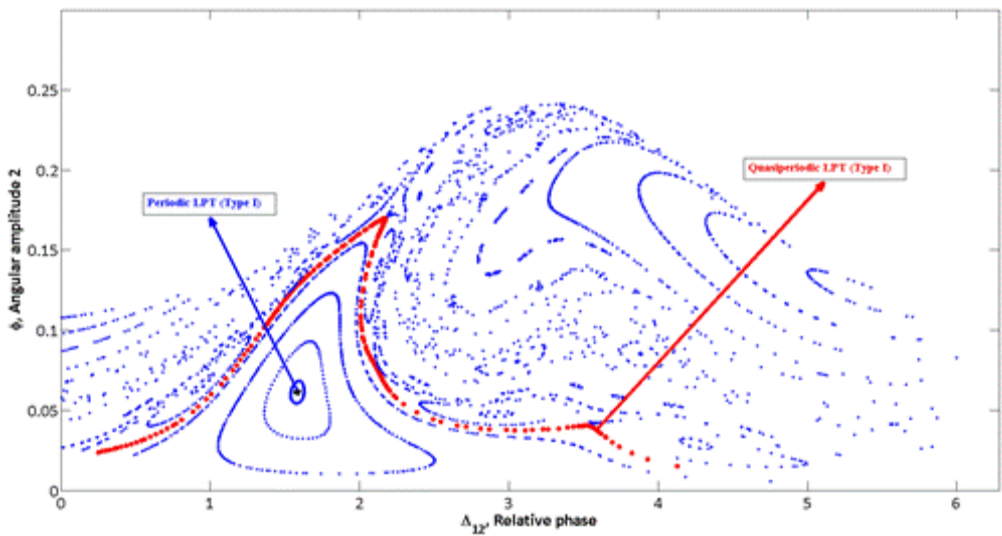

**Fig. 58.** Poincaré section: system parameters: $\theta_0 = 1.23$, $\chi = 0.1632$.

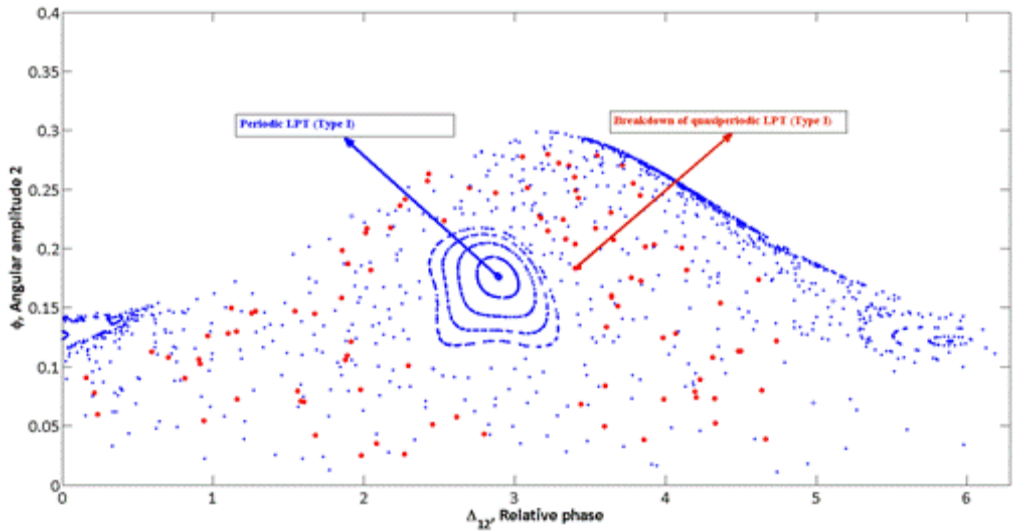

**Fig. 59.** Poincaré section: system parameters: $\theta_0 = 1.23$, $\chi = 0.18$.

Further increase of the strength of $\chi$ results in the termination of the intermittent response with the following formation of the quasi-periodic LPT of the second type (Figs. 60, 61) encircling the new center. This stationary point corresponds to the perfectly synchronous periodic LPT with strong energy exchange between the first two oscillators. As discussed above, the second transition (i.e. breakdown of local energy pulsations spanned over the first two oscillators) is characterized by chaotic energy transport between all oscillators. This spontaneous transition of the second kind is manifested by a sudden blow up of the regular quasi-periodic LPT of the second type.

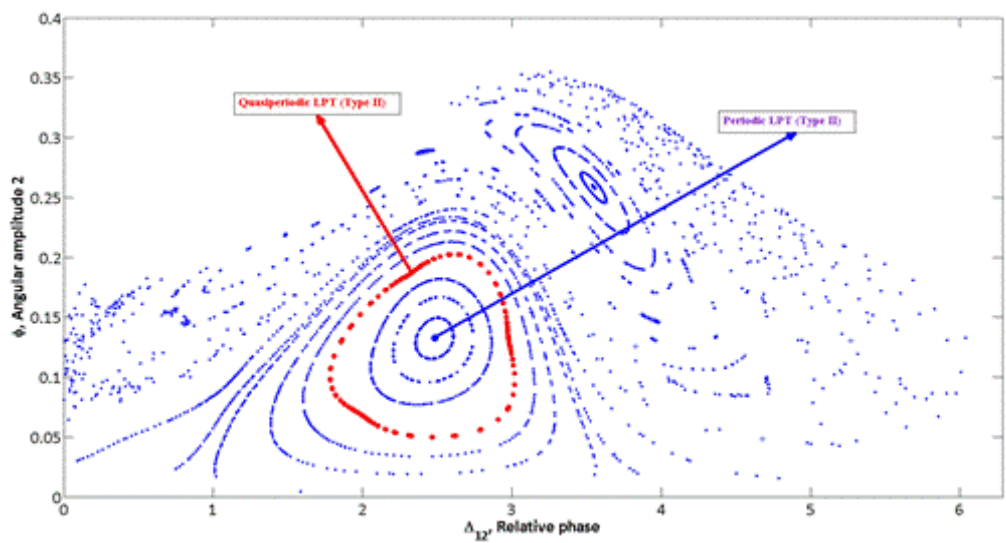

**Fig. 60.** Poincaré section: system parameters: $\theta_0 = 1.23$, $\chi = 0.1991$.

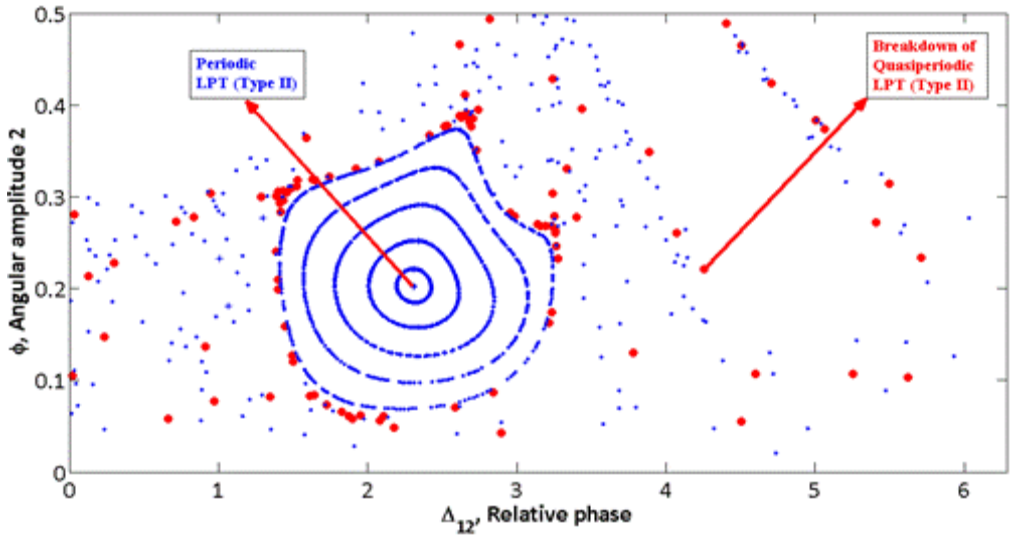

**Fig. 61.** Poincaré section: system parameters: $\theta_0 = 1.23$, $\chi = 0.31$.

We will end this section with some observations and remarks concerning the relation of the quasi-periodic LPTs in the 3DOF system to the general theory of LPTs. In the case of two coupled anharmonic oscillators, emergence of the LPTs (of both types) is determined by an initial excitation applied to the first oscillator. The role of the coupling parameter in the 2DOF and 3DOF systems is similar. Choosing the coupling strength between the two oscillators below a certain threshold entails the formation of the LPT of the first kind with energy localization on the first oscillator and weak energy exchange between the oscillators. However, increasing the strength of coupling above the threshold leads to global bifurcations of the LPT of the first kind manifested by the formation of the LPT of the second kind. The latter falls under category of pure beating with complete energy exchange between the oscillators..

However, the dynamics of the 3DOF system is more complicated than the behavior of its 2DOF counterpart. In particular, the mechanism of transition from a locally pulsating quasi-periodic LPT to strong energy exchange is quite different. The Poincaré sections demonstrate that the first transition occurs due to the entering of the quasi-periodic LPT to a chaotic region bringing about the intermittent response corresponding to almost irreversible energy transfer from the first to the second oscillator. Increasing the coupling strength, we observe the emergence of a new quasi-periodic LPT (of the second type) characterized by the recurrent energy exchange between the first and the second oscillators. The breakdown of strong local energy exchange occurs due to the penetration of the LPT orbit of the second type into a chaotic region. It is clear from the consideration of the Poincaré sections that the birth and the death of regular quasi-periodic LPTs of the first and the second kind goes not through their reconnection (as is in the case of the periodic LPTs) but through the sudden formation of the stable strongly pulsating regime. The mechanism of formation and bifurcations of the synchronous pulsating regimes is beyond the scope of this report.

Thus, the possibility of an extension to multi-particle systems of the LPT concept dealing with intensive energy exchange between the oscillators and its localization on the initially excited oscillator, as well as with the mechanism of its breaking is completely clarified. The reduced 2DOF model plays a key role here, as it does not depend on the number of the weakly interacting oscillators. However, with an increase of the number of particles, another significant possibility of the extension of the LPT concept appears. It is strongly connected with the notion of *effective particle*. Brief presentation of this theory is given below.

## 7.2. Higher dimensional oscillator chains: Brief Discussion

It is important to note that the LPT concept is by no means limited to the analysis of the low-dimensional models considered in the previous sections. It may be a powerful tool in the study of the processes of energy transport and localization in rather complex dynamical models such as finite and infinite oscillatory chains, scalar models, etc. (see, e.g., [57,105,122, 175,176,184,186]). In this section we briefly discuss recent advances in this field.

One of the well-known and intensively studied systems in nonlinear physics is the Fermi–Pasta–Ulam (FPU) oscillatory chain [41]. Long- and short-wavelength approximations of the infinite FPU chain result in the well-known nonlinear integrable PDEs, namely,

Korteweg-de Vries (KdV) and nonlinear Schrodinger (NSE) equations, respectively [166]. In the limit of an infinite chain, these equations possess dynamical solitons (KdV) and envelope solitons (NSE) as their particular solutions. As for the finite FPU chains, their dynamics has been studied in terms of NNMs (see, e.g., [29,34,41,43,49,60,102,150,153,154,166]). In fact, almost all studies of the finite chain dynamics have been concerned with the analysis of stationary processes. We show that the analysis based on the LPT concept is a proper tool to clarify the fundamental problem of energy transport in complicated multidimensional discrete systems.

We recall that the study of FPU oscillator chains was a starting point for the discovery of solitons [88]. The finite periodic homogeneous $\beta - $FPU model is defined by the Hamiltonian

$$H_0 = \frac{1}{2} \sum_{j-1}^{N} [p_j^2 + \frac{1}{2}(q_{j+1} - q_j)^2 + \frac{\beta}{4}(q_{j+1} - q_j)^4], \qquad (7.17)$$

where $q_j$ and $p_j$ are the coordinates and conjugate momenta, respectively, $N$ - the number of particles and ($q_{N+1} = q_1, p_{N+1} = p_1$). As shown in [122], this chain can exhibit complicated dynamical processes such as intense energy exchanges between the two halves of the chain (Fig. 62a) as well as permanent energy localization on one of the halves (Fig. 62b). In the same work [122] it was demonstrated that for each fixed initial energy imparted in one of the halves of the chain there exists a threshold in stiffness nonlinearity, below which a recurrent energy exchange between the two groups of particles (i.e. two halves of the chain) occurs.

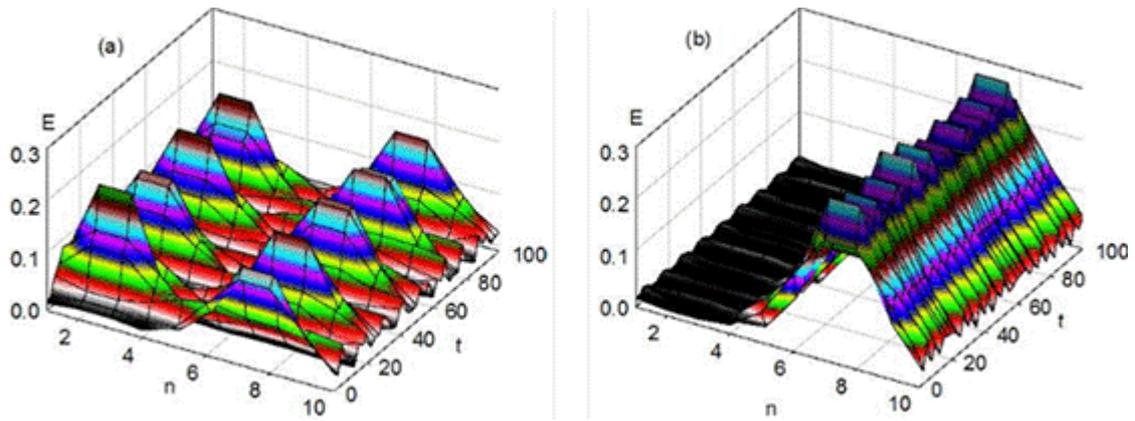

**Fig.62.** 3D diagram of energy distribution vs. time and space in the periodic FPU chain: (a) - complete energy exchanges between the two halves of the chain; (b) - energy localization on a single half of the chain

The regime of complete energy exchange terminates at a certain excitation threshold, and after its crossing the energy exchange change to the energy confinement in one of the halves of the chain. It was shown [122] that despite the complexity of the FPU chain, this dynamical transition from energy exchange to energy localization can be fully predicted using the LPT concept.

In particular, the use of the LPT concept leads to the notion of "effective particles" consisting of more than one real particle. Then the energy exchange in finite oscillatory chain can be simply depicted by a well-known beating response in the system of two weakly coupled effective particles similar to the system of two weakly coupled oscillators.

Also, it was shown [122] that the mechanism of the formation of intense energy exchange in the chain can also be understood through the internal resonant interactions occurring between different groups of the elements in the chain. Thus, with gradually increasing the number of particles in the chain, the distribution of the modal frequencies become more and more dense until the three uppermost frequencies of the spectrum become closely placed, creating the perfect conditions for the resonant interaction between two different groups of particles (exactly, two halves of the chain), which results in the intense energy exchange. It is important to emphasize once again that the regime of strong energy exchanges occurring in the homogeneous chains is possible only for sufficiently long chains (at least 8 particles [1]) such that the modal frequencies become closely spaced.

One of the key changes after the aforementioned dynamical transition (i.e., dynamical transition from energy exchanges to permanent energy localization) is the transformation of small oscillations of the center of the localization area into directed motion (traveling breather). Also, since the domain of strong localization does not correspond to any stationary point in the phase space of the system, the profile of the energy distribution along the chain demonstrates the periodic variations. A further growth of the excitation does not result in any qualitatively changes in the systems behavior, which might increase or decrease the areas corresponding to both partial energy exchange and weak localization. It should be noted that an increase of the particles number in the chain involves the new modes into the resonance interaction with a zone boundary mode because of the spectrum crowding in the high frequency range (Fig. 63). It does not lead to any noticeable changes except some "narrowing" of the effective oscillator. In the limiting case of the infinite chain, the effective oscillator is transformed into the well-known breather. At that, both energy thresholds discussed above tend to zero. This means that the resonant nature of energy localization can be better clarified while dealing with the finite chain.

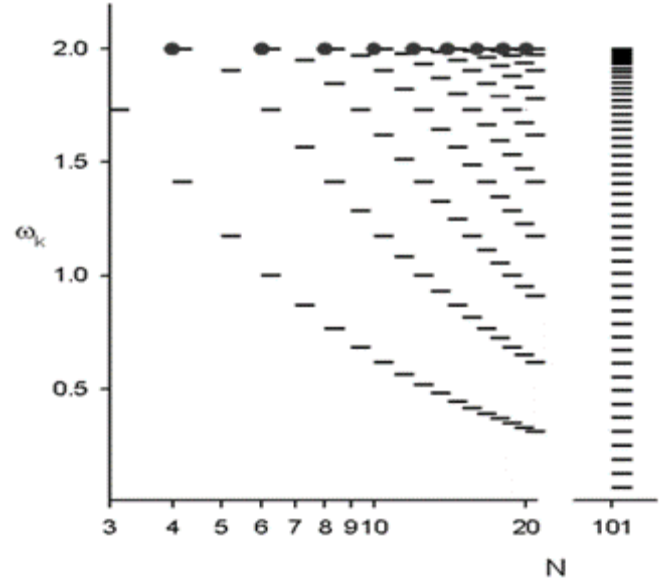

**Fig. 63.** Spectrum of modal frequencies versus number of particles in the FPU chain (N − stands for the number of elements in the chain, $\omega_k$ is a frequency corresponding to the k-th mode.)

The same LPT methodology, successfully applied for the analysis of energy transport in the FPU chains, can be generalized to additional class of nonlinear chains, namely Frenkel-Kontorova (FK) and Klein-Gordon (KG) models [175]. Unlike the FPU chains, the Hamiltonian for these models, along with the gradient component depending on deformations (in this case, quadratic), contains a component depending on the displacements. For example, the Hamiltonian of the systems considered in [175] is defined as follows:

$$H_0 = \frac{1}{2}\sum_{j=1}^{N} p_j{}^2 + \frac{1}{2}\left(q_{j+1} - q_j\right)^2 + V\left(q_j\right),$$ (7.18)

where $V\left(q\right) = \omega_0{}^2\left(1 - \cos(\frac{2\pi q}{d})\right)$ corresponds to the FK model, and $V(q) = \frac{\omega_0^2}{2}q^2 + \frac{\beta}{4}q^4$ corresponds to the KG model.

Contrary to the FPU lattice, the spectrum of the Klein–Gordon model is bounded below by the frequency $\omega_0$, which corresponds to in-phase oscillations of the particles as a whole in the local potential of the substrate. Thus, the spectrum of the Klein–Gordon lattice includes two boundary frequencies corresponding to the wave numbers $k = 0$ and $k = N/2$. Similarly to the case of the $\pi$-mode, the density of the eigenfrequencies near the end grows quickly with increasing number of particles. Therefore, for systems containing more than 10–15 particles, resonance conditions are satisfied near the left end of the spectrum.

Figure 64 depicts the results of simulation for the periodic system with 20 particles in the FK model. The surface of the total particle energy is demonstrated, with the particle numbers and the time in units corresponding to the period of self-oscillations of the lower boundary mode shown along the axes. Excitation levels go from left to right as follows: to the threshold of loss of stability of the boundary mode, after the loss of stability below the localization threshold, and above the localization threshold.

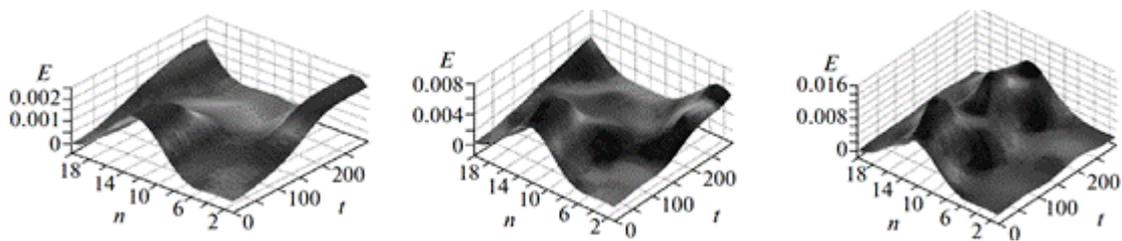

**Fig. 64.** 3D diagram of energy distribution vs. time and space in the periodic F-K model consisting of 20 elements.

It can be clearly seen that, below the threshold of true localization, the energy originally concentrated in the central part of the chain (maximum at approximately $12^{th}$ particle) is transmitted to the particles with the low numbers, while above the localization threshold the energy gets permanently localized in the central part of the chain. At the same time, the simulation performed for the KG model in this spectral region shows the absence of such localization processes. Again, all the aforementioned transitions from energy localization to the recurrent energy transport could be predicted analytically using the same concepts of 'effective particles' and LPT.

Recently a somewhat similar study of the formation of recurrent energy transport as well as the transition to energy localization has been pursued in the asymmetric FPU chain ($\alpha - \beta$ FPU chain) modeling the realistic interatomic interactions [176]. Here, despite the complexity of the model an analytical description of energy exchange between the 'effective particles' (specially introduced for the $\alpha - \beta$ FPU chain) have been successfully developed using the same LPT concept.

Study of resonant energy transport by means of the LPT was also extended to the diatomic FPU chain [186]. The system under consideration represents the FPU dimer chain subjected to periodic boundary conditions and composed of two identical cells of particles (Fig. 65).

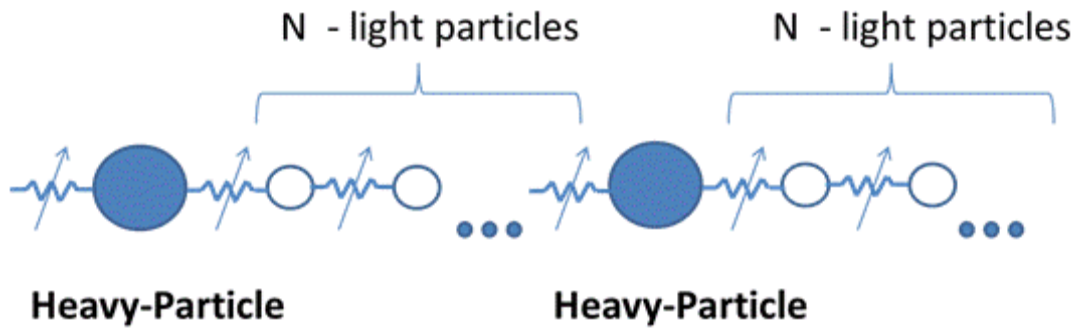

**Fig. 65** Scheme of diatomic FPU chain.

The system under consideration is described by the following set of equations

$$
\begin{aligned}
\ddot{x}_1 \quad &= x_{2N+2} - 2x_1 + x_2 \quad\quad + \varepsilon\alpha\tilde{F}_1 \\
\varepsilon\ddot{x}_2 \quad &= x_1 \quad\ - 2x_2 + x_3 \quad\quad + \varepsilon\alpha\tilde{F}_2 \\
&\quad\ \dots \\
\varepsilon\ddot{x}_{N+1} &= x_N \quad - 2x_{N+1} \quad + x_{N+2} + \varepsilon\alpha\tilde{F}_{N+1} \\
\ddot{x}_{N+2} &= x_{N+1} - 2x_{N+2} \quad + x_{N+3} + \varepsilon\alpha\tilde{F}_{N+2} \\
\varepsilon\ddot{x}_{N+3} &= x_{N+2} - 2x_{N+3} \quad + x_{N+4} + \varepsilon\alpha\tilde{F}_{N+3} \\
&\quad\ \dots \\
\varepsilon\ddot{x}_{2N+2} &= x_{2N+1} - 2x_{2N+2} + x_1 \quad + \varepsilon\alpha\tilde{F}_{2N+2}
\end{aligned}
\tag{7.19}
$$

where $x_k = x_{k+2N+2}$ , $\tilde{F}_j = (x_{j-1} - x_j)^3 - (x_j - x_{j+1})^3$ is the nonlinear part of the inter-particle interaction force, $x_i$ is the displacement of the $i$-th particle of the chain, $\varepsilon$ is the normalized mass ratio between the light and heavy particles of the chain and defined as a small parameter ($0 < \varepsilon << 1$), $\alpha \sim O(1)$ is the parameter of nonlinear stiffness.

Each cell of the chain comprises exactly one heavy particle followed by a group of $N$ light particles of identical masses (i.e. 1:$N$ dimer chain). It was shown [186] that the resonance inter-particle interaction occurring on the optical branch (in the limit of the strong mass mismatch $\varepsilon = m/M \ll 1$) leads to the dynamical transitions undergone by the entire chain.

We now highlight the main features of the regimes corresponding to the optical and acoustic branches. The regimes of the acoustic branch are manifested by the in-phase motion of each pair of neighboring elements. Moreover, the most part of the system energy is carried by the heavy particles. In contrast to the case of the acoustic branch, the regimes of the optical branch are manifested by the anti-phase motion of each pair of neighboring elements, as well as the low amplitude oscillations of the heavy particles. Then, it can be shown that in the limit of the strong mass mismatch $\varepsilon \ll 1$, for any particular regime corresponding to the acoustic branch, the energy carried by the light particles is negligibly small in comparison to the energy carried by the heavy particles. However, in the same limit of $\varepsilon \ll 1$, the regimes of an optical branch are characterized by a negligibly small fraction of energy carried by the heavy particles.

The developed analysis [186] demonstrated the existence of special regimes belonging to the optical branch and leading to energy transport between the groups of light particles when the heavy ones remain nearly stationary. Applying the LPT concept, this study reveals a threshold value of nonlinearity, above which energy imparted in one of the groups, remains permanently localized within the same group without leaking to the other parts of the chain. Moreover, it was demonstrated that the aforementioned regime of strong energy exchange between the groups of light particles exists for an arbitrary number of elements in each of the groups. This result principally differs from those obtained for a homogeneous chain [122], where the system is required to include a sufficient number of particles to ensure the formation of beating phenomenon between the two halves of the chain. Thus, the dynamics of the dimer chain reveals a new mechanism of formation of highly non-stationary regime leading to a massive transport of energy between the two groups of light elements.

The analysis carried out in [186] has shown that the mechanism of formation of strong beats is fully governed by the mass ratio between the light and heavy particles. Thus, decreasing the mass ratio, one can effectively create the closely spaced pairs of modes situated on the optical branch. These pairs are responsible for the resonant interactions between the groups of light particles leading to complete energy exchanges. It is worth emphasizing that formation of beats between the groups of light particles is not the only possible mechanism of a massive energy transport in the FPU dimer chain. One can easily show that, operating on the acoustic branch under the assumption of a strong mass mismatch ($\varepsilon \rightarrow 0$), the FPU dimer chain effectively reduces to a homogeneous FPU chain (i.e., wherein the elements of the reduced chain correspond to the heavy particles of the dimer chain). Thus, according to [122], an inclusion of a sufficient number of heavy particles in the chain will allow for complete energy transport only between the two groups of heavy particles while the energy remaining in the light particles is negligibly small.

## 8. Conclusions

The accepted classification of the problems of mathematical physics (in the oscillation and wave theories) draws first of all a sharp distinction between linear and nonlinear models. This distinction is caused by understandable mathematical reasons, including the inapplicability of the superposition principle in the nonlinear case. However, in-depth physical analysis allows us to introduce another basis for the classification of ordered oscillation problems, focusing on the difference between the stationary (or non-stationary, but non-resonance) and resonance non-stationary processes. In the latter case, the difference between the linear and nonlinear problems is not fundamental, and a specific technique, efficient in the same degree for description of both linear and nonlinear resonance non-stationary processes, has been developed. The existence of an alternative approach in the framework of linear theory seems unexpected. Really, the superposition principle allows us to find a solution describing arbitrary non-stationary oscillations as a combination of linear normal modes, which correspond to stationary processes. However, in the systems of weakly coupled oscillators, in which resonant non-stationary oscillations can occur, another type of fundamental solution exists. It describes strongly modulated non-stationary oscillations characterized by the maximum possible energy exchange between the oscillators or the clusters of the oscillators (effective particles). Such solutions are referred to as Limiting Phase Trajectories (LPTs). This Report demonstrates that the LPT concept suggests a unified approach to the study of such physically different processes as strongly non-stationary energy transfer in a wide range of classical oscillatory systems and quantum dynamical systems with constant and time-varying parameters. Furthermore, this analogy paves the way for simple mechanical simulation of complicated quantum effects.

Finally, we highlight differences between the NNMs and the LPTs that have motivated the introduction and the development of the LPT concept:

| NNM | LPT |
|---|---|
| represents a stationary process independent of initial conditions | represents a nonstationary process dependent of initial conditions |
| is not involved in energy exchange | corresponds to maximum possible energy exchange between different parts of the system |
| can undergo local bifurcation | can undergo global bifurcation |
| can be localized (stationary localization) | can be localized (non-stationary localization) |
| can become an attractor in an active system(synchronization of a traditional type) | can become an attractor in an active system (synchronization of a new type) |
| corresponds to a steady solution in a system subjected to external periodic excitation | corresponds to maximum energy transfer from a source of external periodic excitation |
| is described by smooth sine-type functions | is described by non-smooth functions |
| can be extended to systems with infinite numbers of particles | cannot be extended to systems with infinite numbers of particles but can be considered as a prototype of a breather |

## Acknowledgements

We are grateful to many colleagues for discussing their ideas and related problems with us. Joint research and discussions with O. Gendelman, Yu. Kosevich, C.-H. Lamarque, A. Manevich, M. Kovaleva, A. Musienko, V. Pilipchuk, D. Shepelev, V. Smirnov, A. Vakakis, and many others were especially important. We thank all of them with great pleasure.

Two first authors (L.I.M. & A.K.) acknowledge support for this work received from Russian Foundation for Basic Research (Grant 14-01-00284) and Russian Academy of Sciences (Program No 1). The third author is grateful to Israel Science Foundation (Grant 484 /12) and German-Israeli Foundation (Grant No.2017895) for financial support